\documentclass[amsmath,amssymb,aps,pra,reprint,superscriptaddress,twocolumn,longbibliography]{revtex4-2}
\usepackage{amsmath,amssymb}
\usepackage[english]{babel}
\usepackage{bbold}
\usepackage{upgreek}
\usepackage{dsfont}
\usepackage{graphicx}
\usepackage{comment}
\usepackage{braket}
\usepackage{colortbl}
\usepackage{blindtext}
\usepackage{xr-hyper}
\usepackage{xcolor}
\usepackage[colorlinks]{hyperref}
\definecolor{bluegray}{RGB}{50, 127, 180}
\definecolor{darkblue}{RGB}{23, 40, 94}
\hypersetup{colorlinks=true,linkcolor=bluegray,citecolor=bluegray,urlcolor=darkblue,final}
\usepackage{bm}
\usepackage[utf8]{inputenc}
\allowdisplaybreaks
\usepackage{geometry}
\usepackage{natbib}
\usepackage{physics}
\usepackage[capitalize]{cleveref} 

\addto\captionsenglish{ }

\begin{document}

\title{Quadrupole-induced alignment and sensing with Meissner-levitated micromagnets}

\author{Zhiyuan Wei}
\thanks{These authors contributed equally to this work}
\affiliation{Institute for Complex Quantum Systems, Ulm University, Albert-Einstein-Allee, 89069 Ulm, Germany}

\author{Vanessa Wachter}
\thanks{These authors contributed equally to this work}
\affiliation{Institute for Complex Quantum Systems, Ulm University, Albert-Einstein-Allee, 89069 Ulm, Germany}

\author{Yiqi Wang}
\affiliation{Department of Physics, Harvard University, Cambridge, Massachusetts 02138, USA}

\author{Trisha Madhavan}
\affiliation{Harvard John A. Paulson School of Engineering and Applied Sciences, Harvard University, Cambridge, Massachusetts 02138, USA}

\author{Mikhail D. Lukin}
\affiliation{Department of Physics, Harvard University, Cambridge, Massachusetts 02138, USA}

\author{Benjamin A. Stickler}
\affiliation{Institute for Complex Quantum Systems, Ulm University, Albert-Einstein-Allee, 89069 Ulm, Germany}

\begin{abstract}
Micromagnets levitated above superconductors are promising probes of weak forces, torques, and magnetic fields. Here, we present a comprehensive theoretical description of the coupled translational and rotational dynamics of a Meissner-levitated micromagnet above a planar type-II superconductor, consistent with key features observed in recent experiments. Specifically, we show that an anisotropic magnetic quadrupole moment lifts the rotational degeneracy of the dipole trapping potential, enabling confinement of all six rigid-body degrees of freedom, and that external magnetic fields tune the resulting normal modes and give rise to multistable trapping configurations. The intrinsic angular momentum associated with the magnetization introduces an additional gyroscopic coupling, producing phase-shifted mode mixing even in the absence of mechanical spinning. Using the coupled linear response, we derive noise-equivalent force and magnetic-field sensitivities for displacement readout, accounting for  magnetic noise generated by the superconductor. These results establish how magnetic multipoles and spin-rotation coupling shape the mechanical response and sensing performance of Meissner-levitated micromagnets, providing a framework for precision sensing and probing  surface magnetic fluctuations. 
\end{abstract}

\maketitle

\section{Introduction}\label{Introduction}
Magnetically levitated particles are emerging as a promising platform for precision sensing and tests of fundamental physics~\cite{rossi2026levitodynamics,millen2020quantum,perdriat2021spin, stickler2021quantum}, ranging from diamagnetically levitated objects~\cite{hsu2016cooling,slezak2018cooling,o2019magneto,lewandowski2021high,leng2021mechanical,tian2024feedback,kim2025magnetically,childress2017cavity,brown2023superfluid,romero2012quantum,pino2018onchip,latorre2020chip,latorre2022chip,Gutierrez2023superconducting,Hofer2023highq,schmidt2024remote,hansen2026optical,smit2026superconducting,arrayas2026superconductinglevitatingoscillator,koller2026controlling} to magnetic particles in electric \cite{huillery2020spin}, magnetic \cite{perdriat2023planar,janse2024characterization,janse2026chip,yakymenko2026levitation}, and superconducting traps~\cite{wang2019dynamics,timberlake2019acceleration,vinante2020ultralow,gieseler2020single,ahrens2025levitated,wang2026exploring,wang2026nonlinear,fuwa2023ferromagnetic}. In contrast to optical trapping,  magnetic levitation can provide passive confinement without the absorption and photon-recoil heating associated with trapping light, supporting larger masses, cryogenic operation, and ultralow dissipation~\cite{cirio2012quantum,romero2012quantum,prat2017ultrasensitive,vinante2022levitated}. Such low mechanical dissipation and a direct magnetic response make these systems attractive probes of weak forces, torques, and magnetic fields~\cite{vinante2022levitated,ni2025microscopic,ji2026levitated}. These properties motivate applications in magnetometry~\cite{jackson2016precessing,kumar2017magnetometry,band2018dynamics,vinante2021surpassing,fadeev2021ferromagnetic,kalia2024ultralight,ni2025microscopic,renLevitatedFerromagneticTorsional2026}, acceleration and gravity sensing~\cite{prat2017ultrasensitive,timberlake2019acceleration,lewandowski2021high,timberlake2021probing,fuchs2024measuring,carney2025superconducting,amaral2026magnetic,fadel2026probing}, searches for physics beyond the Standard Model~\cite{wang2019dynamics,zheng2020room,vinante2022levitated,yinExperimentsLevitatedForce2022,higginsMaglevDarkMatter2024,ren2026field,uitenbroek2026first}, and proposals for macroscopic quantum experiments and tests of the quantum nature of gravity~\cite{romero2012quantum,gonzalez2020quantumAcoust,pedernales2020motional,streltsov2021ground,lami2024testing,bulling2026stability}. 

\begin{figure}[htp]
    \centering
    \includegraphics[width=\columnwidth]{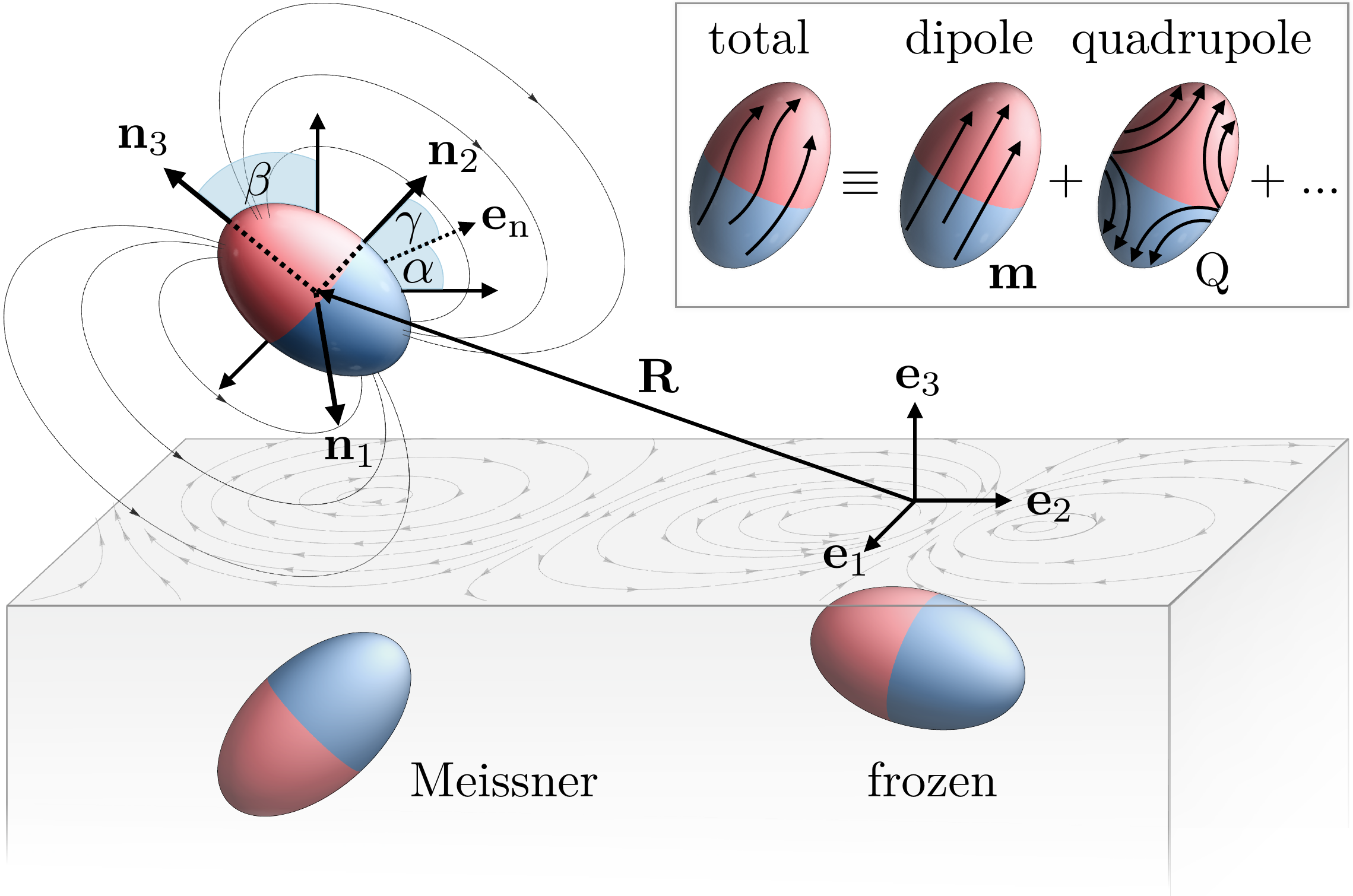}
    \caption{\textbf{Levitation setup.}
    Micromagnet levitating above a type-II superconductor with surface normal $\mathbf{e}_3$.  The particle's configuration is specified by its center-of-mass position  $\mathbf{R}$ and Euler angles $(\alpha,\beta,\gamma)$, which relate the body-fixed principal axes $\mathbf{n}_k$ to the laboratory axes $\mathbf{e}_k$. 
   Reversible screening currents produce the repulsive Meissner interaction, with their exterior field represented by an instantaneous Meissner image. The frozen magnetic field generated by supercurrents around pinned vortices is represented by a fixed inverse mirror image. Inset: schematic magnetic multipole expansion, retaining the magnetic dipole moment $\mathbf{m}$ and quadrupole tensor $\mathrm{Q}$.}
    \label{fig:experimental-setup}
\end{figure}

Meissner-levitated micromagnets are particularly appealing for sensing, with exceptionally low mechanical damping demonstrated experimentally~\cite{vinante2020ultralow,vinante2022levitated,jose2026ultrafast}. In this system, Meissner screening  produces the repulsive lifting force that supports levitation, while flux pinning in type-II superconductors can provide additional restoring forces and torques~\cite{vinante2020ultralow,gieseler2020single,vinante2022levitated,ahrens2025levitated,wang2026nonlinear,wang2026exploring}. The resulting trapping potential couples the magnet's translational and librational motion, allowing for magnetic torque detection through center-of-mass displacement and enabling external magnetic fields to tune the mechanical response. Recent experiments have explored a variety of phenomena in these systems, including gyromagnetic coupling between librational modes~\cite{ahrens2026observation}, full orientational alignment and magnetic tuning of mode frequencies~\cite{wang2026nonlinear}, the generation of mechanical frequency combs~\cite{wang2026nonlinear}, and the influence of flux-pinning dynamics~\cite{wang2026exploring}. Understanding and exploiting these phenomena requires a  quantitative theoretical description of the full rigid-body dynamics near the superconducting surface.

Two distinct features of a levitated magnet are essential to this description. First, its magnetic structure determines the orientational confinement. In particular, the point-dipole interaction is invariant under rotations about the magnetic-moment axis and therefore leaves one rotational degree of freedom unconstrained. Higher magnetic multipoles can break this symmetry, providing a mechanism for the experimentally observed confinement of all six rigid-body degrees of freedom~\cite{wang2026nonlinear}. Second, magnetization distinguishes levitated magnets from nonmagnetic mechanical resonators in a more fundamental way: it carries intrinsic angular momentum that contributes to the total angular momentum of the particle~\cite{vanvleck1951}. The resulting spin-rotation coupling links magnetization to rigid-body motion and underlies the Einstein-de Haas and Barnett effects~\cite{bunker2006molecular,einstein1915experimental,barnett1915magnetization,chudnovsky1994conservation}, which can profoundly modify the rotational dynamics~\cite{rusconi2017quantum,rusconi2017linear,kustura2022stability,rusconi2022spin,wachter2021optical,wachter2026gyroscopically,perdriat2024rotational,sato2023gyromagnetic,jin2024quantum,ma2021torque,delord2020spin,huillery2020spin,steiner2024pentacene}. Remarkably, a magnet can therefore exhibit gyroscopic dynamics even in the absence of mechanical spinning. This intrinsic gyroscopic response forms the basis of proposals for ultrasensitive precessing ferromagnetic magnetometers~\cite{jackson2016precessing,band2018dynamics,ni2025microscopic}, and its signatures have recently been observed in the coupled librations of a nonspinning magnet levitated in a superconducting trap~\cite{ahrens2026observation}.

Thus, magnetic multipoles and intrinsic angular momentum affect the mechanical response through different mechanisms. The former modify the trapping potential, while the latter introduce gyroscopic, velocity-dependent coupling between the mechanical modes. Accounting for both mechanisms is therefore important for a quantitative description of levitated micromagnets as mechanical sensors.

Here, we develop a theoretical description of the full rigid-body dynamics and sensing response of a permanent micromagnet levitated above a type-II superconductor. Including magnetic dipole and quadrupole interactions, we 
show how an anisotropic magnetic quadrupole moment can break the rotational symmetry of the dipole potential, enabling confinement of all six translational and rotational degrees of freedom. We determine the resulting equilibrium configurations and mechanical modes, and show how external magnetic fields tune their frequencies and compositions. This allows control over intermodal hybridization, which we distinguish from the phase-shifted mode mixing generated by spin-rotation coupling. These results provide the basis for calculating the mechanical response of Meissner-levitated micromagnets to weak forces and magnetic fields. To assess their sensing performance, we account for magnetic noise generated by thermal currents and fluctuations of pinned vortices and derive the noise-equivalent force and magnetic-field sensitivities for a specified displacement readout. Our framework thus connects the tunable mechanical dynamics to the sensitivity of Meissner-levitated micromagnets and enables quantitative assessment of their potential for precision sensing and searches for weak interactions. 

The article is structured as follows. In Sec.~\ref{Main results}, we summarize our main results for the rigid-body dynamics and sensing performance of Meissner-levitated micromagnets, with detailed derivations presented in the subsequent sections. Specifically, in Sec.~\ref{Derivation of the Hamiltonian}, we derive the full Hamiltonian and trapping potential. Its minimum configuration and mechanical eigenmodes are discussed in Secs.~\ref{Trapping minimum} and \ref{sec:eigenmodes}. We analyze the linear response and  magnetic noise sources in Secs.~\ref{Noise and sensing} and \ref{Magnetic noise from superconductor}. Finally, we present our conclusions in Sec.~\ref{Closing remarks}. 

\section{Main results}\label{Main results}

We consider a rigid permanent micromagnet of mass $M$ with center-of-mass position ${\bf R}$ levitating in ultrahigh vacuum above a type-II superconductor with surface normal vector ${\bf e}_3$, as seen in \cref{fig:experimental-setup}. The orientation of the magnet is parametrized by three Euler angles $\Omega = (\alpha,\beta,\gamma)$ that rotate the space-fixed reference frame ${\bf e}_k$, $k = 1,2,3$, into the body-fixed principal axes frame, ${\bf n}_k(\Omega) = {\rm R}(\Omega) {\bf e}_k$. The principal body-fixed frame diagonalizes the inertia tensor ${\rm I}(\Omega)$ with moments of inertia $I_k$. In the following, we denote $z = {\bf R} \cdot {\bf e}_3$ and use the $z$-$y'$-$z''$ convention for the Euler angles $(\alpha,\beta,\gamma)$.

The particle's magnetic properties are determined by the body-fixed magnetization field ${\bf M}_0(\boldsymbol{\xi})$, where $\boldsymbol{\xi}$ labels a point inside the body. The magnetization at the spatial point ${\bf r} = {\bf R} + {\rm R}(\Omega) \boldsymbol{\xi}$ is therefore
\begin{equation} \label{eqn:magnetisation-field-trafo}
    {\bf M}({\bf r}) = {\rm R}(\Omega) {\bf M}_0[{\rm R}^\mathsf{T}(\Omega) ({\bf r} - {\bf R})].
\end{equation}
The magnetic multipole moments follow by integration over the particle volume, yielding that the magnetic dipole vector is $\mathbf{m}(\Omega) = {\rm R}(\Omega) {\bf m}_0$ while the traceless and symmetric magnetic quadrupole tensor is $\mathrm{Q}(\Omega) = {\rm R}(\Omega) {\rm Q}_0 {\rm R}^\mathsf{T}(\Omega)$. Here, ${\bf m}_0$ and ${\rm Q}_0$ denote the corresponding dipole and quadrupole moments in the reference orientation $\Omega = 0$, for which ${\bf n}_k = {\bf e}_k$. In the following, we will drop the explicit $\Omega$-dependence of body-fixed tensors and use the subscript $0$ for quantities in the reference orientation.

The magnetization also carries intrinsic spin angular momentum $-{\bf m}/\gamma_{\mathrm{mag}}$, with $\gamma_{\rm mag}$ the gyromagnetic ratio of the magnetic material. Note that we use the convention that the gyromagnetic ratio is positive, $\gamma_{\mathrm{mag}}>0$, for electronic contributions, which is common in the field of levitated magnets \cite{jackson2016precessing,ahrens2026observation}. The spin angular momentum adds to the mechanical angular momentum \cite{rusconiMagneticRigidRotor2016} to give the total angular momentum ${\bf J}$. In the absence of external torques and magnetic fields, the total angular momentum is conserved due to spin-rotation coupling, as manifested in the Einstein-de Haas \cite{einstein1915experimental} and Barnett effects \cite{barnett1915magnetization}.
The Hamiltonian of the magnet can be written as
\begin{equation}\label{eqn:nonlinear-hamiltonian}
H = \dfrac{\mathbf{P}^2}{2M}  + \dfrac{1}{2} \left ( \mathbf{J} + \dfrac{\mathbf{m}}{\gamma_{\mathrm{mag}}} \right ) \cdot \mathrm{I}^{-1}  \left ( \mathbf{J} + \dfrac{\mathbf{m}}{\gamma_{\mathrm{mag}}} \right ) + V,
\end{equation}
where $\mathbf{P}$ is the linear momentum and the angular momentum vector $\mathbf{J}$ is understood in terms of the Euler canonical momenta, explicitly given in Sec.~\ref{Magnetic rigid-body Hamiltonian}. The potential $V$ describes the  conservative force and torque acting on the particle due to the total magnetic field generated by the superconductor and external sources, which we construct in the following.

\subsection{Levitation potential}\label{Levitation potential}
The magnet motion is determined by the total magnetic field acting on the particle, which consists of three contributions: (i)  the image magnetic field ${\bf B}_{\rm im}({\bf r})$ due to Meissner screening, (ii) the frozen magnetic field ${\bf B}_{\rm fr}({\bf r})$ due to trapped flux, and (iii) the external magnetic field ${\bf B}_{\rm ext}({\bf r})$, which can include control fields and magnetic noise.

First, the magnetic field sourced by the particle's magnetic moments induces reversible screening currents in the superconductor. For a planar surface and in the far-field regime, where the particle distance exceeds the London penetration depth $\lambda$, the resulting exterior field 
\begin{equation}
    {\bf B}_{\rm im}({\bf r}) = \mu_0 \int \mathrm{d}^3 \mathbf{r}'\, {\rm G}({\bf r} +2 \lambda {\bf e}_3 - \mathbf{r}') {\bf M}_{\rm im}(\mathbf{r}'),
\end{equation}
is characterized by the magnetostatic dipole Green tensor ${\rm G}({\bf r}) = \grad \otimes \grad 1/4\pi |{\bf r}|$ and the instantaneous Meissner mirror image,
\begin{equation}
{\bf M}_{\rm im}({\bf r}) = {\rm A} {\rm R}(\Omega) {\bf M}_0 [{\rm R}^\mathsf{T}(\Omega) {\rm A}({\bf r} - {\rm A}{\bf R})].
\end{equation}
 Here, we used the rank-two surface reflection tensor ${\rm A} = \mathds{1} - 2{\bf e}_3 \otimes {\bf e}_3$ and $\mu_0$ is the vacuum permeability. In the practically relevant regime $z \gg \lambda$,  the impact of the London penetration depth on the image fields can be neglected.

Second, magnetic flux frozen into a type-II superconductor is carried by pinned Abrikosov vortices, each accompanied by circulating supercurrents. We model the superconductor as a half-space occupying $z<0$, with a planar surface at $z=0$, which is threaded by straight vortices oriented perpendicular to the surface with surface density $n(\boldsymbol{\rho})$. Here, $\boldsymbol{\rho}$ denotes a point on the surface, $\boldsymbol{\rho} \cdot {\bf e}_3 = 0$. Within London theory, the long-wavelength frozen-in magnetic field generated by the pinned vortex configuration in the vacuum region $z>0$, retaining the leading finite-penetration-depth correction, is
\begin{equation}\label{eq:vortex-field-main}
\mathbf B_{\rm fr}(\mathbf{r})\simeq\dfrac{\Phi_0}{2\pi}\int \mathrm{d}^2\boldsymbol{\rho}\,
n(\boldsymbol{\rho})\dfrac{\mathbf{r}+\lambda \mathbf{e}_3-\boldsymbol{\rho}}{\left|\mathbf{r}+\lambda \mathbf{e}_3-\boldsymbol{\rho}\right|^3},
\end{equation} 
where $\Phi_0$ is the superconducting flux quantum. In the far-field regime
$z\gg\lambda $, the shift in $z$ due to the London screening produces only relative corrections of order $\lambda /z$ and can be neglected. The exterior frozen-in field can equivalently be represented by an auxiliary magnetization ${\bf M}_{\rm fr}({\bf r})$ supported inside the superconducting half-space, as illustrated in Fig.~\ref{fig:experimental-setup}. As shown in App.~\ref{app:vortex_field}, this construction reproduces the normal magnetic field at the surface, which uniquely determines the source-free, decaying field in the vacuum region. For a pinned-flux configuration established by the particle's magnetic field during the superconducting transition, we adopt the frozen-image approximation for an ideally hard superconductor. In this approximation, the frozen contribution retains the normal surface field generated by the particle at the transition and can therefore be represented by the inverse mirror image of its magnetization at that time~\cite{kordyukMagneticLevitationHard1998}. Note that this assumes that the particle-induced normal-flux distribution is retained during field cooling, otherwise partial flux expulsion or redistribution would modify the frozen field~\cite{wang2026exploring}. Specifically, for a particle located at $z_{\mathrm{fr}}{\bf e}_{3}$ with orientation $\Omega_{\mathrm{fr}} = (0, \beta_{\mathrm{fr}}, \gamma_{\mathrm{fr}})$ at the superconducting transition, the frozen image is centered at  ${\bf R}_{\rm fr} = -z_{\rm fr} {\bf e}_3$ and is represented by the magnetization  ${\bf M}_{\rm fr}({\bf r})$ at position ${\bf r} = {\bf R}_{{\rm fr}} + \mathrm{A}\mathrm{R}(\Omega_{{\rm fr}})\boldsymbol{\xi}$,
\begin{equation}\label{eqn:def-frozen-magnetisation}
{\bf M}_{\rm fr}({\bf r}) = - \mathrm{A}\mathrm{R}(\Omega_{{\rm fr}}){\bf M}_{0}[{\rm R}^\mathsf{T}(\Omega_{{\rm fr}}){\rm A}({\bf r} - {\bf R}_{{\rm fr}})].
\end{equation}

In the absence of external fields, the image and frozen-in magnetic fields determine the force and torque acting on the particle. Specifically, the frozen field provides the attractive contribution to the particle-surface interaction, whereas the Meissner-image field provides the repulsive interaction responsible for levitation of the magnet. Both contributions are conservative and can therefore be expressed through a potential. The leading contribution to the particle-surface interaction potential arises from the magnetic dipole moment ${\bf m}$,
\begin{align}\label{eqn:dipole-potential}
V_{\rm dd}({\bf R},\Omega) = &  - \dfrac{\mu_{0}}{2} \mathbf{m} \cdot \mathrm{G}(2z {\bf e}_3) \mathrm{A}\mathbf{m} \nonumber \\
&  - \mu_0\mathbf{m} \cdot \mathrm{G} ( \mathbf{R} + z_{\mathrm{fr}} {\bf e}_3 ) \mathbf{m}_{\mathrm{fr}}.
\end{align}
Here, we expanded the localized frozen magnetization in magnetic multipoles, yielding the frozen dipole vector ${\bf m}_{\rm fr} =- {\rm A}{\bf m}(\Omega_{{\rm fr}})$. The first term describes the repulsive interaction between the dipole and its mirror image, whereas the second term gives the attractive interaction between the dipole and the frozen image. The dipole potential \eqref{eqn:dipole-potential} was first derived in Ref.~\cite{kordyukMagneticLevitationHard1998} and further discussed in \cite{gieseler2020single,wang2019dynamics}. However, the dipole potential is invariant under rigid-body rotations around the dipole axis ${\bf m}$ and therefore cannot account for the experimentally observed confinement of rotations about this axis~\cite{wang2026nonlinear}.

The next-order contributions arise from the magnetic quadrupole tensor and comprise dipole-quadrupole interactions, 
\begin{align}\label{eqn:quadrupole-potential} 
V_{\rm dq}({\bf R},\Omega)  = &   -\dfrac{\mu_0}{6} (\mathrm{Q}\grad) \cdot \mathrm{G}(2z {\bf e}_3) \mathrm{A}\mathbf{m} \nonumber \\
& - \dfrac{\mu_0}{6}\left (\mathrm{Q}\grad \right ) \cdot \mathrm{G}(\mathbf{R} + z_{\mathrm{fr}} {\bf e}_3) \mathbf{m}_{\mathrm{fr}} \nonumber \\
& + \dfrac{\mu_0}{6} \left (\mathrm{Q}_{\mathrm{fr}} \grad \right) \cdot \mathrm{G}(\mathbf{R} + z_{\mathrm{fr}}{\bf e}_3) \mathbf{m},
\end{align}
and quadrupole-quadrupole interactions,
\begin{align}\label{eqn:quad-quad-potential} 
V_\mathrm{qq}({\bf R},\Omega)  = &   \dfrac{\mu_0}{72} (\grad\cdot\mathrm{Q}\grad)\mathrm{tr}[\mathrm{G}(2z\mathbf{e}_3)\mathrm{A}\mathrm{Q}\mathrm{A}]\nonumber \\
& +  \dfrac{\mu_0}{36} (\grad\cdot\mathrm{Q}\grad)\mathrm{tr}[\mathrm{G}(\mathbf{R}+z_\mathrm{fr}\mathbf{e}_3)\mathrm{Q}_\mathrm{fr}].
\end{align}
Here, the gradient operator $\grad$ denotes the three-dimensional gradient with respect to the Green-tensor coordinate, with the result evaluated at the indicated displacement arguments only after differentiation. In addition, we denote ${\rm Q}_{{\rm fr}} = -{\rm A}{\rm Q}(\Omega_{{\rm fr}}){\rm A}$. The first term of $V_\mathrm{dq}$ accounts for the interaction between the quadrupole and the image dipole, while the second and third terms describe the interactions between the quadrupole and the frozen dipole and vice versa, respectively. The two terms of $V_\mathrm{qq}$ describe the interactions of the quadrupole with its instantaneous image quadrupole and with the frozen quadrupole, respectively.  Quadrupole components that are not axially symmetric about $\mathbf{m}$ break the rotational symmetry of the dipole potential and thereby enable full orientational alignment of the particle with the space-fixed frame.  Such anisotropic quadrupole components can arise from spatial inhomogeneities of the magnetization, including those induced by particle shapes lacking inversion symmetry.

The full potential is obtained by including the interaction with an external magnetic field, expanded up to quadrupole order,
\begin{equation}\label{eqn:b_ext-dip-quad-potential}
    V_{\rm ext}({\bf R},\Omega) = - \mathbf{m} \cdot \mathbf{B}_{\mathrm{ext}}({\bf R}) - \dfrac{1}{6}\left ( {\rm Q} \grad \right ) \cdot {\bf B}_{\rm ext}({\bf R}),
\end{equation}
and the gravitational potential for gravitational acceleration $g$ along $-{\bf e}_3$, giving
\begin{equation}\label{eqn:total-trapping-potential}
V =  V_{\rm dd} + V_{\rm dq}+ V_\mathrm{qq}+V_{\rm ext} + M g z.
\end{equation}
This potential governs the translational and rotational dynamics of the micromagnet up to quadrupole order. Higher-order magnetic multipole tensors could be incorporated analogously. Here we neglect dipole-octupole interactions despite having the same distance scaling as the quadrupole-quadrupole terms, because at the pure-dipole equilibrium the particle-octupole couplings to the Meissner and frozen image dipoles cancel and the frozen-octupole term is independent of $\gamma$, such that they generate no direct $\gamma$ confinement, as discussed in the following.

\begin{figure*}[htp]
\centering
\includegraphics[trim={1.15cm 18.6cm 1.1cm 1.2cm},clip,width=18.1cm]{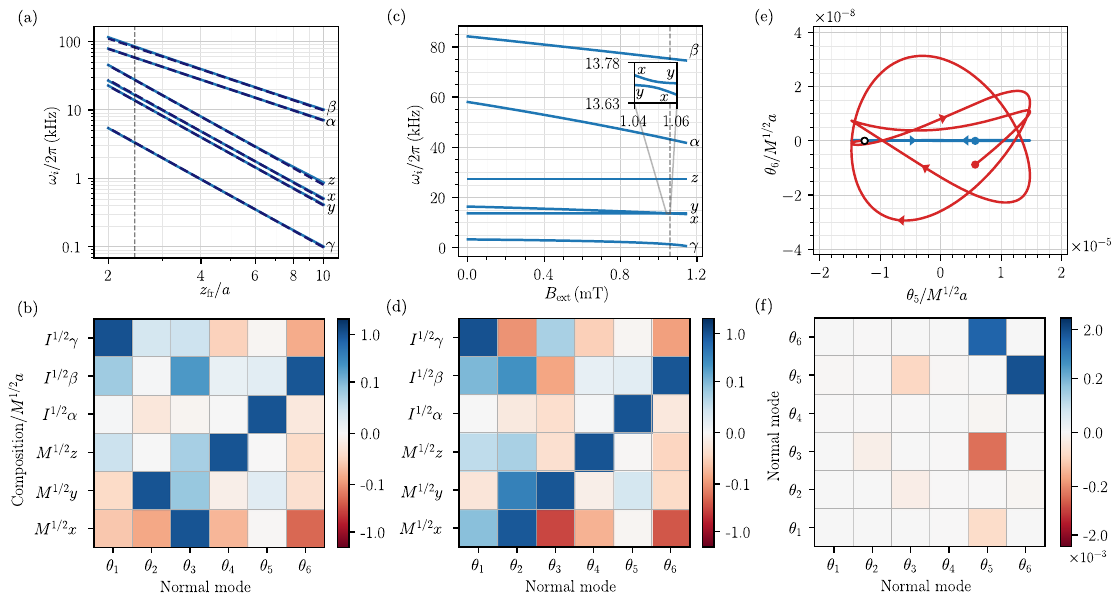}
\caption{\textbf{Normal-mode hybridization and spin-rotation coupling of a spheroidal particle.} (a) Zero-field mode frequencies $\omega_i/2\pi$ as a function of the flux-freezing height $z_\mathrm{fr}$ normalized by the minor semiaxis length $a$.
Solid light-blue curves show numerical results, and dashed dark-blue curves show the analytical approximations given in Eqs.~\eqref{eqn:pure-dipole-omega-yz-alpha}, \eqref{eqn:pure-dipole-omega-x-beta-approx} and \eqref{eqn:omega_gamma}. 
Modes are labeled by their dominant configurational coordinate. The vertical dashed line indicates the value of $z_\mathrm{fr}$ used in (b)-(f). (b) Normal-mode compositions at zero applied field, with modes numbered in order of increasing frequency. Translational and angular components are weighted by $\sqrt{M}$ and $\sqrt{I}$, where $I = (I_1 I_2 I_3)^{1/3}$, respectively, and expressed in units of $\sqrt{M}a$. 
(c) Mode frequencies along the equilibrium branch connected to the zero-field minimum, with $\gamma\simeq\gamma_\mathrm{fr}$, in a homogeneous external field $\mathbf{B}_\mathrm{ext}=-B_\mathrm{ext}\mathbf{e}_{1}$.
The inset resolves the avoided crossing associated with hybridization between the predominantly $x$- and $y$-like modes.
(d) Mode compositions at $B_{\mathrm{ext}} = 1.06\,\mathrm{mT}$ for the same equilibrium branch, illustrating the $x$-$y$ hybridization.
(e) Undamped zero-field trajectories projected onto the $\theta_{5}$-$\theta_{6}$ plane, without (blue) and with (red) spin-rotation coupling. These modes are predominantly associated with $\alpha$ and $\beta$, respectively. Open and filled markers indicate the initial positions and endpoints after the same evolution time, respectively; arrows indicate the direction of motion. (f) Dimensionless mixing coefficients  $T_{ij}$ at zero field, describing the phase-shifted mode admixture induced by spin-rotation coupling. All calculations use a prolate spheroid with principal semiaxes $(a,a,1.1a)$, $a=1.8\,\mu\mathrm{m}$, mass density $7430\,\mathrm{kg}\,\mathrm{m}^{-3}$, remanence field $750\,\mathrm{mT}$, and an effective gyromagnetic ratio specified by $2m_{\mathrm{e}}\gamma_{\mathrm{mag}}/e = 1.28$, where $m_\mathrm{e}$ is the electron mass and $e$ the positive elementary charge~\cite{wang2026nonlinear}. For illustration, we take quadrupole eigenvalues  $Q_1/m_0a = 0.0$ and $Q_2/m_0a = -0.4$ with $m_0=|\mathbf{m}|$, and $\gamma_{\mathrm{fr}} = \pi/4$.}
\label{fig:modes-spin-rotation-transient}
\end{figure*}

\subsection{Complete trapping with quadrupole moment} \label{Complete trapping with quadrupole moment}
The total trapping potential \cref{eqn:total-trapping-potential} can lead to stable three-dimensional trapping and orientational alignment of the magnet. For micron-sized particles in moderate external fields, the dominant contribution to the potential is due to the dipole potential \cref{eqn:dipole-potential}. It gives rise to the force
\begin{equation} \label{eqn:force-pure-dipole-meiss-fixed}
\mathbf{F}_{\mathrm{dd}} = \mu_0 ( {\bf m} \cdot \grad ) [{\rm G}(2z {\bf e}_{3})\mathrm{A}\mathbf{m} + \mathrm{G}({\bf R} + z_{\rm fr} {\bf e}_{3})\mathbf{m}_{\mathrm{fr}}]
\end{equation}
and the torque
\begin{equation} \label{eqn:torque-pure-dipole-meiss-fixed}
\mathbf{N}_\mathrm{dd} = \mu_0 \mathbf{m} \times [\mathrm{G}(2z {\bf e}_{3})\mathrm{A}\mathbf{m} + \mathrm{G}({\bf R} + z_{\rm fr} {\bf e}_{3}) \mathbf{m}_{\mathrm{fr}}],
\end{equation}
which both vanish for ${\bf R} = z_{\rm fr} {\bf e}_3$ and ${\bf m} = -\mathrm{A}{\bf m}_{\rm fr}$, corresponding to the flux-freezing position and dipole orientation. The dipole interaction therefore restores the particle to its flux-freezing configuration and anti-aligns the magnetic dipole moment with the in-plane frozen-image dipole. However, the dipole potential remains invariant under rotations about the dipole axis, leaving rotations about the dipole axis unconstrained. Consequently, the dipole interaction alone traps only five of the six configurational degrees of freedom. Complete orientational alignment requires additional symmetry-breaking interactions. The leading contribution arises from the magnetic quadrupole tensor, which lifts the residual rotational symmetry and generates a restoring potential for rotations about the dipole axis.

Assuming that the dipole moment is aligned with the principal axis, ${\bf m}\parallel{\bf n}_{3}$, the remaining rotational degree of freedom corresponds to rotations about the dipole axis and is parametrized by the Euler angle $\gamma$. Quadrupole-induced confinement of this angle arises through two mechanisms. First, the quadrupole-quadrupole interaction provides a direct $\gamma$-dependent potential already at the dipole equilibrium. Second, the dipole-quadrupole interaction shifts the equilibrium of the other five coordinates, generating an indirect restoring potential in $\gamma$. For the in-plane frozen-dipole configuration considered here, the $\gamma$-dependent terms of the dipole-quadrupole interaction cancel at the dipole equilibrium. The indirect contribution is obtained by perturbatively eliminating the remaining five degrees of freedom. Neglecting gravity and external fields, the sum of the direct and relaxation-induced contributions yields the approximate reduced angular potential 
\begin{align}\label{eq:V_gamma}
 V_\gamma(\gamma) \simeq &\dfrac{ \mu_{0} (Q_1 - Q_2)^{2}}{24 \pi (2z_{\mathrm{fr}})^5}\sin^2(\gamma - \gamma_{{\rm fr}})\nonumber\\
    &\times\left[1+\dfrac{1}{3}\sin^2(\gamma+\gamma_{{\rm fr}})\right],
\end{align}
where a $\gamma$-independent energy offset has been omitted (see Sec.~\ref{Trapping minimum} for details). Here, the quadrupole tensor is assumed to be diagonal in the principal-axes frame with eigenvalues $Q_k$, $k=1,2,3$, and the frozen dipole is oriented parallel to the superconducting surface. The potential is minimized at  $\gamma\simeq\gamma_{\rm fr}\,\pmod{\pi}$, demonstrating that the quadrupole contributions lift the residual rotational symmetry of the dipole potential and complete the orientational trapping of the particle. The equilibrium configuration is further modified by gravity and external magnetic fields. For the parameters considered here, gravity produces only a negligible shift of the equilibrium configuration, whereas external fields provide an additional mechanism for controlling the orientational trapping.

\subsection{Mode hybridization and spin-rotation coupling}
\label{Mode hybridization and spin-rotation coupling}
Expanding the Hamiltonian around the equilibrium configuration yields a set of coupled harmonic oscillators that describe the small-amplitude translational and librational motion of the trapped magnet. The resulting normal modes, with amplitudes $\theta_i$  ($i=1,2,\dots,6$), are linear combinations of the six configurational degrees of freedom and are characterized by their corresponding trapping frequencies $\omega_i$. Fig.~\ref{fig:modes-spin-rotation-transient}(a) shows the frequencies of the predominantly translational and rotational modes as a function of the freezing height $z_\mathrm{fr}$, revealing that for experimentally relevant particle parameters the trapping frequencies span from hundreds of Hz to hundreds of kHz over the range of $z_\mathrm{fr}$. For comparable trapping conditions, the predicted frequencies are consistent with those measured in Ref.~\cite{wang2026nonlinear}. Moreover, the approximate analytical frequencies derived in Sec.~\ref{Transformation to the normal modes} are in good agreement with the eigenfrequencies obtained from the full numerical calculation, confirming the predicted power-law scaling with the normalized levitation height~\cite{gieseler2020single}.

The corresponding mode composition demonstrates that, even in the absence of spin-rotation coupling, the trapping potential generally mixes translational and rotational degrees of freedom [see Fig.~\ref{fig:modes-spin-rotation-transient}(b)]. This hybridization arises from the orientation-dependence of the magnetic forces and thus depends on the trapping equilibrium. Especially prominent hybridization occurs between the $x$ and $\beta$ degrees of freedom, meaning that tilting the magnet shifts its lateral position parallel to the surface. A homogeneous in-plane magnetic field can enhance the hybridization between the $x$ and $y$ modes. The applied field changes both the equilibrium configuration and the local curvature of the trapping potential, providing a mechanism for tuning both the normal-mode frequencies and their compositions. The field-dependent frequencies in Fig.~\ref{fig:modes-spin-rotation-transient}(c) exhibit an avoided crossing, as observed experimentally~\cite{wang2026nonlinear}. The associated $x$-$y$ mode hybridization is also evident from the mode compositions shown in Fig.~\ref{fig:modes-spin-rotation-transient}(d). At sufficiently large field strengths, the system undergoes bifurcations, allowing multiple stable trapping configurations to coexist, as discussed in detail in Sec.~\ref{Trapping minimum}.

In contrast to the hybridization generated by the trapping potential, which arises from mixing among configurational coordinates, spin-rotation coupling originates from a coordinate-momentum interaction. Consequently, it manifests itself as a phase-shifted admixture of otherwise independent normal modes, reflecting the gyroscopic nature of the spin-rotation interaction. This behavior is illustrated by the transient trajectories in Fig.~\ref{fig:modes-spin-rotation-transient}(e), which compare the evolution from the identical initial phase-space coordinates with and without spin-rotation coupling. In its absence, the trajectory remains confined to the initially excited $\theta_5$ direction. When spin-rotation coupling is included, motion is transferred into the $\theta_6$ mode, producing a two-dimensional trajectory and resulting in different endpoints after the same evolution time, which illustrates the phase-shifted admixture of the modes. Thus, the phase-sensitive transient dynamics provide a clear signature of this weak coupling, as recently observed experimentally in Ref.~\cite{ahrens2026observation}.

In the normal-mode basis, the linearized equations of motion for the  mode coordinates $\theta_i$ acquire an antisymmetric velocity-coupling matrix $\Lambda_{ij} \propto 1/\gamma_{{\rm mag}}$ (as detailed in Sec.~\ref{sec:spin_rotation_coupling}), originating from the intrinsic spin angular momentum. The mode coordinates consequently obey 
\begin{equation}\label{eqn:eom-normalmodes-summary}
\ddot{\theta}_{i} + \omega^2_i \theta_{i} - 2\sum^{6}_{j=1}\Lambda_{ij} \dot{\theta}_{j} = 0.
\end{equation}
For nondegenerate, well-separated modes a perturbative solution is
\begin{equation}\label{eqn:modes-mixing-complex-ampl-trans-EdH}
\theta_{i} \simeq \mathrm{Re}\left[ \sum^{6}_{j=1}(\delta_{ij} - \mathrm{i}T_{ij}) \Theta_{j}\mathrm{e}^{-\mathrm{i}\omega_jt} \right],
\end{equation}
where $T_{ij}= 2 \Lambda_{ij} \omega_{j}/(\omega^2_{i} -\omega^2_j)$ ($i\neq j$) quantifies the phase-shifted admixture of coordinate $\theta_i$ into the mode associated with the uncoupled coordinates $\theta_j$, which is displayed in Fig.~\ref{fig:modes-spin-rotation-transient}(f). The complex amplitudes $\Theta_j$ are determined by the specified initial coordinates and velocities. In the absence of spin-rotation coupling, they reduce to $\Theta_j = \theta_j(0) + \mathrm{i} \dot{\theta}_j(0)/\omega_j$. Fig.~\ref{fig:modes-spin-rotation-transient}(f) shows that the spin-induced coupling is weak and mode selective. Its largest matrix elements couple modes $\theta_5$ and $\theta_6$, which for the parameters considered here are predominantly associated with the rotational coordinates $\alpha$ and $\beta$, respectively.

\subsection{Magnetic noise and sensing} 
\label{Magnetic noise and sensing}
The coupled translational and librational motion of the levitated magnet provides a means of detecting weak forces and magnetic fields. An applied signal generally excites several normal modes, while a displacement measurement records a superposition of their responses. The sensitivity therefore depends on both the coupling of the signal to the modes and their contributions to the measured displacement. 

In the linear regime, the driven motion obeys 
\begin{equation} \label{eqn:normalmodes_Langevin}
\ddot{\theta}_{i} + \sum^{6}_{j=1}(\Gamma_{ij}-2 \Lambda_{ij})\dot{\theta}_j + \omega^2_i \theta_i  = \tau_{i}(t)+\eta_i(t),
\end{equation}
where $\tau_i$ and $\eta_i$ are the generalized signal and stochastic forces, respectively. We model the thermal contribution to $\eta_i$ as  approximately Gaussian white noise, while retaining correlations between modes. As detailed in Sec.~\ref{Noise and sensing}, the fluctuation-dissipation theorem then relates the damping matrix $\Gamma_{ij}$ to the cross-spectrum $S_{\eta_i\eta_j}^\mathrm{th}$ of the thermal generalized forces via
\begin{equation}\label{eqn:damping-rate-equipartition}
\Gamma_{ij} = \dfrac{S_{\eta_i\eta_j}^\mathrm{th}}{2k_{\mathrm{B}}T},
\end{equation}
where $k_\mathrm{B}$ denotes the Boltzmann constant and $T$ the temperature of the superconductor. These equations define the susceptibility matrix $\chi_{ij}(\omega)$, which relates the mode displacements to the applied forces. 

The superconducting surface generates magnetic-field fluctuations that exert fluctuating forces and torques on the levitated magnet.
We consider two distinct noise sources, namely thermal current-density fluctuations of the normal electronic component and small-amplitude thermal motion of pinned vortices, as detailed in Sec.~\ref{Magnetic noise from superconductor}. Assuming these contributions are statistically independent, their magnetic-field cross-spectral densities add,
\begin{equation}\label{eq:S_bb_total}
S_{B_k B_\ell}=S_{B_k B_\ell}^\mathrm{th,c}+S_{B_k B_\ell}^\mathrm{th,v}.
\end{equation}
The first term corresponds to thermal current fluctuations of the normal-conducting electrons which are spatially distributed throughout the superconducting material. Screening by the superconductor predominantly restricts their contribution to the exterior noise to a surface layer of thickness set by the London penetration depth $\lambda$. For a laterally extended substrate in the far-field regime $z_\mathrm{eq}\gg\lambda$, where $z_\mathrm{eq}$ denotes the equilibrium height, the local magnetic-field noise is approximately white over the frequency range considered, with nonzero spectral components scaling as 
\begin{equation}
 S_{B_k B_\ell}^\mathrm{th,c}\propto \dfrac{\sigma_\mathrm{c}T\lambda^3}{z_\mathrm{eq}^4},   
\end{equation} 
where $\sigma_\mathrm{c}$ is the conductivity of the normal electronic component and $T$ the temperature of the superconductor. Thermal vortex motion, as specified by the second term in Eq.~\eqref{eq:S_bb_total}, likewise produces approximately white noise at frequencies below the pinning relaxation rates $K_\nu/c_\nu$, where $K_\nu$  and $c_\nu$ are the effective pinning stiffness and viscous drag coefficient of vortex $\nu$. For $N$ independently fluctuating vortices with comparable pinning parameters, localized beneath the particle at lateral distances small compared with $z_\mathrm{eq}$, the leading nonzero components of the noise spectrum scale as 
\begin{equation}
S_{B_k B_\ell}^\mathrm{th,v}\propto \dfrac{N T c_\nu}{K_\nu^2 z_\mathrm{eq}^6}.
\end{equation}
The magnetic-field spectra and their spatial derivatives determine the generalized-force spectra $S_{\eta_i\eta_j}(\omega)$ in the normal-mode basis, including correlations between modes, as derived in Sec.~\ref{Magnetic noise from superconductor}. The damping rates follow from the thermal contribution to these spectra via Eq.~\eqref{eqn:damping-rate-equipartition}. At fixed equilibrium geometry and mode composition, the contributions from thermal current fluctuations and vortex motion scale as $\Gamma_{ii}^\mathrm{c}\propto\sigma_\mathrm{c}\lambda^3$ and $\Gamma_{ii}^\mathrm{v}\propto N c_\nu/K_\nu^2$, respectively, under the assumptions stated above.

\begin{figure}[ht]
    \centering
    \includegraphics[width=\columnwidth]{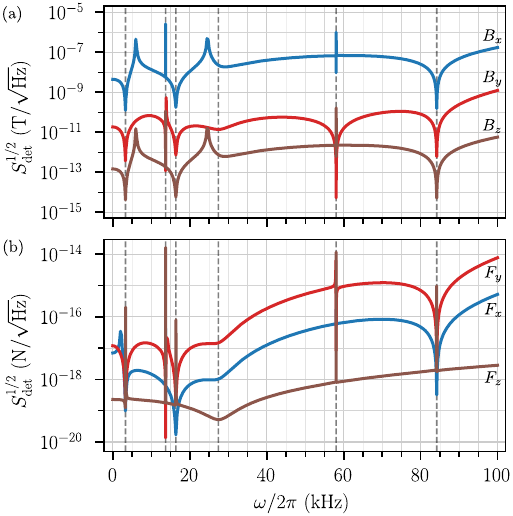}
    \caption{
    \textbf{Magnetic-field and force sensitivities for vertical-displacement readout.} Noise-equivalent amplitude spectra $\sqrt{S_\mathrm{det}(\omega)}$ at zero applied static field for (a) homogeneous magnetic-field signals $B_x$, $B_y$, and $B_z$, and (b) force signals $F_x$, $F_y$, and $F_z$. Blue, red, and brown curves correspond to signals along the laboratory $x$, $y$, and $z$ axes, respectively. Dashed vertical lines mark the normal-mode resonance frequencies.
    The noise model includes current fluctuations of the normal electronic component, thermal motion of pinned vortices, and measurement imprecision modeled as white noise with a displacement power spectral density $S_\mathrm{imp}(\omega)=\hbar/2M\omega_\mathrm{fr}^2$. Particle parameters and zero-field mode compositions are the same as in Fig.~\ref{fig:modes-spin-rotation-transient}. The superconductor parameters used are $T=6\,\mathrm{K}$, $\lambda=100\,\mathrm{nm}$ \cite{pereg-barneaAbsoluteValuesLondon2004}, and $\sigma_\mathrm{c}=1.0\times10^{6}\,\mathrm{S}\,\mathrm{m}^{-1}$~\cite{turner2003observation}. The vortex contribution comprises $N=10$ independently fluctuating vortices with $c_\nu=1.215\times10^{-13}\,\mathrm{N}\,\mathrm{s}\,\mathrm{m}^{-1}$~\cite{tsuchiya2001electronic} and $K_\nu=1.35\times 10^{-2}\,\mathrm{N}\,\mathrm{m}^{-1}$ for an effective length of the coherently moving vortex segment $d_{\mathrm{eff},\nu}=270\,\mathrm{nm}$~\cite{wang2026nonlinear}. With $\Gamma_{ii}=\Gamma_{ii}^\mathrm{c}+\Gamma_{ii}^\mathrm{v}$, the mode quality factors $\omega_i/\Gamma_{ii}$ are $6.8 \times 10^{10}$, $1.6 \times 10^{10}$, $1.3 \times 10^{10}$, $2.1 \times 10^{9}$, $2.5 \times 10^{9}$, and $4.0 \times 10^{9}$ in order of increasing resonance frequency.  
    } 
    \label{fig:superconductor-noise}
\end{figure}

To characterize the sensing performance, we consider a real-valued scalar signal $\psi(t)$ with modal force components $\tau_i(t)=\psi_i\psi(t)$ and a displacement readout
\begin{equation}\label{eqn:output-X-def-with-detector-noise}
X(t) = \sum^{6}_{i=1} X_{i}\theta_{i}(t) + \eta_{\rm imp}(t),
\end{equation}
where the real, time-independent coefficients $\psi_i$ and $X_i$ specify the signal and readout couplings to mode $i$, respectively, and $\eta_\mathrm{imp}$ denotes additive measurement-imprecision noise.  For a signal statistically independent of the noise, the noise-equivalent input spectrum  derived in Sec.~\ref{Noise and sensing} is
\begin{equation}\label{eqn:sensitivity-threshold-Sdet}
S_{\rm det}(\omega) =\dfrac{S_{XX}^\mathrm{noise}(\omega)}{\left| \sum_{i,j=1}^{6}X_i \chi_{ij}(\omega) \psi_j\right|^2}.
\end{equation}
Here, $S_{XX}^\mathrm{noise}$ includes the displacement noise induced by the magnetic-field fluctuations of the superconducting surface and the detector imprecision, taken to be mutually uncorrelated. The quantity $\sqrt{S_\mathrm{det}}$ gives the noise-equivalent amplitude sensitivity for the specified signal and readout. 

Figure~\ref{fig:superconductor-noise} shows the predicted noise-equivalent amplitude sensitivities $\sqrt{S_\mathrm{det}}$ for local magnetic-field and force signals along the three laboratory axes, using a vertical-displacement readout at zero applied static field. The best magnetic-field sensitivity is $4.1\,\mathrm{fT}/\sqrt{\mathrm{Hz}}$ for $B_z$ near the predominantly $\gamma$-like mode resonance at $3.3\,\mathrm{kHz}$. For force signals, the lowest noise-equivalent amplitude sensitivity is $14\,\mathrm{zN}/\sqrt{\mathrm{Hz}}$ for $F_y$ near the predominantly $y$-like resonance at $14\,\mathrm{kHz}$. The mechanical quality factors listed in the figure caption are calculated assuming that dissipation arises solely from thermal current fluctuations of the normal electronic component and small-amplitude motion of pinned vortices. Vortex jumps between pinning configurations can modify the mechanical damping~\cite{wang2026exploring} and may contribute to the lower quality factors of order $10^5$ measured in Ref.~\cite{wang2026nonlinear}.

\section{Derivation of the Hamiltonian}
\label{Derivation of the Hamiltonian}
In this section, we formulate the Hamiltonian dynamics of the levitated magnetic rigid rotor, including spin-rotation coupling associated with the intrinsic angular momentum carried by its magnetization. We then construct the conservative magnetic interaction potential from the Lorentz force and torque and perform a multipole expansion, retaining the leading contributions associated with the magnetic dipole and quadrupole moments.

\subsection{Magnetic rigid-body Hamiltonian}
\label{Magnetic rigid-body Hamiltonian}
The rotation tensor $\mathrm{R}(\Omega)$ maps the laboratory-frame axes $\mathbf{e}_1,\mathbf{e}_2,\mathbf{e}_3$ onto the body-fixed principal axes $\mathbf{n}_1,\mathbf{n}_2,\mathbf{n}_3$, such that $ \mathbf{n}_k(\Omega)=\mathrm{R}(\Omega)\mathbf{e}_k$. The rotation transformation can be parametrized in terms of the Euler angles $\Omega=(\alpha,\beta,\gamma)$ in the \textit{z}-\textit{y}'-\textit{z}'' convention so that the lab-frame representation is
\begin{align}
\mathrm{R}(\Omega) = &\mathrm{R}_{z}(\alpha)\mathrm{R}_{y}(\beta)\mathrm{R}_{z}(\gamma) =
    \begin{pmatrix}
        \cos\alpha & -\sin\alpha & 0\\
        \sin\alpha & \cos\alpha & 0\\
        0 & 0 & 1
    \end{pmatrix}\nonumber\\
     &\begin{pmatrix}
        \cos\beta & 0 & \sin\beta \\
         0 & 1 & 0\\
        -\sin\beta & 0 & \cos\beta
    \end{pmatrix}
    \begin{pmatrix}
        \cos\gamma & -\sin\gamma & 0\\
        \sin\gamma & \cos\gamma & 0\\
        0 & 0 & 1
    \end{pmatrix}.
    \label{eq:rotationMatrix}
\end{align}
 The principal axes, explicitly given in the lab-frame representation in App.~\ref{app:principal_axes}, are defined such that the inertia tensor is diagonal, i.e. $\mathrm{I}=\sum_k I_k \mathbf{n}_k\otimes\mathbf{n}_k$. The particle rotates with angular velocity $\boldsymbol{\omega}$, which is defined through $\dot{\mathbf{n}}_k=\boldsymbol{\omega}\times\mathbf{n}_k$, and can be expressed as 
\begin{equation}\label{eqn:angular-vel-def}
    \boldsymbol{\omega}=\dot{\alpha}\mathbf{u}_\alpha +\dot{\beta}\mathbf{u}_\mathrm{\beta}+\dot{\gamma}\mathbf{u}_\gamma.
\end{equation} 
Here we introduced the rotation-axes vectors 
\begin{equation}\label{eqn:m_mu-vectors-def}
\mathbf{u}_\alpha= \mathbf{e}_3,\qquad \mathbf{u}_\beta=\mathbf{e}_\mathrm{n},\qquad\mathbf{u}_\gamma = \mathbf{n}_3,
\end{equation}
with the nodal axis $\mathbf{e}_\mathrm{n}=-\sin\alpha\mathbf{e}_1+\cos\alpha\mathbf{e}_2$, which form  a non-orthogonal basis. Its dual basis is given by
\begin{align}\label{eqn:m_mu-vectors-dual}
\mathbf{u}^\alpha&=\dfrac{1}{\sin^2\beta}(\mathbf{e}_3-\cos\beta\mathbf{n}_3),\nonumber\\
\mathbf{u}^\beta&=\mathbf{e}_\mathrm{n},\nonumber\\
\mathbf{u}^\gamma&=\dfrac{1}{\sin^2\beta}(\mathbf{n}_3-\cos\beta\mathbf{e}_3),
\end{align}
such that the orthogonality relations $\mathbf{u}^\mu\cdot\mathbf{u}_\nu=\delta_{\mu\nu}$, $\mu,\nu\in\Omega$ are satisfied.
The rotational dynamics of the magnetic particle follows from the Newtonian rigid-rotor equations of motion, supplemented by the intrinsic spin angular momentum associated with the magnetic dipole moment. The corresponding equation of motion is
\begin{equation}
    \dfrac{\mathrm{d}}{\mathrm{d}t}\left(\mathrm{I}\boldsymbol{\omega}-\dfrac{\mathbf{m}}{\gamma_{\mathrm{mag}}}\right)=\mathbf{N},
\end{equation}
where $\mathbf{N}$ denotes the conservative Lorentz torque due to the total magnetic field acting on the particle (see App.~\ref{Lorentz force and torque}). Together with the center-of-mass equation of motion $M\ddot{\mathbf{R}}=\mathbf{F}$, where $\mathbf{F}$ is the corresponding Lorentz force, these equations can be formulated in terms of Euler-Lagrange equations. The torque and force can be traced back to the scalar potential $V$ through $\mathbf{F}=-\boldsymbol{\nabla}_\mathbf{R} V$ as well as 
\begin{equation}\label{eq:torque_potential}
    \mathbf{N}\cdot\mathbf{u}_\mu=-\partial_\mu V,\qquad\mu\in\Omega,
\end{equation} 
as we show in Sec.~\ref{Meissner-image potential}. The corresponding Lagrangian is
\begin{equation}\label{eqn:nonlinear-lagrangian}
L = \dfrac{1}{2} M \mathbf{\dot{R}}^2 +\dfrac{1}{2} \boldsymbol{\omega} \cdot \mathrm{I} \boldsymbol{\omega} -\dfrac{\mathbf{m}}{\gamma_{\mathrm{mag}}}  \cdot \boldsymbol{\omega} - V.
\end{equation}
From the Lagrangian, the canonical momentum associated with the center-of-mass motion follows as $\mathbf{p}=\partial L/\partial \dot{\mathbf{R}}$, while the canonical momenta associated with the Euler angles are $p_\alpha$, $p_\beta$, $p_\gamma$. These define the total angular momentum vector through
\begin{equation}\label{eq:J_in_p}
    \mathbf{J}=\sum_{\mu\in\Omega}\mathbf{u}^\mu p_\mu,
\end{equation}
from which it follows that
\begin{equation}\label{eqn:EdH-total-momentum}
\mathbf{J} =\dfrac{\partial L}{\partial\boldsymbol{\omega}} =\mathrm{I} \boldsymbol{\omega} - \dfrac{\mathbf{m}}{\gamma_{\mathrm{mag}}}.
\end{equation}
A Legendre transformation then yields the kinetic terms in the Hamiltonian in Eq.~\eqref{eqn:nonlinear-hamiltonian}. To derive the remaining interaction potential $V$, we connect the magnetic fields generated by the superconducting currents to effective Meissner and frozen-image fields, and determine the leading multipole contributions in the following sections.

\subsection{Meissner and frozen image fields}\label{Meissner and frozen image fields}
Magnetic flux trapped in the type-II superconductor during the superconducting transition is carried by pinned Abrikosov vortices accompanied by circulating supercurrents. Together with the reversible Meissner screening currents induced by the instantaneous particle field, these currents generate the superconducting contribution to the magnetic field acting on the particle.

We consider straight pinned vortices oriented perpendicular to the superconducting surface with surface density
\begin{equation}
\label{eqn:vortex-density-theory}
n(\boldsymbol{\rho}) = \sum_\nu w_\nu\delta^{(2)}\left(\boldsymbol{\rho}-\boldsymbol{\rho}_\nu\right),
\end{equation}
where $w_{\nu}=\pm1$ denotes the winding number of the singly quantized vortex $\nu$, and $\boldsymbol{\rho}_\nu$ is the transverse position at which it intersects the surface. Solving the Maxwell-London equations with the appropriate interface 
conditions gives the exterior field generated by the pinned vortex configuration, as detailed in App.~\ref{app:vortex_field}. At $z\gg\lambda$, the exterior vortex field reduces to Eq.~\eqref{eq:vortex-field-main}, retaining the leading finite-penetration-depth correction. Using the effective frozen magnetization introduced in Sec.~\ref{Levitation potential}, we write the exterior field as 
\begin{equation}\label{eqn:frozen-in-bfield-general}
\mathbf{B}_{\mathrm{fr}}(\mathbf{r}) \simeq \mu_{0} \int \mathrm{d}^3\mathbf{r}' \, \mathrm{G}(\mathbf{r} - \mathbf{r}')\mathbf{M}_{\mathrm{fr}}(\mathbf{r}'),\quad z>0.
\end{equation}
For the long-wavelength vortex field, including the leading penetration-depth correction, an explicit choice is
\begin{equation}
\mathbf{M}_\mathrm{vort}(\mathbf{r})=\dfrac{2\Phi_0}{\mu_0} n(\boldsymbol{\rho})\Theta(-\lambda -z)\mathbf{e}_3,
\label{eq:effective-vortex-magnetization}
\end{equation} 
where $\Theta$ denotes the step function restricting the auxiliary source to $z<-\lambda $. Substituting $\mathbf{M}_\mathrm{fr}=\mathbf{M}_\mathrm{vort}$ into the preceding integral reproduces Eq.~\eqref{eq:vortex-field-main}, as verified in App.~\ref{app:vortex_field}. 

In the following, we neglect the penetration-depth correction and use the ideal-screening limit for both image fields. Assuming negligible flux expulsion during field cooling, we use the frozen-image approximation introduced in Sec.~\ref{Levitation potential} and represent $\mathbf{M}_\mathrm{fr}$ by the inverse mirror magnetization defined in Eq.~\eqref{eqn:def-frozen-magnetisation}.  The Green-tensor integral evaluated for this distribution gives the frozen-image magnetic field,
\begin{align}\label{eqn:frozen-in-bfield-mfr}
\mathbf{B}_{\mathrm{fr}}(\mathbf{r}) = -\mu_{0} \int \mathrm{d}^3\boldsymbol{\xi} \, \mathrm{G}&\left[\mathbf{r} -   \mathbf{R}_{\mathrm{fr}}-\mathrm{A}\mathrm{R}(\Omega_{{\rm fr}})\boldsymbol{\xi}\right]\nonumber\\
&\mathrm{A}\mathrm{R}(\Omega_{{\rm fr}})\mathbf{M}_0(\boldsymbol{\xi}).
\end{align}

The pinned-vortex contribution is taken to remain fixed after field cooling in the static trapping model. By contrast, subsequent changes in the particle configuration induce reversible Meissner screening currents. In the same planar ideal-screening limit, their exterior field is represented by an instantaneous mirror image magnetization below the superconducting surface that follows the particle position and orientation. The instantaneous Meissner image is centered at $\mathrm{A}\mathbf{R}$ and is obtained by reflecting the particle magnetization distribution across the superconducting surface. The  magnetic field generated by the Meissner image then follows as 
\begin{equation}\label{eqn:meissner-bfield-general}
\mathbf{B}_{\mathrm{im}}(\mathbf{r}) = \mu_{0} \int \mathrm{d}^3\boldsymbol{\xi} \, \mathrm{G}(\mathbf{r} - \mathrm{A}\mathrm{R}(\Omega) \boldsymbol{\xi} -  \mathrm{A}\mathbf{R})  \mathrm{A}\mathrm{R}(\Omega)\mathbf{M}_{0}(\boldsymbol{\xi}).
\end{equation}
We use this decomposition of the magnetic field into Meissner and frozen contributions as the starting point for deriving the magnetic interaction potential. In the following derivation, we suppress explicit position  and orientation arguments, ${\bf R}$ and $\Omega$, where no ambiguity arises.

\subsection{Magnetic interaction potential}
\label{Meissner-image potential}
We derive the interaction potential from the Lorentz force and torque expressions given in App.~\ref{Lorentz force and torque}. For magnetic fields whose spatial distribution is independent of the particle configuration, the force and torque follow from a scalar potential, as derived in App.~\ref{app:Potential in static external field}. Since the frozen magnetization $\mathbf{M}_\mathrm{fr}$ associated with the field of the pinned-flux configuration does not depend on the instantaneous particle position or orientation, its contribution to the interaction potential follows directly from Eq.~\eqref{eqn:frozen-in-bfield-general} as 
\begin{equation}\label{eqn:frozen-image-potential}
V_{\mathrm{fr}} = - \mu_{0} \int \mathrm{d}^3\boldsymbol{\xi}\,\mathrm{d}^3\mathbf{r}' \,\mathbf{M}_{0}(\boldsymbol{\xi})\cdot\mathrm{R}^\mathsf{T}\mathrm{G}(\mathrm{R}\boldsymbol{\xi} + \mathbf{R} - \mathbf{r}') \mathbf{M}_{\mathrm{fr}}(\mathbf{r}').
\end{equation}
Again, if the frozen magnetization is described by the inverse-mirror model in  Eq.~\eqref{eqn:def-frozen-magnetisation}, the potential due to the frozen image can be written in the form
\begin{align}\label{eqn:frozen-image-potential_mfr}
V_{\mathrm{fr}}  =  & \mu_{0} \int \mathrm{d}^3\boldsymbol{\xi}\,\mathrm{d}^3\boldsymbol{\xi}'\, \mathbf{M}_{0}(\boldsymbol{\xi}) \cdot\mathrm{R}^\mathsf{T}\mathrm{G}[\mathrm{R}\boldsymbol{\xi} -\mathrm{A}\mathrm{R}(\Omega_{{\rm fr}}) \boldsymbol{\xi}'\nonumber\\
&  + \mathbf{R} - \mathbf{R}_{\mathrm{fr}} ]\mathrm{A}\mathrm{R}(\Omega_{{\rm fr}}) \mathbf{M}_0(\boldsymbol{\xi}').
\end{align}

The Meissner contribution differs qualitatively from the frozen field because the image magnetization distribution depends explicitly on the instantaneous particle configuration. Consequently, the corresponding interaction potential cannot be obtained directly from the standard expression for a non-dynamical external field. We derive the Meissner contribution starting from the Lorentz force in Eq.~\eqref{eqn:Lorentz-force-final}. To account for the explicit configuration dependence of the Meissner image, we temporarily treat the particle position $\mathbf{R}$ and the image position, denoted by an auxiliary position $\mathbf{R}'$, as independent variables during differentiation. The gradient $\grad_{\mathbf{R}}$ acts only on the particle coordinate $\mathbf{R}$,  while $\mathbf{R}'$ is held fixed. After each differentiation, the physical configuration is recovered by setting $\mathbf{R}'=\mathbf{R}$. This identification is understood throughout all following derivations. 

Substituting Eq.~\eqref{eqn:meissner-bfield-general} into Eq.~\eqref{eqn:Lorentz-force-final} yields the force generated by the Meissner image,
\begin{align}
\label{eqn:meissner-force-initial}
\mathbf{F}_{\mathrm{im}}&=\mu_0\int\mathrm{d}^3\boldsymbol{\xi}\,\mathrm{d}^3\boldsymbol{\xi}'\,\left[\mathrm{R}\mathbf{M}_0(\boldsymbol{\xi})\right]\cdot\grad_{\mathbf{R}}\nonumber\\
&\left[\mathrm{G}\left(\mathrm{R}\boldsymbol{\xi}-\mathrm{A}\mathrm{R}\boldsymbol{\xi}'+\mathbf{R}-\mathrm{A}\mathbf{R}'\right)\mathrm{A}\mathrm{R}\mathbf{M}_0(\boldsymbol{\xi}')\right].
\end{align}
Using the total symmetry property of the field-gradient tensor $\grad \otimes \mathrm{G}(\mathbf{r})$, the gradient can be commuted through the integral such that the force follows from the auxiliary potential
\begin{align}
\label{eqn:meissner-phi-RR1-def}
& U(\mathbf{R},\mathbf{R}')=-\mu_0\int\mathrm{d}^3\boldsymbol{\xi}\,\mathrm{d}^3\boldsymbol{\xi}'\,\mathbf{M}_0(\boldsymbol{\xi})\nonumber\\
&\quad\cdot\mathrm{R}^\mathsf{T}\mathrm{G}\left(\mathrm{R}\boldsymbol{\xi}-\mathrm{A}\mathrm{R}\boldsymbol{\xi}'+\mathbf{R}-\mathrm{A}\mathbf{R}'\right)\mathrm{A}\mathrm{R}\mathbf{M}_0(\boldsymbol{\xi}'),
\end{align}
according to
\begin{equation}
\label{eqn:meissner-force-gradoutside}
\mathbf{F}_{\mathrm{im}}=-\grad_{\mathbf{R}}U(\mathbf{R},\mathbf{R}').
\end{equation}
We now show that the physical interaction potential associated with the Meissner image is given by $U(\mathbf{R},\mathbf{R})/2$. Since the physical configuration is obtained by identifying $\mathbf{R}'=\mathbf{R}$, the derivative of $U(\mathbf{R},\mathbf{R})$ follows from the chain rule as
\begin{equation}
\label{eqn:meissner-phi-RR1-grad}
\grad_{\mathbf{R}}U(\mathbf{R},\mathbf{R})=\grad_{\mathbf{R}}U(\mathbf{R},\mathbf{R}')+\grad_{\mathbf{R}'}U(\mathbf{R},\mathbf{R}').
\end{equation}
The first term is directly related to the force through Eq.~\eqref{eqn:meissner-force-gradoutside}. To evaluate the second term, we use the reflection symmetry of the Green tensor,
\begin{equation}
\label{eqn:G-trans-properties}
\mathrm{G}(-\mathrm{A}\mathbf{R})=\mathrm{A}\mathrm{G}(\mathbf{R})\mathrm{A},
\end{equation}
together with the symmetry of the Green tensor $\mathrm{G}^\mathsf{T}=\mathrm{G}$. The force can then be rewritten as
\begin{align}
\label{eqn:meissner-force-reflection}
\mathbf{F}_{\mathrm{im}}&=\mu_0\grad_{\mathbf{R}}\int\mathrm{d}^3\boldsymbol{\xi}\,\mathrm{d}^3\boldsymbol{\xi}'\,\mathbf{M}_0(\boldsymbol{\xi}')\nonumber\\&\cdot\mathrm{R}^\mathsf{T}\mathrm{G}\left(\mathrm{R}\boldsymbol{\xi}'-\mathrm{A}\mathrm{R}\boldsymbol{\xi}+\mathbf{R}'-\mathrm{A}\mathbf{R}\right)\mathrm{A}\mathrm{R}\mathbf{M}_0(\boldsymbol{\xi}).
\end{align}
Interchanging the integration variables $\boldsymbol{\xi}\leftrightarrow\boldsymbol{\xi}'$ together with the auxiliary coordinates $\mathbf{R}\leftrightarrow\mathbf{R}'$ gives 
\begin{equation}\label{eqn:meissner-force-reflection-eval}
   \mathbf{F}_{\mathrm{im}}=-\grad_{\mathbf{R}'}U(\mathbf{R},\mathbf{R}') .
\end{equation}
Furthermore, since $U$ depends on $\mathbf{R}$ and $\mathbf{R}'$ only through the combination $\mathbf{R}-\mathrm{A}\mathbf{R}'$, the gradients satisfy
\begin{equation}\label{eqn:meissner-force-dRdR1}
\grad_{\mathbf{R}'}U(\mathbf{R},\mathbf{R}')=-\mathrm{A}\grad_{\mathbf{R}}U(\mathbf{R},\mathbf{R}')=\mathrm{A}\mathbf{F}_{\mathrm{im}}.
\end{equation}
Combining these relations yields $\mathbf{F}_{\mathrm{im}}=-\mathrm{A}\mathbf{F}_{\mathrm{im}}$. Since the components of the Meissner force parallel to the surface change signs under this transformation and hence vanish, the Meissner force acts only normal to the superconducting surface.
Substituting this relation into Eq.~\eqref{eqn:meissner-phi-RR1-grad} gives the Meissner force
\begin{equation}
\label{eqn:meissner-force-symmetrised}
\mathbf{F}_{\mathrm{im}}=-\dfrac{1}{2}\grad_{\mathbf{R}}U(\mathbf{R},\mathbf{R}),
\end{equation}
such that the Meissner contribution is generated by the interaction potential $V_\mathrm{im}=
U(\mathbf{R},\mathbf{R})/2$. The potential corresponding to the Meissner-image magnetization is hence explicitly given as
\begin{align}
\label{eqn:meissner-image-potential}
V_\mathrm{im}=&-\dfrac{\mu_0}{2}\int\mathrm{d}^3\boldsymbol{\xi}\,\mathrm{d}^3\boldsymbol{\xi}'\,\mathbf{M}_0(\boldsymbol{\xi})\nonumber\\
&\cdot\mathrm{R}^\mathsf{T}\mathrm{G}\left[\mathrm{R}\boldsymbol{\xi}-\mathrm{A}\mathrm{R}\boldsymbol{\xi}'+
(\mathbb 1-\mathrm{A})\mathbf{R}\right]\mathrm{A}\mathrm{R}\mathbf{M}_0(\boldsymbol{\xi}').
\end{align}
Compared to the frozen-image contribution, the Meissner potential acquires an additional factor of $1/2$, originating from the fact that the Meissner image field is generated self-consistently by the instantaneous particle configuration.

\subsection{Torque from Meissner potential}
\label{Torque from Meissner potential}
Having derived the Meissner-image potential, we now verify that it generates the correct torque according to Eq.~\eqref{eq:torque_potential}. As in the force derivation, we treat the image position $\mathbf{R}'$ as independent of the particle position $\mathbf{R}$ during differentiation. For compactness, we hide the argument of the Green tensor $\mathrm{G}=\mathrm{G}(\mathrm{R}\boldsymbol{\xi} - \mathrm{A}\mathrm{R}  \boldsymbol{\xi}' + \mathbf{R}-\mathrm{A}\mathbf{R}')$ in the following analysis.
The rotational derivative acts on
three orientation-dependent contributions: the particle magnetization, the image magnetization, and the Green-tensor argument, 
\begin{align}
& - \partial_\mu V_\mathrm{im} = \dfrac{\mu_0}{2}\int\mathrm{d}^3\boldsymbol{\xi} \, \mathrm{d}^3\boldsymbol{\xi}' \left\{ [(\partial_\mu\mathrm{R}) \mathbf{M}_0(\boldsymbol{\xi})] \cdot \mathrm{G}\mathrm{A}\mathrm{R} \mathbf{M}_0(\boldsymbol{\xi}') \right. \nonumber\\
& \: + \, [\mathrm{R}\mathbf{M}_0(\boldsymbol{\xi})]\cdot\mathrm{G}\mathrm{A}(\partial_\mu\mathrm{R})\mathbf{M}_0(\boldsymbol{\xi}')  + \, \left[(\partial_\mu\mathrm{R}\boldsymbol{\xi})-\mathrm{A}(\partial_\mu\mathrm{R}\boldsymbol{\xi}')\right] \nonumber \\
& \quad \phantom{\dfrac{1}{2}} \left. \cdot \, \grad_{\mathbf{R}} \left[ \mathbf{M}_0(\boldsymbol{\xi}) \cdot \mathrm{R}^\mathsf{T}\mathrm{G}\mathrm{A}\mathrm{R} \mathbf{M}_0(\boldsymbol{\xi}') \right] \right\}.
\label{eqn:dVim_full_rot_derivative}
\end{align}
Here $\partial_\mu$ denotes differentiation with respect to the Euler angles $\mu=\alpha,\beta,\gamma$. The terms generated by differentiating the image magnetization and the image contribution to the Green-tensor argument can be mapped onto the corresponding particle terms after interchanging the dummy variables $\boldsymbol{\xi}\leftrightarrow\boldsymbol{\xi}'$ and using the symmetry and reflection properties of the Green tensor, as shown in App.~\ref{app:torque_derivation_details}. The factor $1/2$ in the Meissner potential is thereby canceled, and Eq.~\eqref{eqn:dVim_full_rot_derivative} reduces to
\begin{align}
-\partial_\mu& V_\mathrm{im}=\mu_0\int\mathrm{d}^3\boldsymbol{\xi}\,\mathrm{d}^3\boldsymbol{\xi}'\{[(\partial_\mu\mathrm{R})\mathbf{M}_0(\boldsymbol{\xi})]\cdot\mathrm{G}\mathrm{A}\mathrm{R}\mathbf{M}_0(\boldsymbol{\xi}')\nonumber\\
&+(\partial_\mu\mathrm{R}\boldsymbol{\xi})\cdot [(\mathrm{R}\mathbf{M}_0(\boldsymbol{\xi})) \cdot\grad_{\mathbf{R}}\otimes \mathrm{G}\mathrm{A}\mathrm{R}\mathbf{M}_0(\boldsymbol{\xi}')]\}.
\label{eqn:dVim_symmetrized}
\end{align}

Using the relation for the derivative of the
rotation tensor with respect to the Euler angles $\partial_\mu\mathrm{R}=\mathbf{u}_\mu\times \mathrm{R}$, we obtain
\begin{align}
-\partial_\mu &V_\mathrm{im}=\mu_0\int\mathrm{d}^3\boldsymbol{\xi}\,\mathrm{d}^3\boldsymbol{\xi}'\{[\mathbf{u}_\mu\times\mathrm{R}\mathbf{M}_0(\boldsymbol{\xi})]\cdot\mathrm{G}\mathrm{A}\mathrm{R}\mathbf{M}_0(\boldsymbol{\xi}')\nonumber\\
&+(\mathbf{u}_\mu\times\mathrm{R}\boldsymbol{\xi})\cdot\left[\mathbf{M}_0(\boldsymbol{\xi})\cdot\mathrm{R}^\mathsf{T}\grad_{\mathbf{R}}\right]\mathrm{G}\mathrm{A}\mathrm{R}\mathbf{M}_0(\boldsymbol{\xi}')\}.
\label{eqn:dVim_torque_form}
\end{align}
On the other hand, inserting the Meissner-image field \cref{eqn:meissner-bfield-general} into the Lorentz torque expression 
 in Eq.~\eqref{eqn:Lorentz-torque-final} gives the image torque  
\begin{align}\label{eqn:Meissner-torque-initial-expression}
\mathbf{N}_{\mathrm{im}} & = \mu_0 \int \mathrm{d}^3\boldsymbol{\xi}\,\mathrm{d}^3\boldsymbol{\xi}' \left\{ [\mathrm{R}\mathbf{M}_{0}(\boldsymbol{\xi})] \times \mathrm{G} \mathrm{A}\mathrm{R}  \mathbf{M}_{0}(\boldsymbol{\xi}') \right.\nonumber\\
& \quad + \left. (\mathrm{R}\boldsymbol{\xi}) \times [\mathbf{M}_{0}(\boldsymbol{\xi}) \cdot \mathrm{R}^\mathsf{T} \grad_{\mathbf{R}}] \mathrm{GAR} \mathbf{M}_{0}(\boldsymbol{\xi}') \right\}.
\end{align}
Taking the scalar product with $\mathbf{u}_\mu$ and using the cyclic property $\mathbf{u}_\mu\cdot(\mathbf{a}\times\mathbf{b})=(\mathbf{u}_\mu\times\mathbf{a})\cdot\mathbf{b}$, we find that Eq.~\eqref{eqn:Meissner-torque-initial-expression} corresponds to Eq.~\eqref{eqn:dVim_torque_form}, that is, $\mathbf N_{\mathrm{im}}\cdot\mathbf{u}_\mu =-\partial_\mu V_\mathrm{im}$.
This confirms that the Meissner-image potential reproduces the correct generalized torque.

\subsection{Multipole expansion of the interaction potential}
\label{Multipole expansion of potential}

We now perform a multipole expansion of the magnetic interaction potential derived above by assuming that the particle is small compared to the characteristic length scale on which the magnetic fields vary, such that the fields can be Taylor expanded about the particle center-of-mass position $\mathbf{R}$, 
\begin{equation}\label{eqn:B-Taylor}
\mathbf{B}(\mathbf{R}+\mathrm{R}\boldsymbol{\xi} )=\sum_{n=0}^\infty\dfrac{1}{n!}[(\mathrm{R}\boldsymbol{\xi}) \cdot\grad]^n\mathbf{B}(\mathbf{R}).
\end{equation}

We first consider the Meissner-image interaction potential in Eq.~\eqref{eqn:meissner-image-potential}. The field expansion in Eq.~\eqref{eqn:B-Taylor} enters as the expansion of the Green tensor. Introducing the image separation vector
$\mathbf{d}(\mathbf{R})=(\mathbb{1}-\mathrm{A})\mathbf{R}$ and assuming $|\boldsymbol{\xi}|,|\boldsymbol{\xi}'|\ll|\mathbf{d}|$, we expand the Green tensor as 
\begin{align}
\mathrm{G}(\mathrm{R}\boldsymbol{\xi}-\mathrm{A}\mathrm{R}\boldsymbol{\xi}'+\mathbf{d})&\simeq\mathrm{G}(\mathbf{d})+[(\mathrm{R}\boldsymbol{\xi}-\mathrm{A}\mathrm{R}\boldsymbol{\xi}')\cdot\grad]\mathrm{G}(\mathbf{d})\nonumber\\
&-[(\mathrm{R}\boldsymbol{\xi})\cdot\grad][(\mathrm{A}\mathrm{R}\boldsymbol{\xi}')\cdot\grad]\mathrm{G}(\mathbf{d}).
\end{align}
Here, we retained terms of zeroth and first order in the internal coordinates, which describe dipole-dipole and dipole-quadrupole interactions, together with the mixed second-order term corresponding to the quadrupole-quadrupole interaction. The pure second-order terms, corresponding to dipole-octupole interactions, are neglected. The retained interaction potential can therefore be expressed in terms of the magnetic dipole moment
\begin{equation}\label{eqn:def-dipole-moment}
{\bf m}_0 = \int \mathrm{d}^3\boldsymbol{\xi} \, \mathbf{M}_0(\boldsymbol{\xi}),
\end{equation}
and the traceless symmetric quadrupole tensor
\begin{align}\label{eqn:def-quadrupole-moment}
{\rm Q}_0 = & 3\int \mathrm{d}^3 \boldsymbol{\xi} \left\{ \mathbf{M}_0( \boldsymbol{\xi}) \otimes  \boldsymbol{\xi}  +   \boldsymbol{\xi}  \otimes \mathbf{M}_0( \boldsymbol{\xi})\vphantom{\dfrac{2}{3}} \right. \nonumber\\ 
& \left. - \dfrac{2}{3} [\mathbf{M}_0( \boldsymbol{\xi}) \cdot  \boldsymbol{\xi}] \mathbb{1} \right\}.
\end{align}

Substituting the truncated Green-tensor expansion into the Meissner potential in Eq.~\eqref{eqn:meissner-image-potential} and using the laboratory-frame dipole moment $\mathbf{m}=\mathrm{R}\mathbf{m}_0$ yields 
\begin{align}\label{eq:V_im_expanded}
&V_{\rm im}\simeq-\dfrac{\mu_0}{2}\mathbf{m}\cdot\mathrm{G}(\mathbf{d})\mathrm{A}\mathbf{m}\nonumber\\
&\;-\dfrac{\mu_0}{2}\int\mathrm{d}^3\boldsymbol{\xi}\,\mathbf{M}_0(\boldsymbol{\xi})\cdot\mathrm{R}^\mathsf{T}
\left[(\mathrm{R}\boldsymbol{\xi})\cdot\grad\right]\mathrm{G}(\mathbf{d})\mathrm{A}\mathbf{m}\nonumber\\
&\;+\dfrac{\mu_0}{2}\int\mathrm{d}^3\boldsymbol{\xi}'\,\mathbf{m}\cdot\left[(\mathrm{A}\mathrm{R}\boldsymbol{\xi}')\cdot
\grad\right]\mathrm{G}(\mathbf{d})\mathrm{A}\mathrm{R}\mathbf{M}_0(\boldsymbol{\xi}')\nonumber\\
&+\dfrac{\mu_0}{2}\int\mathrm{d}^3\boldsymbol{\xi}\,\mathrm{d}^3\boldsymbol{\xi}'\,\mathbf{M}_0(\boldsymbol{\xi})\cdot\mathrm{R}^\mathsf{T}[(\mathrm{R}\boldsymbol{\xi})\cdot\grad]\nonumber\\
&\hphantom{\dfrac{\mu_0}{2}\int\mathrm{d}^3}\times[(\mathrm{A}\mathrm{R}\boldsymbol{\xi}')\cdot\grad]\mathrm{G}(\mathbf{d})\mathrm{A}\mathrm{R}\mathbf{M}_0(\boldsymbol{\xi}').
\end{align}
The first line describes the dipole-dipole interaction between the particle and its image. The second and third terms are first order in the internal coordinates $\boldsymbol{\xi}$ and $\boldsymbol{\xi}'$ and represent the two dipole-quadrupole cross terms, while the final term is bilinear in $\boldsymbol{\xi}$ and $\boldsymbol{\xi}'$ and describes the quadrupole-quadrupole interaction. As shown in App.~\ref{app:meissner_quadrupole}, the two first-order terms are identical, such that their sum removes the factor $1/2$ from the first-order contribution. The combined first-order term is linear in $\boldsymbol{\xi}$ and can be written as the contraction of the Green-tensor derivative with the first moment tensor of the magnetization distribution,  
\begin{equation}
\int \mathrm{d}^3\boldsymbol{\xi}\,\mathbf{M}_0(\boldsymbol{\xi})\otimes\boldsymbol{\xi}.
\end{equation}
Since the Green-tensor derivative is symmetric and traceless for $\mathbf{d}\neq 0$, only the symmetric traceless part of this first moment contributes to the interaction energy, which is  represented by the quadrupole tensor in Eq.~\eqref{eqn:def-quadrupole-moment}. Using the  laboratory-frame quadrupole moment according to $\mathrm{Q}=\mathrm{R}\mathrm{Q}_0\mathrm{R}^\mathsf{T}$, we obtain the quadrupole contribution
\begin{equation}\label{eq:V_im_quad}
V_{\rm im}^{\rm dq}=-\dfrac{\mu_0}{6}(\mathrm{Q}\grad)\cdot\mathrm{G}(\mathbf{d})\mathrm{A}\mathbf{m}.
\end{equation}

The mixed second-order term contains a product of two first-moment tensors, associated with the particle and its reflected image. The second derivative of the Green tensor is likewise totally symmetric and traceless, so only the symmetric traceless part of each tensor contributes. These parts are $\mathrm{Q}/6$ and $\mathrm{A}\mathrm{Q}\mathrm{A}/6$, respectively. As derived in App.~\ref{app:meissner_quadrupole}, their contraction can be written as
\begin{equation}\label{eq:V_imQQ}
V_\mathrm{im}^\mathrm{qq}=\dfrac{\mu_0}{72} (\grad\cdot\mathrm{Q}\grad)\mathrm{tr}[\mathrm{G}(\mathbf{d})\mathrm{A}\mathrm{Q}\mathrm{A}].
\end{equation}

Similarly, we consider the interaction potential generated by the frozen magnetization, with its multipole expansion taken about the reference point $\mathbf{R}_{\mathrm{fr}}$. By defining the laboratory-frame displacement $\boldsymbol{\xi}_\mathrm{fr}=\mathbf{r}-\mathbf{R}_\mathrm{fr}$, the frozen dipole and quadrupole moments, $\mathbf{m}_\mathrm{fr}$ and $\mathrm{Q}_\mathrm{fr}$, are defined analogously to Eqs.~\eqref{eqn:def-dipole-moment} and \eqref{eqn:def-quadrupole-moment}, with $\boldsymbol{\xi}$ replaced by $\boldsymbol{\xi}_\mathrm{fr}$ and $\mathbf{M}_{0}$ replaced by $\mathbf{M}_{\mathrm{fr}}(\boldsymbol{\xi}_{\rm fr} + {\bf R}_{\rm fr})$. Introducing $\mathbf{d}_\mathrm{fr}=\mathbf{R}-\mathbf{R}_{\rm fr}$ and assuming $|\boldsymbol{\xi}|,|\boldsymbol{\xi}_\mathrm{fr}|\ll|\mathbf{d}_\mathrm{fr}|$, we expand the Green tensor in Eq.~\eqref{eqn:frozen-image-potential} about $\mathbf{d}_\mathrm{fr}$, as in the Meissner-image case. Retaining the zeroth- and first-order terms together with the mixed second-order term, the frozen-magnetization potential becomes
\begin{align}
&V_{\rm fr}=-\mu_0\mathbf{m}\cdot\mathrm{G}(\mathbf{d}_{\rm fr})\mathbf{m}_{\rm fr}\nonumber\\
&-\mu_0\int \mathrm{d}^3\boldsymbol{\xi}\,\mathbf{M}_0(\boldsymbol{\xi})\cdot
\mathrm{R}^\mathsf{T}\left[(\mathrm{R}\boldsymbol{\xi})\cdot\grad\right]\mathrm{G}(\mathbf{d}_{\mathrm{fr}})\mathbf{m}_{\mathrm{fr}}\nonumber\\
&+\mu_0\mathbf{m}\cdot\int \mathrm{d}^3\boldsymbol{\xi}_{\mathrm{fr}}\,
\left[(\boldsymbol{\xi}_{\mathrm{fr}}\cdot\grad)\mathrm{G}(\mathbf{d}_{\mathrm{fr}})\right]\mathbf{M}_{\mathrm{fr}}(\mathbf{R}_\mathrm{fr}+\boldsymbol{\xi}_{\mathrm{fr}})\nonumber\\
&+\mu_0\int\mathrm{d}^3\boldsymbol{\xi}\,\mathrm{d}^3\boldsymbol{\xi}_\mathrm{fr}\,\mathbf{M}_0(\boldsymbol{\xi})\cdot\mathrm{R}^\mathsf{T}[(\mathrm{R}\boldsymbol{\xi})\cdot\grad]\nonumber\\
&\hphantom{\dfrac{\mu_0}{2}\int\mathrm{d}^3}\times(\boldsymbol{\xi}_\mathrm{fr}\cdot\grad)\mathrm{G}(\mathbf{d}_\mathrm{fr})\mathbf{M}_\mathrm{fr}(\mathbf{R}_\mathrm{fr}+\boldsymbol{\xi}_\mathrm{fr}).
\end{align}
The first line corresponds to interaction between the particle dipole and the frozen-source dipole, while the second and third terms describe the interaction between the particle quadrupole and the frozen dipole, and between the frozen quadrupole and the particle dipole, respectively. The final term can be identified with the interaction between the particle quadrupole and the frozen quadrupole. Applying the same symmetry and tracelessness arguments as in the Meissner-image case yields
\begin{equation}\label{eq:V_fr_quad}
V_\mathrm{fr}^{\mathrm{dq}}=-\dfrac{\mu_0}{6}(\mathrm{Q}\grad)\cdot\mathrm{G}(\mathbf{d}_{\rm fr})\mathbf{m}_{\rm fr}+
\dfrac{\mu_0}{6}(\mathrm{Q}_{\rm fr}\grad)\cdot\mathrm{G}(\mathbf{d}_{\rm fr})\mathbf{m},
\end{equation}
and
\begin{equation}\label{eq:V_frQQ}
V_\mathrm{fr}^\mathrm{qq}=\dfrac{\mu_0}{36} (\grad\cdot\mathrm{Q}\grad)\mathrm{tr}[\mathrm{G}(\mathbf{d}_\mathrm{fr})\mathrm{Q}_\mathrm{fr}].
\end{equation}
Combining the retained Meissner and frozen contributions therefore gives
\begin{equation}
 V_{\rm fr}+V_{\rm im}\simeq V_{\rm dd}+V_{\rm dq}+V_\mathrm{qq},
\end{equation}
where the individual contributions reproduce the dipole-dipole, dipole-quadrupole, and quadrupole-quadrupole potentials given in Eqs.~\eqref{eqn:dipole-potential}, \eqref{eqn:quadrupole-potential}, and \eqref{eqn:quad-quad-potential}.

Finally, when the pinned-flux configuration is generated by the micromagnet's field  during the superconducting transition, the associated frozen magnetization is modeled by the inverse-mirror distribution in Eq.~\eqref{eqn:def-frozen-magnetisation}. By parameterizing the frozen-source displacement as $\boldsymbol{\xi}_\mathrm{fr}=\mathrm{A}\mathrm{R}(\Omega_\mathrm{fr})\boldsymbol{\xi}$, the corresponding frozen moments reduce to 
\begin{equation}
    {\bf m}_{{\rm fr}} = - {\rm A}{\bf m}(\Omega_{{\rm fr}}),\quad{\rm Q}_{{\rm fr}} = -{\rm A}{\rm Q}(\Omega_{{\rm fr}}){\rm A},
\end{equation}
as stated in Sec.~\ref{Levitation potential}.

\section{Trapping minimum}\label{Trapping minimum}

\subsection{Zero-field minimum }
\label{Weak-field trapping}

Having derived the approximate potential, we now determine the equilibrium trapping configuration $q_{\rm eq}$. Here and in the following we denote configuration space coordinates by $q = (q_1,q_2,q_3,q_4,q_5,q_6) \equiv (x,y,z,\alpha,\beta,\gamma)$. We first consider the in-plane freezing configuration $\beta_{\rm fr} \simeq \pi/2$ in the case that the frozen image is generated by particle magnetization, and neglect external magnetic fields and gravity. Choosing the coordinate frame such that ${\bf R}_{\rm fr} = -z_{\rm fr} {\bf e}_3$ and ${\bf m}_{\rm fr}$ points in direction $-{\bf e}_1$, the equilibrium configuration of the pure dipole potential \eqref{eqn:dipole-potential} is given by
\begin{equation}
\label{eqn:weak-field-qdip}
q_\mathrm{dip} = (0,0, z_{\mathrm{fr}}, 0, \pi/2),
\end{equation}
while the angle $\gamma$ stays undetermined. This follows from the dipole force and torque in Eqs.~\eqref{eqn:force-pure-dipole-meiss-fixed} and \eqref{eqn:torque-pure-dipole-meiss-fixed}, which vanish for $\mathbf{R}=z_\mathrm{fr}\mathbf{e}_3$ and  $\mathbf{m}=-\mathrm{A}\mathbf{m}_\mathrm{fr}$, in agreement with Refs.~\cite{wang2019dynamics,gieseler2020single}. 

An anisotropic quadrupole moment lifts this rotational degeneracy of the dipole potential and generates an effective trapping potential for $\gamma$. We assume that the quadrupole tensor is diagonal in the principal-axes frame, with distinct transverse eigenvalues, $Q_1\neq Q_2$,
\begin{equation}
 \mathrm{Q} = \sum^{3}_{k=1}Q_{k} \mathbf{n}_{k}\otimes\mathbf{n}_{k}.
 \label{eq:Q_tensor_diag}
\end{equation}
The quadrupole-quadrupole interaction~\eqref{eqn:quad-quad-potential} contributes directly at the dipole equilibrium. Defining its reduced angular potential by subtracting a fixed reference energy, $V_\mathrm{qq}^\mathrm{red}=V_\mathrm{qq}(q_\mathrm{dip},\gamma)-V_\mathrm{qq}(q_\mathrm{dip},\gamma_\mathrm{fr})$,
we obtain
\begin{align}\label{eq:VQQ_red}
V_\mathrm{qq}^\mathrm{red}(q_\mathrm{dip},\gamma)=&\dfrac{ \mu_{0} (Q_1 - Q_2)^{2}}{6 \pi (2z_{\mathrm{fr}})^5}\sin^2(\gamma - \gamma_{{\rm fr}})\nonumber\\
&\times\left[1+\dfrac{3}{16}\sin^2(\gamma+\gamma_\mathrm{fr})\right].
\end{align}

The dipole-quadrupole potential, on the other hand, is independent of $\gamma$ at $q_\mathrm{dip}$ since the first two terms in Eq.~\eqref{eqn:quadrupole-potential} cancel and the third term is independent of $\gamma$. Nevertheless, its gradients with respect to the other five coordinates depend on $\gamma$. Relaxation of these coordinates therefore generates an additional, indirect angular potential.  To estimate the corrected equilibrium, we expand the dipole equilibrium as $q_{\mathrm{est}} = (q_\mathrm{dip} + \Delta q,\gamma)$. Separating the energy at the dipole configuration, we define the displacement-dependent energy $\Delta U (\Delta q,\gamma)=V(q_\mathrm{dip}+\Delta q,\gamma)-V(q_\mathrm{dip},\gamma)$. Expanding about the dipole equilibrium gives
\begin{align}\label{eqn:V-q_eq_taylor-expand}
\Delta U(\Delta q,\gamma)\simeq &\sum_{i=1}^{5}
\partial_i V (q_\mathrm{dip},\gamma)\Delta q_{i} \nonumber\\
& +\dfrac{1}{2}\sum_{i,j=1}^{5}(K_\mathrm{dip})_{ij}\Delta q_{i}\Delta q_{j},
\end{align}
where $\partial_i=\partial/\partial q_i$ and
\begin{equation}
    (K_\mathrm{dip})_{ij}=\partial_i\partial_j V_\mathrm{dd}(q_\mathrm{dip}).
\end{equation}
The leading shifts $\Delta q_i$ are obtained by retaining gradients linear in the quadrupole moments, so that both energy contributions in Eq.~\eqref{eqn:V-q_eq_taylor-expand} are quadratic in those moments. The displacement dependence of $V_\mathrm{qq}$ contributes only at higher orders to the energy and is omitted in this expansion. 

Minimizing $\Delta U$ \eqref{eqn:V-q_eq_taylor-expand} with respect to the five coordinate shifts gives
\begin{align}
    \partial_i V(q_\mathrm{dip},\gamma)+\sum_{j=1}^{5}(K_\mathrm{dip})_{ij}\Delta q_{j} & = 0, 
\end{align}
and hence
\begin{equation}\label{eqn:dipole-correction-equil}
\Delta q_{i}(\gamma)=-\sum_{j=1}^{5}(K_\mathrm{dip}^{-1})_{ij}\partial_j V(q_\mathrm{dip},\gamma).
\end{equation}
Evaluating $\Delta U$ at these shifts yields the relaxation-induced angular potential
\begin{equation}\label{eqn:effective potential-gamma}
V_{\mathrm{rel}}(\gamma)=-\dfrac{1}{2}\sum_{i,j=1}^{5}\partial_i V(q_\mathrm{dip},\gamma)(K_\mathrm{dip}^{-1})_{ij}\partial_j V(q_\mathrm{dip},\gamma).
\end{equation}
For the diagonal quadrupole tensor considered here, neglecting gravity and external fields and omitting an additive constant, this becomes to the retained perturbative order
\begin{align}
V_\mathrm{rel}(\gamma) \simeq &-\dfrac{ \mu_{0} (Q_1 - Q_2)^{2}}{8 \pi (2z_{\mathrm{fr}})^5}\sin^2(\gamma - \gamma_{{\rm fr}})\nonumber\\
    &\times\left[1+\dfrac{5}{36}\sin^2(\gamma+\gamma_{{\rm fr}})\right].
\end{align}
Adding the direct and relaxation-induced contributions yields the reduced angular potential $V_\gamma$ given in Eq.~\eqref{eq:V_gamma}.
The estimated equilibrium angle is determined by
\begin{equation}\label{eqn:minimisation-conditions-gamma}
\partial_{\gamma}V_\gamma(\gamma_{\mathrm{est}})=0,
\end{equation}
with minima at $\gamma_\mathrm{est}\simeq\gamma_\mathrm{fr}\,\pmod\pi$. Gravity and external magnetic fields can further modify the equilibrium configuration and are included in the numerical minimization of the full trapping potential.

\subsection{Impact of external homogeneous magnetic fields}
\label{Impact of external magnetic field}
\begin{figure*}[htp]
    \centering
    \includegraphics[trim={1.1cm 22.7cm 3.2cm 1.0cm},clip,width=18cm]{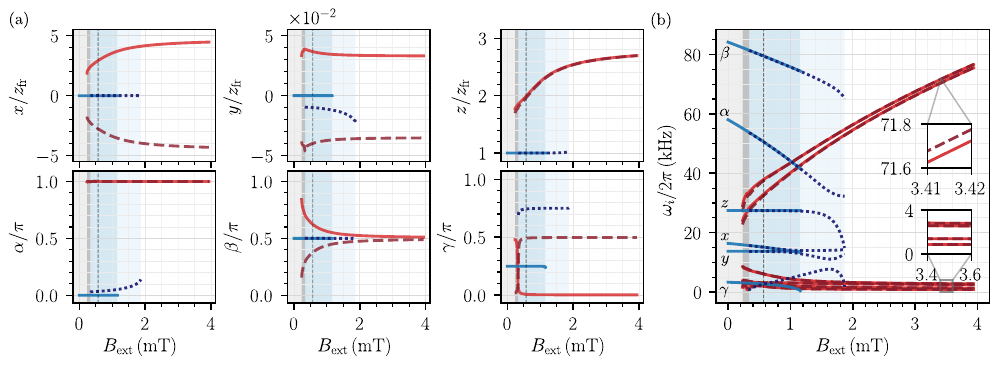}
    \caption{\textbf{Equilibrium configurations and normal-mode frequencies in an opposing magnetic field.}
    (a) Equilibrium configurations as a function of the field magnitude for $\mathbf{B}_\mathrm{ext}=-B_\mathrm{ext}\mathbf{e}_{1}$.
    Blue curves denote the approximately centered weak-field equilibria, with solid and dotted curves corresponding to the branches near $\gamma_{\mathrm{fr}}$ and $\gamma_{\mathrm{fr}} + \pi/2$, respectively.
    Red solid and dashed curves correspond to the positive- and negative-$x$ strong-field equilibria, respectively.  The background shading indicates successive intervals with one, three, four, three, and two locally stable minima as the  field increases, counted modulo the symmetry $\gamma\rightarrow\gamma+\pi$. The vertical dashed line marks the energy crossover at $B_{\mathrm{ext}} = 0.57\,\mathrm{mT}$: the global minimum lies on a weak-field branch below this field and on a strong-field branch above it. Energy differences within each pair of branches are small compared with the energy separation between the two families away from the crossover.
    (b) Mode frequencies $\omega_i/2\pi$ as a function of the homogeneous applied field $B_\mathrm{ext}$. Curve colors and line styles identify the same equilibrium branches as in (a). The upper inset resolves the small difference between the $\beta$-like mode frequencies on the two off-axis branches. The slightly lower-frequency libration is predominantly $\alpha$-like on both branches. 
    The lower inset shows that the corresponding frequencies of the remaining four modes nearly coincide between the two off-axis branches. Particle and flux-freezing parameters are those of Fig.~\ref{fig:modes-spin-rotation-transient}, with $z_\mathrm{fr}=4.4\,\mu\mathrm{m}$.
    }
    \label{fig:strong-field}
\end{figure*}

A homogeneous magnetic field applied parallel to the superconducting surface can qualitatively modify the trapping landscape. Although a homogeneous field exerts no direct force on the magnetic dipole, it can change its equilibrium orientation and thereby alter the position-dependent interaction with the frozen image. We consider a field applied opposite to the dipole-limit zero-field equilibrium orientation, namely  $\mathbf{B}_{\mathrm{ext}} = - B_\mathrm{ext} \mathbf{e}_{1}$. The field dependence of the equilibrium configuration is shown in Fig.~\ref{fig:strong-field}(a). For $\gamma_\mathrm{fr}=\pi/4$, the equilibrium branch connected to the zero-field configuration with $\gamma\simeq\gamma_\mathrm{fr}$ changes only negligibly over the initial weak-field range (light gray shading). Throughout this section, symmetry-related configurations under $\gamma\rightarrow\gamma+\pi$ are counted as a single equilibrium branch.

Increasing the opposing field stabilizes two off-axis equilibria whose dipole moments approach alignment with the applied field. Their emergence and the associated change in positional confinement can be understood with a reduced dipole model in the limit of complete alignment. Neglecting gravity and the quadrupole interactions, we restrict the configuration space to the planar subspace  $y=0$, $\beta=\pi/2$. On this subspace, the dipole and external-field contributions to the total trapping potential~\eqref{eqn:total-trapping-potential} reduce to 
\begin{equation}\label{eq:potential_reduced_strongB}
V=m_0\left(B_\mathrm{ext}+\dfrac{\mu_0 m_0[2x^2-(z+z_{\rm fr})^2]}{4\pi[x^2 + (z+z_{\rm fr})^2]^{5/2}}\right)\cos\alpha+\dfrac{\mu_{0} m_{0}^2}{(4z)^3\pi},
\end{equation}
where $m_0=|\mathbf{m}|$. At the dipole-limit weak-field equilibrium $x=0$ and $z=z_\mathrm{fr}$, the two orientation-dependent contributions compete: the frozen-image interaction favors $\alpha=0$, whereas the opposing external field favors $\alpha=\pi$. As the external field becomes dominant, the strong-field equilibrium branches approach alignment with the applied field, $\alpha\rightarrow\pi$. In the reduced model, reversing the magnetic dipole moment changes the sign of the position-dependent dipole interaction with the frozen image, making the centered configuration $x=0$ unstable to lateral displacements.  Instead, at fixed $\alpha=\pi$, the reduced potential exhibits two off-axis minima along $\mathbf{e}_1$, located at $x=\pm x_\mathrm{eq}$. Within this reduced model, these equilibria are related by the reflection symmetry $x\rightarrow -x$ and are energetically degenerate. The external-field contribution supplies positive angular stiffness around the reversed orientation and can thereby stabilize the off-axis equilibria. The reduced potential illustrates the mechanism for off-axis confinement, while the equilibrium configurations and their stability are determined from the full six-dimensional potential, which also allows lateral shifts along $y$ and out-of-plane tilts of the dipole.

The full numerical calculation shows that the two off-axis branches become stable near $B_\mathrm{ext}=0.23\, \mathrm{mT}$, while the approximately centered branch connected to zero field remains stable. Three stable minima therefore coexist over the dark gray interval in Fig.~\ref{fig:strong-field}. An additional bifurcation occurs near $B_\mathrm{ext}=0.34\, \mathrm{mT}$, where a second approximately centered minimum appears, producing an interval with four coexisting minima (dark blue shading). This additional branch has $\gamma\simeq\gamma_\mathrm{fr}+\pi/2$ and exhibits a small lateral displacement along $y$, accompanied by a small shift in the angle $\alpha$. Since the homogeneous field has no direct $\gamma$ dependence, its effect on the $\gamma$ confinement is mediated by the quadrupole interaction and the relaxation of the other configurational coordinates. 

At $B_\mathrm{ext}=0.57\,\mathrm{mT}$, the lowest-energy centered and off-axis equilibria become degenerate in potential energy. Above this field, the global minimum lies on an off-axis branch, while the centered equilibria persist as local minima until their respective stability boundaries. With increasing field, the minimum connected to $B_{\mathrm{ext}}=0$ first loses stability, leaving three stable minima: the approximately centered branch near $\gamma_{\mathrm{fr}}+\pi/2$ and the two off-axis branches (light blue shading). At even larger fields, the remaining approximately centered minimum also loses stability, leaving only the two off-axis strong-field equilibria, whose dipole orientations approach $\alpha=\pi$.

The stability boundaries are determined from the stiffness matrix of the full trapping potential in the six-dimensional configuration space, with components
\begin{equation}
 K_{ij} = \partial_i \partial_j V(q_{\mathrm{eq}}),
\end{equation}
evaluated along each equilibrium branch. Harmonic stability as a potential minimum requires this matrix to be positive definite. At a stability boundary, its smallest eigenvalue vanishes, indicating a direction of zero quadratic curvature.  The corresponding critical fields depend on the dipole and quadrupole interactions and on the field-dependent equilibrium shifts.  For the quadrupole configuration considered here, the quadrupole interaction also weakly breaks the ideal $x\rightarrow -x$ symmetry of the reduced model,  rendering the two strong-field equilibria slightly inequivalent. The resulting multistability and rapid variation of the equilibrium configuration near the bifurcations may provide access to nonlinear motion and transitions between trapping wells.

\section{Eigenmodes and approximate frequencies}
\label{sec:eigenmodes}

\subsection{Hamiltonian in quadratic form}
\label{Hamiltonian in quadratic form}
Having determined the equilibrium configuration of the levitated magnet, we now analyze its small-amplitude dynamics within the harmonic approximation. To determine the normal modes, we first express the Hamiltonian~\eqref{eqn:nonlinear-hamiltonian} in terms of the generalized coordinates $q$ and their conjugate momenta $p=(p_x,p_y,p_z, p_\alpha, p_\beta, p_\gamma)$.  Using Eq.~\eqref{eq:J_in_p}, the rotational kinetic energy can be expressed in terms of the canonical Euler momenta $p_\mu$ and the components  $\mathbf{m}\cdot\mathbf{u}_\mu/\gamma_\mathrm{mag}$ of the intrinsic angular momentum associated with the magnetization. The full Hamiltonian then takes the compact form
\begin{equation}\label{eq:H_phase_space_compact}
 H=\dfrac{1}{2}\!\sum_{i,j=1}^6\!\left(p_i\!+\!\dfrac{\mathbf{m}\cdot\mathbf{u}_i}{\gamma_\mathrm{mag}}\right)\!\left[\mathrm{M}^{-1}(q)\right]_{ij}\!\left(p_j\!+\!\dfrac{\mathbf{m}\cdot\mathbf{u}_j}{\gamma_\mathrm{mag}}\right)\!+\!V,  
\end{equation}
where we use the convention $\mathbf{m}\cdot\mathbf{u}_i=0$ for $i\in\{1,2,3\}$, while $i=4,5,6$ correspond, respectively, to  $\mu,\nu=\alpha,\beta,\gamma$. The components of the inverse mass matrix $\mathrm{M}$ in the full configuration space are defined by
\begin{equation}\label{eqn:mass-matrix-def}
   \left[\mathrm{\mathrm{M}}^{-1}(q)\right]_{ij}= \begin{cases}
       \dfrac{1}{M}\delta_{ij}, & i,j \in\{1,2,3\},\\
     \mathbf{u}^i\cdot\mathrm{I}^{-1}\mathbf{u}^{j}, & i,j \in\{4,5,6\},\\
      0, & \text{otherwise},
   \end{cases} 
\end{equation}
where in the rotational block,  $i,j=4,5,6$ correspond to $\mu,\nu=\alpha,\beta,\gamma$, respectively.

The equilibrium momentum $p_\mathrm{eq}$ follows from $\partial H/\partial p_i=0$, which yields $ p_{\mathrm{eq},i}=-\mathbf{m}(q_\mathrm{eq})\cdot\mathbf{u}_i(q_\mathrm{eq})/\gamma_\mathrm{mag}$. At this point, the first derivatives of the kinetic Hamiltonian term with respect to the configuration coordinates vanish. The equilibrium configuration $q_\mathrm{eq}$ is therefore determined solely by the stationary condition, $\partial_i V(q_\mathrm{eq}) = 0$. Consequently, $(q_{\mathrm{eq}},p_{\mathrm{eq}})$ is a stationary point of the Hamiltonian dynamics, satisfying $\dot q_i=\dot p_i=0$, although the equilibrium values of the canonical Euler momenta need not vanish.

Introducing small deviations from equilibrium $\delta q_i = q_i -q_{\mathrm{eq},i}$ and $\delta p_i = p_i - p_{\mathrm{eq},i}$, and expanding the Hamiltonian~\eqref{eq:H_phase_space_compact} to second order in these deviations gives
\begin{align}\label{eqn:linear-ham-phase-space-labframe}
&H_\mathrm{lin}  =  \dfrac{1}{2}\!\sum^{6}_{i,j = 1} \! \left[\delta p_{i} + \sum^{6}_{k = 1}\partial_k\left(\dfrac{\mathbf{m}\cdot\mathbf{u}_i}{\gamma_\mathrm{mag}}\right)\delta q_{k}\right]\left(\mathrm{M}^{-1}\right)_{ij}  \nonumber\\ 
&\times\left[\delta p_{j} + \sum^{6}_{\ell=1}\partial_\ell\left(\dfrac{\mathbf{m}\cdot\mathbf{u}_j}{\gamma_\mathrm{mag}}\right)\delta q_{\ell}\right]+\dfrac{1}{2}\sum_{i,j=1}^6 \delta q_{i} K_{ij}  \delta q_{j},
\end{align}
where we have omitted the constant equilibrium potential energy and all quantities are evaluated at $q_\mathrm{eq}$, i.e., ${\rm M}^{-1} = {\rm M}^{-1}(q_{\rm eq})$.

\subsection{Normal modes in the absence of spin-rotation coupling}
\label{Transformation to the normal modes}

In the absence of spin-rotation coupling, that is, neglecting the terms proportional to 1/$\gamma_\mathrm{mag}$ in the kinetic Hamiltonian, such that
\begin{equation}
H_\mathrm{0}  =  \dfrac{1}{2}\!\sum^{6}_{i,j = 1} \!\left[ \delta p_{i}\left(\mathrm{M}^{-1}\right)_{ij} \delta p_{j}+\delta q_{i} K_{ij}  \delta q_{j}\right],
\end{equation}
the Hamiltonian can be diagonalized in terms of normal-mode coordinates $\theta_i$ and their conjugate momenta $\zeta_{i}$, 
\begin{equation}\label{eq:H_diagonal}
 H_0=\dfrac{1}{2}\sum_{i=1}^6\left(\zeta_i^2+\omega_i^2\theta_i^2\right). 
\end{equation}
Here, $\omega_i$ denote the corresponding eigenfrequencies, for which analytical expressions in the dipole-dominated limit are discussed below. The normal-mode coordinates $\theta_i$, which have dimensions of length times the square root of mass, and their conjugate momenta $\zeta_{i}$ are defined by the canonical transformation
\begin{equation}\label{eqn:symp-trans-W1}
\theta_{i} = \sum^{6}_{j,k=1} W_{ij} M_{jk}\delta q_{k}, \qquad \zeta_{i} = \sum^{6}_{j=1} W_{ij} \delta p_{j}, 
\end{equation}
where the mass-normalized transformation matrix $\mathrm{W}$ diagonalizes the stiffness matrix $\mathrm{K}$ and the mass matrix $\mathrm{M}$,
\begin{equation}
   \sum_{k,\ell=1}^6 W_{ik} M_{k\ell}W_{j\ell}=\delta_{ij},\quad  \sum_{k,\ell=1}^6 W_{ik} K_{k\ell}W_{j\ell}=\omega_i^2\delta_{ij},
\end{equation}
with generalized eigenvalues $\omega_i^2$.

The normal mode frequencies can be approximated analytically for spheroidal particles with semiaxes $(a,a,c)$ in the dipole-dominated flux-pinning limit. Their principal moments of inertia are $I_{1} =I_{2} = M(a^2+c^2)/5$ and  $I_{3} = 2Ma^2 /5$. Then, the motion along $y$, $z$, and $\alpha$ decouple from the remaining degrees of freedom and therefore  constitute three independent normal modes. Their eigenfrequencies are
\begin{equation}\label{eqn:pure-dipole-omega-yz-alpha}
\dfrac{\omega_{y}}{\omega_{\mathrm{fr}}} = \dfrac{a}{2z_{\mathrm{fr}}}, \quad \dfrac{\omega_{z}}{\omega_{\mathrm{fr}}} = \dfrac{a}{z_{\mathrm{fr}}}, \quad  \dfrac{\omega_{\alpha}}{\omega_{\mathrm{fr}}} = \sqrt{\dfrac{5a^2}{3 (a^2+c^2)}},
\end{equation}
where we defined the frequency scale
\begin{equation}\label{eqn:scaling-omega_f}
\omega_{\mathrm{fr}} = \sqrt{\dfrac{3 \mu_{0} m_{0}^2}{32 \pi Ma^2 z^3_{\mathrm{fr}}}}.
\end{equation}

The two remaining dipole-trapped coordinates, $x$ and $\beta$, are coupled by  the dipole potential and form a pair of hybridized normal modes with frequencies
\begin{align}\label{eqn:pure-dipole-omega-x-beta}
\dfrac{\omega^2_{\pm}}{\omega^2_{\mathrm{fr}}} = \dfrac{5a^2}{3(a^2+c^2)} &\left[\vphantom{ \sqrt{1 + \left (\dfrac{9 (a^2+c^2)}{40z^2_{\mathrm{fr}}} \right )^2}}1 + \dfrac{9(a^2+c^2)}{40z^2_{\mathrm{fr}}}\right. \nonumber\\
&\left.\pm  \sqrt{1 + \left[\dfrac{9 (a^2+c^2)}{40z^2_{\mathrm{fr}}} \right]^2} \right].
\end{align}
In the limit $\sqrt{a^2+c^2} \ll z_{\mathrm{fr}}$, the mixing becomes weak, such that the lower-frequency mode is predominantly translational along $x$, whereas the higher-frequency mode is predominantly librational in $\beta$. Their frequencies reduce to
\begin{equation}\label{eqn:pure-dipole-omega-x-beta-approx}
\dfrac{\omega_{+}}{\omega_{\mathrm{fr}}} \simeq \sqrt{\dfrac{10a^2}{3(a^2+c^2)}}, \qquad \dfrac{\omega_{-}}{\omega_{\mathrm{fr}}} \simeq \sqrt{\dfrac{3}{8}}\dfrac{a}{z_{\mathrm{fr}}}.
\end{equation}

Neglecting hybridization between the angle $\gamma$ and the other modes, the corresponding mode frequency is estimated from the  curvature of the combined reduced angular potential $V_\gamma$~\cref{eq:V_gamma} at $\gamma\simeq\gamma_\mathrm{fr}$, which gives
\begin{equation}\label{eqn:VQuad-freq-est}
\dfrac{\omega_{\gamma}}{\omega_{\mathrm{fr}}} \simeq \sqrt{\dfrac{5}{72}\left[1+\dfrac{1}{3}\sin^2(2\gamma_\mathrm{fr})\right]}\dfrac{|Q_{1} - Q_{2}|}{m_{0}z_{\mathrm{fr}}}.
\end{equation}
For $\gamma_\mathrm{fr}=\pi/4$, as used here, this reduces to 
\begin{equation}\label{eqn:omega_gamma}
\dfrac{\omega_{\gamma}}{\omega_{\mathrm{fr}}} \simeq \sqrt{\dfrac{5}{54}}\dfrac{|Q_{1} - Q_{2}|}{m_{0}z_{\mathrm{fr}}}.
\end{equation}
If the quadrupole eigenvalues scale as $Q_{k} \sim m_{0}a$, appropriate when the quadrupole arises from magnetization inhomogeneities on the particle length scale, the predominantly $\gamma$ mode exhibits the same $z_\mathrm{fr}^{-5/2}$  frequency scaling as the leading translational modes. 

For the approximately centered weak-field equilibria, the opposing field reduces the angular restoring stiffness, producing the decreasing high-frequency branches in Fig.~\ref{fig:strong-field}(b). Over their common stability interval, the branches near $\gamma_\mathrm{fr}$ and $\gamma_\mathrm{fr}+\pi/2$ have nearly coincident higher-frequency modes, whereas the differences are more pronounced for the lower-frequency modes. The branch near $\gamma_\mathrm{fr}+\pi/2$  remains stable up to a larger opposing field. At the stability boundary of each branch, a normal-mode frequency vanishes in the absence of spin-rotation coupling as the trapping potential develops a direction of zero quadratic curvature.

In the strong-field limit, the off-axis equilibria approach the in-plane orientation $\alpha\simeq \pi$, $\beta\simeq \pi/2$. Evaluated at this limiting orientation, the external-field contribution to the stiffness matrix is diagonal in the $\alpha$- and $\beta$-subspace, namely 
\begin{equation}\label{eqn:stiffness-K-ext}
(K_{\rm ext})_{44} = (K_{\rm ext})_{55} = m_0 B_{\rm ext}.
\end{equation}
When this contribution dominates the dipole and quadrupole contributions to the angular stiffness, the $\alpha$ and $\beta$ degrees of freedom are only weakly coupled to the remaining configurational coordinates. The corresponding normal modes are therefore predominantly librational in $\alpha$ and $\beta$. For the spheroidal particle considered above, these two modes are approximately degenerate to leading order, with  frequencies 
\begin{equation}\label{eqn:omega-alpha-beta-strong-DC}
\dfrac{\omega_{\alpha,\mathrm{ext}}}{\omega_\mathrm{fr}} \simeq \dfrac{\omega_{\beta,\mathrm{ext}}}{\omega_\mathrm{fr}} \simeq \sqrt{\dfrac{160\pi B_\mathrm{ext}z_\mathrm{fr}^3a^2}{3\mu_0 m_0(a^2+c^2)}}.
\end{equation}
The two increasing high-frequency branches in the numerical results approach the predicted $\sqrt{B_\mathrm{ext}}$ scaling. The frequencies associated with the two off-axis equilibria at $x_\mathrm{eq}<0$ and $x_\mathrm{eq}>0$ nearly coincide because the two configurations are related by an approximate reflection symmetry. The small residual frequency differences reflect the weak breaking of this symmetry by the quadrupole interaction.

\subsection{Quadratic spin-rotation Hamiltonian}
\label{sec:spin_rotation_coupling}
We now retain the spin-rotation coupling and treat it as a weak perturbation of the trapping Hamiltonian. In the normal-mode basis introduced above, this interaction takes the form of a gyroscopic coupling between the otherwise independent harmonic modes, thereby inducing mode hybridization and modifying the system dynamics. 

It is convenient to separate the transformed spin-rotation coupling matrix into its symmetric $\sigma_{ij}=\sigma_{ji}$ and antisymmetric $\Lambda_{ij}=-\Lambda_{ji}$ parts, defined as 
\begin{align}\label{eqn:linear-ham-W1-tensors}
\sigma_{ij}&= -\dfrac{1}{2}\sum^{6}_{k,\ell =1} W_{ik} \left[ \partial_k\left(\dfrac{\mathbf{m}\cdot\mathbf{u}_\ell} {\gamma_\mathrm{mag}}\right)+\partial_\ell\left(\dfrac{\mathbf{m}\cdot\mathbf{u}_k}{\gamma_\mathrm{mag}}\right)\right] W_{j\ell},\nonumber\\
\Lambda_{ij}&= -\dfrac{1}{2}\sum^{6}_{k,\ell =1} W_{ik} \left[ \partial_k\left(\dfrac{\mathbf{m}\cdot\mathbf{u}_\ell}{\gamma_\mathrm{mag}}\right)-\partial_\ell\left(\dfrac{\mathbf{m}\cdot\mathbf{u}_k}{\gamma_\mathrm{mag}}\right)\right] W_{j\ell},
\end{align}
where all quantities are evaluated at the equilibrium configuration. In terms of these matrices, the linearized Hamiltonian in Eq.~\eqref{eqn:linear-ham-phase-space-labframe} becomes
\begin{equation}
H_\mathrm{lin} = \dfrac{1}{2} \sum^{6}_{i = 1}
\left[\zeta_{i} - \sum_{j=1}^6\left(\sigma_{ij}- \Lambda_{ij}\right)\theta_{j}\right]^2 + \dfrac{1}{2} \sum^{6}_{i = 1} \omega_i^2 \theta_i^2. 
\end{equation}
Since $\sigma_{ij}$ is symmetric, it can be absorbed into the canonically shifted momenta
\begin{equation} \label{eqn:zeta-pi-shift}
    \pi_i=\zeta_i-\sum_{j=1}^6\sigma_{ij}\theta_j.
\end{equation}
This transformation leaves the coordinates unchanged and preserves the canonical Poisson brackets, $\{\theta_i,\pi_j\}=\delta_{ij}$ and $\{\pi_i,\pi_j\}=0$, such that the Hamiltonian takes the equivalent form
\begin{equation}\label{eq:H_lin_SpinRotation}
H_\mathrm{lin} = \dfrac{1}{2} \sum^{6}_{i = 1}
\left(\pi_{i} + \sum_{j=1}^6\Lambda_{ij}\theta_{j}\right)^2 + \dfrac{1}{2} \sum^{6}_{i = 1} \omega_i^2 \theta_i^2. 
\end{equation}
Combining the corresponding Hamilton equations by eliminating the canonical momenta $\pi_i$ gives the second-order equations of motion in Eq.~\eqref{eqn:eom-normalmodes-summary} without further approximation. To first order in the spin-rotation coupling, the same equations of motion follow from the Hamiltonian
\begin{equation}\label{eqn:linear-hamiltonian-EdHmod}
H_\mathrm{lin} \approx \dfrac{1}{2} \sum^{6}_{i= 1} \left(\pi^2_{i} +  \omega^2_{i} \theta_{i}^2 \right)+ \sum^{6}_{i,j= 1}\pi_{i} \Lambda_{ij} \theta_{j},
\end{equation}
which suffices for the perturbative analysis below.

To determine the effect of the spin-rotation coupling on the normal modes, we seek perturbative 
solutions in the form of Eq.~\eqref{eqn:modes-mixing-complex-ampl-trans-EdH}, where $T_{ij}$ is the mode-mixing matrix that describes the first-order correction to the uncoupled mode eigenvectors. Substitution  into Eq.~\eqref{eqn:eom-normalmodes-summary} and retention of terms to first order in $\Lambda_{ij}$ gives
\begin{equation}\label{eqn:Tij-relations}
(\omega^2_{i} - \omega^2_{j})(\delta_{ij} - \mathrm{i}T_{ij}) = -2\mathrm{i}\omega_{j}\Lambda_{ij}.
\end{equation}
For $i\neq j$, the off-diagonal mixing coefficients are therefore
\begin{equation}\label{eq:Tij_offdiagonal}
  T_{ij} = \dfrac{2\omega_{j}\Lambda_{ij}}{\omega^2_{i} - \omega^2_{j}}.  
\end{equation}
The diagonal elements are not determined by Eq.~\eqref{eqn:Tij-relations} and correspond to the arbitrary phases of the mode eigenvectors. We choose these phases such that the dominant component of each eigenvector is real and positive, yielding $T_{ii}=0$. The perturbative treatment is valid provided that the spin-rotation coupling remains small compared to the frequency separation of the uncoupled modes, $|2\omega_j\Lambda_{ij}|\ll|\omega^2_{i} - \omega^2_{j}|$. The resulting eigenvectors describe the weak, phase-shifted hybridization of well-separated trap modes induced by the spin-rotation coupling.

\section{Linear response and displacement sensing}
\label{Noise and sensing}
\subsection{Generalized forces and driven response}
\label{Driven response}
Having characterized the equilibrium configurations and normal modes of the trapped micromagnet, we now examine their linear response to weak external forces and torques in the presence of noise. This response  provides the basis for the sensitivity analysis developed below. 

In configuration space, the generalized force combines the Cartesian force components with the torque components in the Euler basis, $N_{\mu} = {\bf N} \cdot{\bf u}_{\mu}$, according to $f = (F_x, F_y, F_z,N_{\alpha}, N_{\beta}, N_{\gamma})$. Using the inverse normal-mode transformation of Eq.~\eqref{eqn:symp-trans-W1}, the generalized force conjugate to $\theta_i$ is
\begin{equation}\label{eqn:generalized-forces-normalmode}
\tau_{i} = \sum^{6}_{j=1}W_{ij}f_j.
\end{equation}
We extend the linearized Hamiltonian dynamics following from Eq.~\eqref{eq:H_lin_SpinRotation} by including the deterministic generalized forces $\tau_{i}(t)$, stochastic generalized forces $\eta_i(t)$, and phenomenological damping, resulting in the Langevin equations given in Eq.~\eqref{eqn:normalmodes_Langevin}.
We take the stochastic generalized forces to be stationary and of zero mean, $\mathbb{E}[\eta_i(t)]=0$.  For a finite observation time $t_0$, we define their  Fourier amplitudes by 
\begin{equation}\label{eqn:forward-ft-def}
\eta_{i,t_0}(\omega) = \int_{-t_0/2}^{t_0/2} \mathrm{d}t\,\mathrm{e}^{\mathrm{i}\omega t} \eta_i(t), 
\end{equation}
and the corresponding two-sided cross-spectral-density matrix is
\begin{equation}\label{eq:S_psi}
S_{\eta_i\eta_j}(\omega) =\lim_{t_0\to\infty} \dfrac{1}{t_0}\mathbb{E}\left[\eta_{i,t_0}(\omega)\eta_{j,t_0}^\ast(\omega)\right].
\end{equation}
All spectral densities introduced below follow this convention. For the equilibrium thermal contribution, we adopt a Gaussian white-noise model with correlations  $\mathbb{E}[\eta_i^\mathrm{th}(t)\eta_j^\mathrm{th}(t')]=S_{\eta_i\eta_j}^\mathrm{th}\delta(t-t')$. Here, $S_{\eta_i\eta_j}^\mathrm{th}$ denotes the frequency-independent thermal force spectrum in the normal-mode basis. For a mode coupled to a classical, equilibrium thermal bath at temperature $T$, the fluctuation–dissipation relation connects the thermal generalized-force spectrum to the damping matrix, as derived in App.~\ref{app:damping_rates}. With the two-sided spectral-density convention used here, the relation takes the form given in Eq.~\eqref{eqn:damping-rate-equipartition}. The damping matrix and the corresponding thermal-force spectrum are real and symmetric. The full spectrum $S_{\eta_i\eta_j}(\omega)$ can also include additional frequency-dependent forcing.

In Fourier space, the linear response is
\begin{equation}\label{eq:linear_response_theta}
\theta_i(\omega)=\sum_j\chi_{ij}(\omega)[\tau_j(\omega)+\eta_j(\omega)], 
\end{equation}
where the inverse susceptibility matrix is defined by
\begin{equation}
[\upchi^{-1}(\omega)]_{ij}=(\omega_i^2-\omega^2)\delta_{ij}-\mathrm{i}\omega\Gamma_{ij}+2\mathrm{i}\omega\Lambda_{ij}.
\end{equation}
Since the stochastic forces have zero mean, the mean displacement is determined by the deterministic signal,
\begin{equation}
\mathbb{E}[\theta_i(\omega)]=\sum_{j=1}^6\chi_{ij}(\omega)\tau_j(\omega).
\end{equation}
The cross-spectral density of the displacement fluctuations about this mean is
\begin{equation}
S_{\theta_i\theta_j}(\omega)=\sum_{k,\ell=1}^6\chi_{ik}(\omega)S_{\eta_k\eta_\ell}(\omega)\chi_{j\ell}^\ast(\omega).
\end{equation}

\subsection{Noise-equivalent sensitivity}
\label{Sensitivity bounds}
Neglecting measurement backaction, we use the scalar signal $\psi(t)$ with power spectral density $S_{\psi\psi}$ and the displacement readout introduced in Sec.~\ref{Magnetic noise and sensing} to derive the noise-equivalent input spectrum given in Eq.~\eqref{eqn:sensitivity-threshold-Sdet}.

Assuming that the applied signal, generalized force noise, and measurement-imprecision noise are mutually uncorrelated, the power spectral density of the displacement readout follows from Eqs.~\eqref{eq:linear_response_theta} and \eqref{eqn:output-X-def-with-detector-noise} as
\begin{equation}\label{eqn:psd-detector-output-X}
S_{XX}(\omega) =\left| \sum_{i,j=1}^{6}X_i \chi_{ij}(\omega) \psi_j\right|^2 S_{\psi\psi}(\omega)+S_{XX}^\mathrm{noise}(\omega),
\end{equation}
where the displacement-noise spectrum is
\begin{equation}
S_{XX}^\mathrm{noise}(\omega)=\sum_{i,j,k,\ell=1}^6 X_i\chi_{ik}(\omega) S_{\eta_{k}\eta_{\ell}}(\omega)\chi_{j\ell}^\ast(\omega)X_j+ S_{\rm imp}(\omega).
\end{equation}
Here, $S_{\rm imp}(\omega)$ denotes the measurement-imprecision noise spectrum. The force-noise contribution retains correlations between different modes through the full matrix $S_{\eta_{k}\eta_{\ell}}(\omega)$, whose  magnetic contribution generated by the superconductor is derived in Sec.~\ref{Magnetic noise from superconductor}. 

Equating the signal-induced contribution to the readout spectrum with $S_{XX}^\mathrm{noise}(\omega)$ yields the noise-equivalent input spectrum in Eq.~\eqref{eqn:sensitivity-threshold-Sdet}. The spectral signal-to-noise ratio is $S_{\psi\psi}(\omega)/S_{\rm det}(\omega)$, hence, $S_{\psi\psi}(\omega)\geq S_\mathrm{det}(\omega)$ corresponds to a spectral signal-to-noise ratio of at least unity.

For concreteness, we consider a linear readout of the vertical displacement $\delta z$, namely $X(t)=\delta z(t)+\eta_\mathrm{imp}(t)$.  The inverse normal-mode transformation in Eq.~\eqref{eqn:symp-trans-W1} gives $\delta z=\sum_{i=1}^6 W_{i3}\theta_i$, and therefore $X_i=W_{i3}$. For a signal consisting of a force $F_x(t)$ along the laboratory-frame $x$ direction, Eq.~\eqref{eqn:generalized-forces-normalmode} yields  $\tau_i(t) = W_{i1}F_x(t)$. Analogous relations apply to forces along the remaining laboratory-frame directions, while the modal coupling coefficients for magnetic-field signals follow from the generalized forces and torques induced by the corresponding field components.

\section{Magnetic noise}
\label{Magnetic noise from superconductor}
We now determine the generalized force noise acting on the magnet due to magnetic-field fluctuations generated by the superconductor. We first formulate the coupling to an arbitrary fluctuating field in the laboratory-frame configuration coordinates and then transform the resulting generalized-force cross-spectral density matrix to the normal-mode basis introduced in the previous section.

Let $\delta\mathbf{B}(\mathbf{R},t)$ denote the fluctuating magnetic field evaluated at the center-of-mass position of the magnet. Up to magnetic-quadrupole order, its interaction with the magnet is described analogously to Eq.~\eqref{eqn:b_ext-dip-quad-potential} by 
\begin{equation}
 V_{\delta B}(q,t)=-\sum_{k=1}^3 m_k(q)\delta\!B_k-\dfrac{1}{6}\sum_{k,\ell=1}^3 Q_{k\ell}(q)\partial_k\delta\!B_\ell,
\end{equation}
where $k,\ell=1,2,3$ label the $x,y,z$ components, respectively. The generalized noise force conjugate to $q_i$ is $h_i(q,t)=-\partial_i V_{\delta B}(q,t)$. Evaluating this expression at the equilibrium configuration gives
\begin{align}\label{eq:lab-frame-noise-forcing}
h_{i}(q_{\mathrm{eq}},t) &=   \sum^{3}_{k = 1} \left[ (\partial_{i}m_k)\delta\!B_k + m_{k}\partial_{i}\delta\!B_k\right] \nonumber \\
& + \dfrac{1}{6} \sum^{3}_{k,\ell = 1} \left[(\partial_{i}Q_{k\ell})\partial_{k}\delta\!B_\ell + Q_{k\ell} \partial_{i} \partial_{k}\delta\!B_\ell \right],
\end{align}
where all quantities on the right-hand side are evaluated at $q=q_\mathrm{eq}$. For $i=4,5,6$, the partial derivatives acting on the field vanish because the noise field is independent of the particle orientation. Conversely, for $i=1,2,3$, derivatives of the magnetic moments vanish as they are independent of the particle's center-of-mass position. 

Since the generalized noise in Eq.~\eqref{eq:lab-frame-noise-forcing} depends on the fluctuating magnetic field and its spatial derivatives, we introduce the magnetic-field cross-spectral density 
\begin{equation}
\mathbb{E}[\delta\!B_k(\mathbf{r},\omega)\delta\!B_\ell^\ast(\mathbf{r}',\omega')]=2\pi\delta(\omega-\omega')S_{B_k B_\ell}(\mathbf{r},\mathbf{r}',\omega),
\end{equation}
assuming statistically stationary fluctuations. The cross-spectral densities involving field gradients are obtained by differentiating the corresponding spatial arguments, namely,
\begin{align}\label{eq:field-gradient_spectra}
S_{\partial_s\!B_k B_\ell}(\mathbf{r},\mathbf{r}',\omega)& =  \partial_{r_s} S_{B_k B_\ell}(\mathbf{r},\mathbf{r}',\omega),\nonumber\\
S_{B_k\partial_t\!B_\ell}(\mathbf{r},\mathbf{r}',\omega)& =  \partial_{r'_t} S_{B_k B_\ell}(\mathbf{r},\mathbf{r}',\omega),\nonumber\\
 S_{\partial_s\!B_k \partial_t\!B_\ell}(\mathbf{r},\mathbf{r}',\omega)& =  \partial_{r_s} \partial_{r'_t} S_{B_k B_\ell}(\mathbf{r},\mathbf{r}',\omega),
\end{align}
where $s,t=1,2,3$. Spectra involving the second spatial derivatives appearing in the quadrupole force are obtained analogously by applying the corresponding additional derivatives to the first spatial argument, the second, or both. All derivatives are taken before setting $\mathbf{r}=\mathbf{r}'=\mathbf{R}_\mathrm{eq}$. Combining these field and field-gradient spectra according to Eq.~\eqref{eq:lab-frame-noise-forcing} gives the laboratory-frame generalized-force cross-spectral density matrix with components $S_{h_{i}h_{j}}(\omega)$. The corresponding generalized force noise in the normal-mode basis is obtained using the same transformation as for the deterministic generalized forces in Eq.~\eqref{eqn:generalized-forces-normalmode}. Its cross-spectral-density matrix is therefore
\begin{equation}\label{eqn:psd-noise-lab-to-normalmodes}
S_{\eta_{i}\eta_{j}}(\omega) = \sum^{6}_{k,\ell=1} W_{ik}W_{j\ell} S_{h_{k}h_{\ell}}(\omega).
\end{equation}
In the following, we evaluate the magnetic-field cross-spectral density for thermal current-density fluctuations of normal-conducting electrons (th,c), and small-amplitude thermal motion of pinned vortices (th,v), from which subsequently $S_{h_i h_j}$ and $S_{\eta_i\eta_j}$ are obtained.

\subsection{Thermal current-density fluctuations}
\label{Thermal conduction electrons}
We first consider magnetic noise generated by thermal fluctuations of the dissipative current carried by the normal electronic component of the superconductor. We write $\delta\mathbf{j}_\mathrm{c}(\mathbf{r},t)=\mathbf{j}_\mathrm{c}(\mathbf{r},t)-\mathbb{E}[\mathbf{j}_\mathrm{c}(\mathbf{r},t)]$. Within a local and isotropic conductivity model, the current fluctuations have zero mean $\mathbb{E}[\mathbf{j}_\mathrm{c}(\mathbf{r},t)]=0$ and, in the classical regime, obey~\cite{henkelMagnetostaticFieldNoise2005}
\begin{equation}\label{eqn:j-fourier-noise-3d}
\mathbb{E}[\mathbf{j}_\mathrm{c}(\mathbf{r}, \omega) \otimes \mathbf{j}_\mathrm{c}^{\ast}(\mathbf{r}', \omega')]= 4\pi k_\mathrm{B}T \sigma_\mathrm{c} \delta^{(3)}(\mathbf{r} - \mathbf{r}')\delta(\omega-\omega')\mathbb{1}.
\end{equation}
Here, we assumed the conductivity to be real and frequency independent, $\sigma_\mathrm{c}=\mathrm{Re}[\sigma_\mathrm{c}(\omega)]$, as relevant in the frequency range considered below. 

Solving the Maxwell-London equations for a superconducting half-space yields the magnetic field $\delta \mathbf{B}_\mathrm{c}(\mathbf{r},\omega)$ induced by the current fluctuations in the vacuum region.  The derivation is given in App.~\ref{Magnetic field generated by superconductor noise currents}, where the field is first represented in terms of its transverse spatial Fourier components and subsequently transformed to real space. In the far-field regime, $z\gg \lambda$, the field can be written as 
\begin{equation}\label{eq:B_fluctuation_electron}
\delta \mathbf{B}_\mathrm{c}(\mathbf{r},\omega)=2\mu_0\lambda\int\mathrm{d}^2\boldsymbol{\rho}'\,\mathrm{G}(\mathbf{r}-\boldsymbol{\rho}')[\mathbf{e}_3\times\delta\mathbf{J}_\mathrm{c}(\boldsymbol{\rho}',\omega)],
\end{equation}
where
\begin{equation}\label{eq:depth_current_electrons}
\delta\mathbf{J}_\mathrm{c}(\boldsymbol{\rho},\omega) =\int_{-\infty}^0\mathrm{d}z'\,\mathrm{e}^{z'/\lambda}(\mathbb{1}-\mathbf{e}_3\otimes\mathbf{e}_3)\delta\mathbf{j}_\mathrm{c}(\boldsymbol{\rho},z',\omega)  
\end{equation}
is the exponentially weighted, depth-integrated in-plane current fluctuation. The corresponding field cross-spectral-density matrix consequently becomes 
\begin{align}
\mathrm{S}^\mathrm{th,c}_{B B}(\mathbf{r},\mathbf{r}',\omega)=&4\mu_0^2\sigma_\mathrm{c}k_\mathrm{B}T\lambda^3\int\mathrm{d}^2\boldsymbol{\rho}''\,[\mathrm{G}(\mathbf{r}-\boldsymbol{\rho}'')\nonumber\\
&\cdot(\mathbb{1}-\mathbf{e}_3\otimes\mathbf{e}_3)\mathrm{G}(\mathbf{r}'-\boldsymbol{\rho}'')].
\end{align}
At coincident observation points $\mathbf{r}=\mathbf{r}'$, this evaluates to
\begin{equation}\label{eq:S_bb_electrons}
\mathrm{S}^\mathrm{th,c}_{B B}(\mathbf{r})=\dfrac{3\mu_0^2\sigma_\mathrm{c}k_\mathrm{B}T\lambda^3}{32\pi z^4}(\mathbb{1}+\mathbf{e}_3\otimes\mathbf{e}_3),
\end{equation}
and thus scales as $z^{-4}$ over the relevant frequency range. The field-gradient and gradient-gradient spectra follow by differentiating the noncoincident field spectrum according to Eq.~\eqref{eq:field-gradient_spectra}. Their far-field forms can be written as
\begin{equation}\label{eq:S_dbb_eletron}
S^\mathrm{th,c}_{B_k \partial_\ell\!B_s}(\mathbf{r})=\dfrac{3\mu_0^2\sigma_\mathrm{c}k_\mathrm{B}T\lambda^3}{8\pi z^5}E_{k\ell s},
\end{equation}
and
\begin{equation}\label{eq:S_dbdb_eletron}
S^\mathrm{th,c}_{\partial_k B_\ell \partial_s\!B_t}(\mathbf{r})=\dfrac{15\mu_0^2\sigma_\mathrm{c}k_\mathrm{B}T\lambda^3}{16\pi z^6}E_{k\ell st}.
\end{equation}
The dimensionless tensors $E_{k\ell s}$ and $E_{k\ell st}$ are given in App.~\ref{app:noise-correlations}. Spectra involving second spatial derivatives of the magnetic field, including their mixed correlations with the field and its first derivatives, are obtained by further differentiation of the same noncoincident field spectrum.

\subsection{Thermal vortex motion}
\label{Thermal vortex motion}
In addition to thermal current-density fluctuations of the normal electronic component, pinned vortices undergo thermally driven motion within their pinning potentials. For simplicity, we approximate the London-weighted near-surface part of each vortex, which dominates the field outside the superconductor, as a straight segment that moves rigidly in the transverse direction~\cite{blatter1994vortices}. The instantaneous in-plane position of vortex $\nu$ is written as
\begin{equation}
\boldsymbol{\rho}_\nu(t)=\bar{\boldsymbol{\rho}}_\nu + \delta\boldsymbol{u}_\nu^\mathrm{th}(t), \qquad \mathbb{E}[\delta\boldsymbol{u}_\nu^\mathrm{th}(t)]=0,
\end{equation} 
where $\bar{\boldsymbol{\rho}}_\nu$ denotes its mean position and $\delta\boldsymbol{u}_\nu^\mathrm{th}(t)$ the thermal displacement, which are restricted to the plane of the superconducting surface. Locally, the pinning potential is taken to be harmonic, and the inertia of the vortex segment is neglected. Its overdamped motion is then governed by the pinning stiffness and viscous drag~\cite{blatter1994vortices}, 
\begin{equation}\label{eq:Langevin_u}
c_\nu\delta\dot{\mathbf{u}}_\nu^\mathrm{th}+K_\nu\delta\mathbf{u}_\nu^\mathrm{th}=\eta_\nu,
\end{equation}
with $\eta_\nu$ denoting a zero-mean thermal Langevin force. The effective stiffness is related to the Labusch parameter $\alpha_\mathrm{L}$ per unit vortex length according to $K_\nu=\alpha_\mathrm{L}d_{\mathrm{eff},\nu}$, where $d_{\mathrm{eff},\nu}$ is the effective length of the coherently moving vortex segment. 

Assuming that the thermal forces acting on different vortices are statistically independent and isotropic in the surface plane, the displacement fluctuations in the quasistatic regime, $\omega\ll K_\nu/c_\nu$, obey 
\begin{equation}
\mathbb{E}[\delta \mathbf{u}_{\nu}^{\mathrm{th}}(\omega)\otimes\delta \mathbf{u}_{\mu}^{\mathrm{th}\ast}(\omega')]= \dfrac{4\pi k_{\mathrm{B}}Tc_\nu}{K_{\nu}^2}\delta_{\nu\mu}\delta(\omega-\omega')(\mathbb{1}-\mathbf{e}_3\otimes\mathbf{e}_3).
\end{equation} 

As detailed in App.~\ref{app:noise-correlations}, the transverse vortex displacement produces a corresponding lateral translation of its magnetic-field profile. Neglecting the leading finite-penetration-depth correction in $\lambda$, the field of vortex $\nu$ at its mean position can be expressed as 
\begin{equation}
\mathbf{B}_\nu(\mathbf{r})=2\Phi_0\int_{-\infty}^0\mathrm{d}z'\mathrm{G}(\mathbf{r}-\bar{\boldsymbol{\rho}}_\nu-z'\mathbf{e}_3)\mathbf{e}_3.
\end{equation}
Expanding this field to first order in the thermal displacement $\delta\mathbf{u}_\nu^\mathrm{th}$ gives
\begin{equation}
\delta\mathbf{B}_\nu^\mathrm{th}(\mathbf{r},\omega)=-\left[\delta\mathbf{u}_{\nu}^\mathrm{th}(\omega)\cdot\grad_{\boldsymbol{\rho}}\right]\mathbf{B}_\nu(\mathbf{r}).
\end{equation}
The total fluctuating field produced by $N$ vortices is
\begin{equation}
\delta\mathbf{B}^\mathrm{th}(\mathbf{r},\omega)=\sum_{\nu=1}^N \delta\mathbf{B}_\nu^\mathrm{th}(\mathbf{r},\omega).
\end{equation}
Because the thermal motion of different vortices is assumed to be independent, their noise powers add, such that the 
resulting  magnetic-field cross-spectral density becomes
\begin{equation}\label{eq:S_bb_vortex_thermal}
S_{B_kB_\ell}^{\mathrm{th,v}}(\mathbf{r},\mathbf{r}',\omega)\approx  \sum_{\nu=1}^N\dfrac{2k_{\mathrm B}T c_\nu}{K_\nu^2}\left[\grad_{\boldsymbol{\rho}}B_{\nu,k}(\mathbf{r})\cdot \grad_{\boldsymbol{\rho}'}B_{\nu,\ell}(\mathbf{r}')\right],
\end{equation}
where the scalar product is restricted to the two in-plane derivative components. If all contributing vortices have approximately equal pinning parameters and approximately identical magnetic couplings to the observation points, this expression reduces to $N$ times the single-vortex spectrum. For a fixed group of vortices whose lateral distances from the particle's vertical projection are small compared with $z_\mathrm{eq}$, the leading nonzero components of the magnetic-field noise spectrum therefore scale as $z_\mathrm{eq}^{-6}$. The field-gradient spectra entering the generalized-force correlations are again obtained by differentiating this expression with respect to $\mathbf{r}$ and $\mathbf{r}'$.

\section{Conclusions}\label{Closing remarks}
In conclusion, we have characterized the coupled translational and rotational dynamics of a micromagnet levitated above a type-II superconductor, providing a theoretical basis for interpreting ongoing experiments and developing this platform for sensing. We derived the interaction potential in terms of the magnet's spatial magnetization distribution, extending the description beyond existing point-dipole models~\cite{gieseler2020single,kordyukMagneticLevitationHard1998}. Its multipole expansion reveals how an anisotropic magnetic quadrupole moment can lift the rotational degeneracy of the dipole potential, providing a mechanism for the experimentally observed confinement of all six rigid-body degrees of freedom~\cite{wang2026nonlinear, wang2026exploring}. 

The resulting translational and librational motion exhibits strong intermodal mixing, with frequencies and mode compositions tunable through external magnetic fields. This coupling and tunability enable the possibility of driving the mechanical motion beyond the linear regime via time-dependent fields. At larger opposing fields, the trapping landscape bifurcates, giving rise to coexisting stable equilibria and rapid changes of the equilibrium configuration. Together with the intrinsically nonlinear trapping potential, this tunability opens the possibility of driving the motion beyond the linear regime with time-dependent fields, giving access to nonlinear intermodal dynamics, including the generation of mechanical frequency combs demonstrated in our recent work~\cite{wang2026nonlinear}. In addition to these potential-induced interactions, we have shown how the intrinsic angular momentum of the magnet produces gyroscopic coupling between mechanical modes. The associated phase-shifted librational motion provides a signature of spin-rotation coupling, making Meissner-levitated micromagnets suitable probes of this interaction, as recently observed experimentally~\cite{ahrens2026observation}. 

The sensitivity to magnetic forces and torques further offers opportunities for probing magnetic defects and fluctuations near surfaces. Here, we have evaluated the contributions of thermal currents and small-amplitude fluctuations of pinned vortices to the force and magnetic-field sensitivities. Near the mechanical resonances, the predicted sensitivity is limited by these substrate fluctuations, such that the levitated magnet directly resolves the magnetic noise of the superconductor. Discrete vortex jumps between pinning configurations, which can give rise to flux creep, constitute an additional source of magnetic fluctuations beyond the scope of the present analysis and will be addressed separately. Such stochastic fluctuations also carry information about the superconducting substrate. As demonstrated experimentally~\cite{wang2026exploring}, a levitated magnet can probe flux dynamics involving only a few vortices, providing access to microscopic rearrangements underlying flux creep. 

Finally, for the parameters used in Fig.~\ref{fig:superconductor-noise}, we predict substrate-limited resonant sensitivities of $4.1\,\mathrm{fT}/\sqrt{\mathrm{Hz}}$ for magnetic fields and $14\,\mathrm{zN}/\sqrt{\mathrm{Hz}}$ for forces, and broadband sensitivities of $150\,\mathrm{fT}/\sqrt{\mathrm{Hz}}$ and $230\,\mathrm{zN}/\sqrt{\mathrm{Hz}}$  over a range of $5\,\mathrm{Hz}$ to  $500\,\mathrm{Hz}$. Near resonance, the sensitivity is limited by the magnetic noise of the superconducting substrate, whereas the broadband values are set by the assumed displacement imprecision and thus improve directly with the readout. Larger levitation heights, lower temperatures, and lower normal-fluid conductivity further reduce the substrate-limited floor. These figures support earlier proposals to use Meissner-levitated micromagnets for searches for dark-matter candidates, such as axions~\cite{ahrens2025levitated,higginsMaglevDarkMatter2024}, dark photons~\cite{kalia2024ultralight}, ultraheavy particles~\cite{uitenbroek2026first}, chameleon fields~\cite{yinExperimentsLevitatedForce2022}, and exotic spin- and velocity-dependent couplings~\cite{renLevitatedFerromagneticTorsional2026}. The combination of tunable coupled motion, intrinsic spin-rotation coupling, and sensitivity to local magnetic fluctuations makes this platform promising both for precision measurements and for studying the dynamics of magnetic systems.

\section*{Acknowledgements}
The work done in this paper was supported by the Deutsche Forschungsgemeinschaft (German Research Foundation), grant number 521602913. B.A.S. and V.W. acknowledge funding by the Carl-Zeiss-Foundation through the project QPhoton. V.W. is supported by the Financial Support Programmes for Female Academics, from the Office for Gender Equality, Ulm University. B.A.S. is supported by the DFG No. 510794108. M.D.L., Y.W., and T.M. acknowledge support by the NSF Center for Ultracold Atoms, Amazon Web Services (grant No. A60290), NSF (grant No. OMA-2121044), DOE Quantum Systems Accelerator Center (grant No. DE-AC02-05CH11231), and the Air Force Office of Scientific Research (grant No. FA9550-23-1-0333). Y.W. acknowledges support from the HQI Postdoctoral Fellowship Program. T.M. acknowledges support from the NSF Graduate Research Fellowship Program (grant No. 2140743).

\appendix

\section{Principal axes}\label{app:principal_axes}
Using $ \mathbf{n}_k(\Omega)=\mathrm{R}(\Omega)\mathbf{e}_k$, the principal axes in terms of the Euler angles read
\begin{subequations} \label{eq:labAxes_bodyFrame}
\begin{align}
    \mathbf{n}_1&=\begin{pmatrix}
        \cos\alpha\cos\beta\cos\gamma-\sin\alpha\sin\gamma\\ 
        \sin\alpha\cos\beta\cos\gamma+\cos\alpha\sin\gamma \\
        -\sin\beta\cos\gamma
    \end{pmatrix},\\
    \mathbf{n}_2&=\begin{pmatrix}
        -\cos\alpha\cos\beta\sin\gamma-\sin\alpha\cos\gamma\\
        -\sin\alpha\cos\beta\sin\gamma+\cos\alpha\cos\gamma\\
        \sin\beta\sin\gamma
    \end{pmatrix},\\
    \mathbf{n}_3&=\begin{pmatrix}
    \cos\alpha\sin\beta\\
   \sin\alpha \sin\beta\\
    \cos\beta
    \end{pmatrix},
\end{align}
\end{subequations}
where vector elements refer to the space-fixed basis $\mathbf{e}_j$, i.e. the lab frame.
Knowing the components of the principal axes in the lab frame then allows determination of the lab-frame components of the inertia tensor $\mathrm{I}$, and consequently the mass-matrix referred to in \cref{eqn:mass-matrix-def}.

\section{Lorentz force and torque}\label{Lorentz force and torque}

The Lorentz force acting on the magnetic particle by an external magnetic field $\mathbf{B}(\mathbf{r})$ is given by
\begin{equation}\label{eqn:Lorentz-force-general}
\mathbf{F}
= \int \mathrm{d}^3 \mathbf{r}\,(\grad\times\mathbf{M})\times \mathbf{B},
\end{equation}
dropping the explicit positional $\mathbf{r}$-dependence for simplicity.
We use the vector identity
\begin{align}\label{eqn:Lorentz-force-vec-identity-01}
(\grad \times \mathbf{M}) \times \mathbf{B} & = \grad \cdot (\mathbf{B} \otimes \mathbf{M}) - \grad(\mathbf{M} \cdot \mathbf{B}) \\
& \quad  - (\grad \cdot \mathbf{B}) \mathbf{M} + (\grad \otimes \mathbf{B}) \cdot \mathbf{M},
\end{align}
noting that the first two terms are total differentials.
By Gauss' theorem these reduce to surface integrals over the boundary of the particle~\cite{jacksonClassicalElectrodynamics2021b}, and vanish for a localized magnetization and continuous field.
The third term vanishes from the Maxwell equation $\grad \cdot \mathbf{B} = 0$.
The final term can be rewritten as
\begin{equation}\label{eqn:Lorentz-force-vec-identity-02}
(\grad \otimes \mathbf{B}) \cdot \mathbf{M}= \mathbf{M}\cdot (\grad \otimes \mathbf{B}) + \mathbf{M} \times (\grad \times \mathbf{B}).
\end{equation}
In the magnetostatic regime where there are no free currents, $\grad \times \mathbf{B} = \mathbf{0}$.
Hence the force reduces to
\begin{equation}\label{eqn:Lorentz-force-final}
\mathbf{F}  = \int \mathrm{d}^3 \mathbf{r}\, \mathbf{M} \cdot (\grad \otimes \mathbf{B}) = \int \mathrm{d}^3 \mathbf{r}\,(\mathbf{M} \cdot \grad)\mathbf{B}.
\end{equation}
This implies that the force felt by the particle is determined by the field gradient $\grad \otimes \mathbf{B}$, which is traceless.
As \cref{eqn:Lorentz-force-vec-identity-02} is true even if the magnetization $\mathbf{M}(\mathbf{r})$ is replaced by an arbitrary vector field, the field gradient is also symmetric.

The total torque acting on the magnetized body follows from the Lorentz force density as 
\begin{equation}\label{eqn:Lorentz-torque-general}
\mathbf{N} = \int \mathrm{d}^3 \mathbf{r}\,(\mathbf{r} -\mathbf{R})\times [(\grad \times \mathbf{M}) \times \mathbf{B}].
\end{equation}
Using the results in \cref{eqn:Lorentz-force-vec-identity-01,eqn:Lorentz-force-vec-identity-02} we expand to
\begin{align}\label{eqn:Lorentz-torque-defN1}
\mathbf{N} & = \int\mathrm{d}^3 \mathbf{r} \, (\mathbf{r} - \mathbf{R}) \times (\mathbf{M}\cdot\grad)\mathbf{B} \\
& \quad + \int\mathrm{d}^3 \mathbf{r} \, (\mathbf{r} - \mathbf{R}) \times [\grad \cdot (\mathbf{B} \otimes \mathbf{M}) - \grad(\mathbf{M} \cdot \mathbf{B})].
\end{align}
The torque density in the second integral can be evaluated with the product rule and the gradient identity $\grad \otimes (\mathbf{r} - \mathbf{R}) = \mathbb{1}$ to obtain
\begin{align}\label{eqn:Lorentz-torque-density-intermediate}
& (\mathbf{r} - \mathbf{R}) \times [\grad \cdot (\mathbf{B} \otimes \mathbf{M}) - \grad(\mathbf{M} \cdot \mathbf{B})] \\
&\quad  = \grad \cdot \{\mathbf{B} \otimes [(\mathbf{r} - \mathbf{R}) \times \mathbf{M}]\} + \mathbf{M} \times \mathbf{B} \\
& \qquad + \grad \times [(\mathbf{r} - \mathbf{R}) \, (\mathbf{M} \cdot \mathbf{B})].
\end{align}
The first and third terms are total differentials and vanish upon integration. Substituting this expression back into \cref{eqn:Lorentz-torque-defN1} the total torque can be simplified into
\begin{equation}\label{eqn:Lorentz-torque-final}
\mathbf{N}= \int\mathrm{d}^3\mathbf{r}\,\mathbf{M}\times\mathbf{B}+ \int\mathrm{d}^3\mathbf{r}\,(\mathbf{r}-\mathbf{R})\times(\mathbf{M}\cdot\grad)\mathbf{B}. 
\end{equation}
The first term corresponds to the \emph{magnetic torque}, while the second term represents an \emph{orbital torque} associated with spatial variations of the magnetic field.

\section{Potential in static external field}
\label{app:Potential in static external field}

For the static external magnetic field $\mathbf{B}({\bf r})$, we derive the corresponding magnetic potential $V_{\mathrm{B}}$.
Beginning from the Lorentz force \cref{eqn:Lorentz-force-final} and the permanent magnet rotation \cref{eqn:magnetisation-field-trafo} we obtain
\begin{equation}\label{eqn:force-to-pot-ext-init}
\mathbf{F} = \int \mathrm{d}^3 \boldsymbol{\xi}\, [ \grad_{\mathbf{R}} \otimes \mathbf{B}({\rm R}\boldsymbol{\xi} + \mathbf{R})] \cdot [\mathrm{R} \mathbf{M}_{0}(\boldsymbol{\xi})]
\end{equation}
where the symmetry of the field gradient allows for its transposition.
The field has an \emph{implicit} dependence on the center-of-mass position, and we can equivalently use the gradient $\grad_{\mathbf{R}}$ in place of the spatial gradient $\grad$.
Since the volume of integration is independent of the center of mass, the gradient commutes through the integral and we can write the force in terms of a potential, $\mathbf{F} = - \grad_{\mathbf{R}} V_{\mathrm{B}}$. This potential is given by
\begin{equation}\label{eqn:force-to-pot-ext-final}
V_{\mathrm{B}} = - \int \mathrm{d}^3 \boldsymbol{\xi}\, [\mathrm{R}\mathbf{M}_{0}(\boldsymbol{\xi})] \cdot  \mathbf{B}({\rm R}\boldsymbol{\xi} + \mathbf{R}).
\end{equation}

Having derived the potential, we show that differentiation with respect to the rotational orientation corresponds to the torque, such that $\mathbf{N} \cdot\mathbf{u}_{\mu} =-\partial_\mu V_{\mathrm{B}}$.
With respect to \cref{eqn:force-to-pot-ext-final}, the partial differentials with respect to the Euler angles are
\begin{align}\label{eqn:pot-ext-to-torque-init}
- \partial_{\mu} V_{\mathrm{B}}
& = \int \mathrm{d}^3\boldsymbol{\xi} \, (\partial_{\mu} \mathrm{R})\mathbf{M}_{0}(\boldsymbol{\xi}) \cdot \mathbf{B}(\mathrm{R}\boldsymbol{\xi} + \mathbf{R}) \\
& \quad +  \int \mathrm{d}^3\boldsymbol{\xi}\, \mathrm{R}\mathbf{M}_{0}(\boldsymbol{\xi}) \cdot \partial_{\mu} \mathbf{B}(\mathrm{R}\boldsymbol{\xi} + \mathbf{R}).
\end{align}
Here, the volume of integration is body-fixed and hence independent of the rotational orientation, allowing the commutation of the derivative through the integral.
The integrand of the first integral represents the orientational dependence of the magnetization field of the magnet.
The integrand of the second integral represents the implicit dependence of the field on the rotational orientation, and therefore applying the differential chain rule we obtain
\begin{equation}\label{eqn:ext-field-rot-diff}
\partial_{\mu} \mathbf{B}(\mathrm{R}\boldsymbol{\xi} + \mathbf{R}) = [ \grad \otimes  \mathbf{B}(\mathrm{R}\boldsymbol{\xi} + \mathbf{R})] \cdot (\partial_{\mu} \mathrm{R})\boldsymbol{\xi}.
\end{equation}
Again using the property $\partial_{\mu} \mathrm{R} = \mathbf{u}_{\mu} \times \, \mathrm{R}$, we can then evaluate \cref{eqn:pot-ext-to-torque-init} as
\begin{align}\label{eqn:pot-ext-to-torque-dR}
& - \partial_{\mu} V_{\mathrm{B}} = \int \mathrm{d}^3\boldsymbol{\xi}[\mathbf{u}_{\mu} \times \, \mathrm{R} \mathbf{M}_{0}(\boldsymbol{\xi})] \cdot \mathbf{B}(\mathrm{R}\boldsymbol{\xi} + \mathbf{R}) \nonumber \\
& \quad + \int \mathrm{d}^3\boldsymbol{\xi} \left\{\mathrm{R} \mathbf{M}_{0}(\boldsymbol{\xi}) \cdot [ \grad \otimes  \mathbf{B}(\mathrm{R}\boldsymbol{\xi} + \mathbf{R})]
\cdot \, (\mathbf{u}_{\mu} \times \, \mathrm{R}\boldsymbol{\xi}) \right\}.
\end{align}
Using the cyclic property of the scalar triple product as in \cref{Torque from Meissner potential}, we finally obtain
\begin{align} \label{eqn:pot-ext-to-torque-final}
- \partial_{\mu} V_{\mathrm{B}} 
= & \, \mathbf{u}_{\mu} \cdot \int \mathrm{d}^3\boldsymbol{\xi} \left\{ [{\rm R}\mathbf{M}_{0}(\boldsymbol{\xi})] \times \mathbf{B}({\rm R}\boldsymbol{\xi} + \mathbf{R}) \right. \nonumber \\
& \quad + \left. ({\rm R}\boldsymbol{\xi}) \times [\mathbf{M}_{0}(\boldsymbol{\xi}) \cdot {\rm R}^\mathsf{T}\grad ] \, \mathbf{B}({\rm R}\boldsymbol{\xi} + \mathbf{R}) \right\}.
\end{align}
With comparison to \cref{eqn:Lorentz-torque-final}, we can identify the integral as the torque acting on the magnet due to the external magnetic field.
As such we have shown that the orientation derivatives of the potential indeed correspond to the torque acting on the magnet.

\section{Derivation of Meissner torque}
\label{app:torque_derivation_details}

Here, we show how the rotational derivative of the Meissner-image potential reduces to Eq.~\eqref{eqn:dVim_symmetrized}. The rotational derivative of $V_\mathrm{im}$ has two types of contributions. First, $\partial_\mu$ acts on the rotation matrices multiplying the magnetization vectors. Second, it acts on the Green tensor through the rotated coordinates in its argument $\mathbf{r}'=\mathrm{R}\boldsymbol{\xi}-\mathrm{A}\mathrm{R}\boldsymbol{\xi}'+\mathbf{R}-\mathrm{A}\mathbf{R}'$. Accordingly, we split Eq.~\eqref{eqn:dVim_full_rot_derivative} into two contributions
\begin{equation}
-\partial_\mu V_\mathrm{im}=N_{1,\mu}+N_{2,\mu},
\end{equation}
with
\begin{align}
N_{1,\mu}&=\dfrac{\mu_0}{2}\int\mathrm{d}^3\boldsymbol{\xi}\,\mathrm{d}^3\boldsymbol{\xi}'\{[(\partial_\mu\mathrm{R})\mathbf{M}_0(\boldsymbol{\xi})]\cdot\mathrm{G}\mathrm{A}\mathrm{R}\mathbf{M}_0(\boldsymbol{\xi}')\nonumber\\
&\quad+[\mathrm{R}\mathbf{M}_0(\boldsymbol{\xi})]
\cdot\mathrm{G}\mathrm{A}(\partial_\mu\mathrm{R})\mathbf{M}_0(\boldsymbol{\xi}')\},
\label{eqn:app_N1mu_start}
\end{align}
and
\begin{align}
N_{2,\mu}&=\dfrac{\mu_0}{2}\int\mathrm{d}^3\boldsymbol{\xi}\,\mathrm{d}^3\boldsymbol{\xi}'\left[(\partial_\mu\mathrm{R}\boldsymbol{\xi})-\mathrm{A}(\partial_\mu\mathrm{R}\boldsymbol{\xi}')\right]\cdot\grad_{\mathbf{R}}\nonumber\\
&\quad\left[\mathbf{M}_0(\boldsymbol{\xi})\cdot\mathrm{R}^\mathsf{T}\mathrm{G}\mathrm{A}\mathrm{R}\mathbf{M}_0(\boldsymbol{\xi}')\right].
\label{eqn:app_N2mu_start}
\end{align}
The first term in $N_{1,\mu}$ corresponds to the rotational derivative of the particle magnetization, while the second term corresponds to the rotational derivative of the image magnetization. Similarly, the two terms in $N_{2,\mu}$ arise from differentiating the two rotated coordinates $\mathrm{R}\boldsymbol{\xi}$ and $\mathrm{A}\mathrm{R}\boldsymbol{\xi}'$ in the argument of the Green tensor.

We first consider $N_{1,\mu}$. The second term in Eq.~\eqref{eqn:app_N1mu_start} can be transformed by interchanging the dummy variables $\boldsymbol{\xi}\leftrightarrow\boldsymbol{\xi}'$, $\mathbf{R}\leftrightarrow\mathbf{R}'$,
and using the symmetry of the Green tensor together with its reflection property~\eqref{eqn:G-trans-properties}.
Under this transformation, the Green-tensor argument is mapped onto itself, while the scalar product is transposed. Consequently, the two terms of $N_{1,\mu}$ become identical, and hence
\begin{equation}
N_{1,\mu}=\mu_0\int\mathrm{d}^3\boldsymbol{\xi}\,\mathrm{d}^3\boldsymbol{\xi}'\,
[(\partial_\mu\mathrm{R})\mathbf{M}_0(\boldsymbol{\xi})]\cdot
\mathrm{G}\mathrm{A}\mathrm{R}\mathbf{M}_0(\boldsymbol{\xi}').
\label{eqn:app_N1mu_final}
\end{equation}

We now turn to $N_{2,\mu}$.
The Green tensor depends on the auxiliary positions only through the combination $\mathbf{R}-\mathrm{A}\mathbf{R}'$.
As such the gradients with respect to these two coordinates satisfy
\begin{equation}
\grad_{\mathbf{R}'}\otimes\mathrm{G}=-\mathrm{A}\grad_{\mathbf{R}}\otimes\mathrm{G},
\end{equation}
again suppressing the argument of the Green tensor.
Therefore, the second term in Eq.~\eqref{eqn:app_N2mu_start} can equivalently be written as a derivative with respect to the auxiliary image coordinate, giving
\begin{align}
N_{2,\mu}&=\dfrac{\mu_0}{2}\int\mathrm{d}^3\boldsymbol{\xi}\,\mathrm{d}^3\boldsymbol{\xi}'\left[(\partial_\mu\mathrm{R}\boldsymbol{\xi})\cdot\grad_{\mathbf{R}}+(\partial_\mu\mathrm{R}\boldsymbol{\xi}')\cdot\grad_{\mathbf{R}'}\right]\nonumber\\
&\quad \left[\mathbf{M}_0(\boldsymbol{\xi})\cdot\mathrm{R}^\mathsf{T}\mathrm{G}\mathrm{A}\mathrm{R}\mathbf{M}_0(\boldsymbol{\xi}')\right].
\label{eqn:app_N2mu_RR1}
\end{align}
The second contribution in Eq.~\eqref{eqn:app_N2mu_RR1} can then be mapped onto the first one by the same interchange of dummy variables and auxiliary coordinates used above, again employing the symmetry and reflection identities of the Green tensor. The two terms are therefore equal, and the factor $1/2$ is canceled.
Using the total symmetry of the Green tensor derivative, we obtain the result
\begin{align}
N_{2,\mu}&=\mu_0\int\mathrm{d}^3\boldsymbol{\xi}\,\mathrm{d}^3\boldsymbol{\xi}'\,(\partial_\mu\mathrm{R}\boldsymbol{\xi})\nonumber\\
&\quad\cdot\{[\mathrm{R}\mathbf{M}_0(\boldsymbol{\xi})]
\cdot\grad_{\mathbf{R}}\otimes\mathrm{G}\}\mathrm{A}\mathrm{R}\mathbf{M}_0(\boldsymbol{\xi}').
\label{eqn:app_N2mu_final}
\end{align}
Combining Eqs.~\eqref{eqn:app_N1mu_final} and \eqref{eqn:app_N2mu_final}, we obtain Eq.~\eqref{eqn:dVim_symmetrized}.

\section{Derivation of the Meissner dipole-quadrupole and quadrupole-quadrupole contributions}
\label{app:meissner_quadrupole}

Here, we derive the quadrupole contribution to the Meissner-image potential. Starting from Eq.~\eqref{eq:V_im_expanded}, we write the first-order contribution as
\begin{equation}
V_\mathrm{im}^{(1)}=\dfrac{\mu_0}{2}U_1+\dfrac{\mu_0}{2}U_2,
\end{equation}
with
\begin{subequations}
\begin{equation}
U_1= - \int \mathrm{d}^3\boldsymbol{\xi}\,\mathbf{M}_0(\boldsymbol{\xi})\cdot\mathrm{R}^\mathsf{T}
\left[(\mathrm{R}\boldsymbol{\xi})\cdot\grad\right]\mathrm{G}(\mathbf{d})\mathrm{A}\mathbf{m},
\end{equation}
and
\begin{equation}
U_2=\int \mathrm{d}^3\boldsymbol{\xi}'\,\mathbf{m}\cdot\left[
(\mathrm{A}\mathrm{R}\boldsymbol{\xi}')\cdot\grad\right]\mathrm{G}(\mathbf{d})\mathrm{A}\mathrm{R}\mathbf{M}_0(\boldsymbol{\xi}').
\end{equation}
\end{subequations}
Under reflection by $-\mathrm{A}$, the gradient of the Green tensor transforms as
\begin{equation}\label{eqn:Gprime-trans-properties}
\grad \otimes \mathrm{G}(-\mathrm{A} \mathbf{r}) = - (\mathrm{A} \grad) \otimes [ \mathrm{A}\mathrm{G}(\mathbf{r})\mathrm{A}].
\end{equation}
Using this transformation property the second term $U_{2}$ can be transformed as
\begin{equation}
U_2= - \int \mathrm{d}^3\boldsymbol{\xi}'\,(\mathrm{A}\mathbf{m}) \cdot\left[
(\mathrm{R}\boldsymbol{\xi}') \cdot \grad \right] \mathrm{G}(-\mathrm{A}\mathbf{d})\mathrm{R}\mathbf{M}_0(\boldsymbol{\xi}').
\end{equation}
Since the image separation vector satisfies $- \mathrm{A}\mathbf{d}=\mathbf{d}$, using the total symmetry of the Green tensor derivative to rearrange contractions we obtain $U_2=U_1$.
Consequently, $V_\mathrm{im}^{(1)}= \mu_0 U_1$, 
and therefore
\begin{equation}
V_\mathrm{im}^{(1)}=-\mu_0\int \mathrm{d}^3\boldsymbol{\xi}\,\mathbf{M}_0(\boldsymbol{\xi})\cdot\mathrm{R}^\mathsf{T}\left[(\mathrm{R}\boldsymbol{\xi})\cdot\grad\right]\mathrm{G}(\mathbf{d})\mathrm{A}\mathbf{m}.
\label{eq:app_Vim_first_order}
\end{equation}

To identify the corresponding multipole moment, we rewrite Eq.~\eqref{eq:app_Vim_first_order} in component form as
\begin{equation}
V_\mathrm{im}^{(1)}=-\mu_0 \!\! \sum^{3}_{i,j,k=1} \!\int\! \mathrm{d}^3\boldsymbol{\xi}\,[\mathrm{R}{\bf M}_0(\boldsymbol{\xi})]_{i}(\mathrm{R}\boldsymbol{\xi})_j\,\partial_j G_{ik}(\mathbf{d})(\mathrm{A}\mathbf{m})_k.
\end{equation}
Introducing the first moment tensor
\begin{equation}
\tilde{q}_{ij}=\int \mathrm{d}^3\boldsymbol{\xi}\,[\mathrm{R}{\bf M}_0(\boldsymbol{\xi})]_{i}(\mathrm{R}\boldsymbol{\xi})_j ,
\end{equation}
the first-order contribution becomes
\begin{equation}
V_\mathrm{im}^{(1)}=-\mu_0 \sum^{3}_{i,j,k=1} \tilde{q}_{ij}\,\partial_j G_{ik}(\mathbf{d})(\mathrm{A}\mathbf{m})_k.
\end{equation}

In the source-free region occupied by the particle, the Green-tensor derivative $\partial_j G_{ik}$ is traceless and totally symmetric, allowing interchange of the indices $i$ and $j$.
Therefore, only the symmetric traceless part of $\tilde{q}_{ij}$ contributes to the interaction energy.
Comparing with the definition of the quadrupole tensor in Eq.~\eqref{eqn:def-quadrupole-moment}, one finds that the replacement
\begin{equation}
\tilde{q}_{ij}\rightarrow\dfrac{1}{6}Q_{ij}
\end{equation}
gives equivalent results with respect to contraction with $\partial_j G_{ik}$. Substituting this relation yields
\begin{equation}
V_\mathrm{im}^{(1)}=-\dfrac{\mu_0}{6} \sum^{3}_{i,j,k=1} Q_{ij}\,\partial_j G_{ik}(\mathbf{d})(\mathrm{A}\mathbf{m})_k,
\end{equation}
which can be written in compact form as Eq.~\eqref{eq:V_im_quad}.

Analogously, we derive the quadrupole-quadrupole contribution to the Meissner-image potential by writing the mixed second-order contribution in Eq.~\eqref{eq:V_im_expanded} in component form, 
\begin{align}
V_\mathrm{im}^\mathrm{qq}=&\dfrac{\mu_0}{2}\sum_{i,j,k,\ell=1}^{3}\int \mathrm{d}^3\boldsymbol{\xi}\,\mathrm{d}^3\boldsymbol{\xi}'\,[\mathrm{R}\mathbf{M}_0(\boldsymbol{\xi})]_i(\mathrm{R}\boldsymbol{\xi})_j\nonumber\\
&\quad[\mathrm{A}\mathrm{R}\mathbf{M}_0(\boldsymbol{\xi}')]_k(\mathrm{A}\mathrm{R}\boldsymbol{\xi}')_\ell\,\partial_j\partial_\ell G_{ik}(\mathbf{d}).
\end{align}
Using the first-moment tensor $\tilde{q}_{ij}$ introduced above and defining its reflected counterpart,
\begin{equation}
\tilde{q}^\mathrm{im}_{k\ell}=\int \mathrm{d}^3\boldsymbol{\xi}'\,[\mathrm{A}\mathrm{R}\mathbf{M}_0(\boldsymbol{\xi}')]_k(\mathrm{A}\mathrm{R}\boldsymbol{\xi}')_\ell,
\end{equation}
we obtain
\begin{equation}
V_\mathrm{im}^\mathrm{qq}=\dfrac{\mu_0}{2}\sum_{i,j,k,\ell=1}^{3}\tilde{q}_{ij}\tilde{q}^\mathrm{im}_{k\ell}\,\partial_j\partial_\ell G_{ik}(\mathbf{d}).
\label{eq:app_Vim_QQ_moments}
\end{equation}
In matrix notation, the reflected first-moment tensor satisfies $\tilde{\mathrm{q}}^\mathrm{im}=\mathrm{A}\tilde{\mathrm{q}}\mathrm{A}$.

Since the second derivative of the Green tensor is totally symmetric and traceless for $\mathbf{d}\neq0$, only the
symmetric traceless parts of the two first-moment tensors contribute to Eq.~\eqref{eq:app_Vim_QQ_moments}. As in the dipole-quadrupole case, the replacements
\begin{equation}
\tilde{q}_{ij}\rightarrow\dfrac{1}{6}Q_{ij},\qquad\tilde{q}^\mathrm{im}_{k\ell}\rightarrow\dfrac{1}{6}(\mathrm{A}\mathrm{Q}\mathrm{A})_{k\ell}
\end{equation}
give equivalent results within this contraction. Substituting these relations yields
\begin{equation}
V_\mathrm{im}^\mathrm{qq}=\dfrac{\mu_0}{72}\sum_{i,j,k,\ell=1}^{3}Q_{ij}(\mathrm{A}\mathrm{Q}\mathrm{A})_{k\ell}\,\partial_j\partial_\ell G_{ik}(\mathbf{d}).
\end{equation}
To obtain the compact form, we use the derivative symmetry $\partial_j\partial_\ell G_{ik}=\partial_i\partial_j G_{k\ell}$. The contraction over $k$ and $\ell$ can then be written as a trace, giving
\begin{equation}
V_\mathrm{im}^\mathrm{qq}=\dfrac{\mu_0}{72}\sum_{i,j=1}^{3}Q_{ij}\,\partial_i\partial_j\mathrm{tr}\!\left[\mathrm{G}(\mathbf{d})\mathrm{A}\mathrm{Q}\mathrm{A}\right].
\end{equation}
Since the quadrupole tensor is independent of the Green-tensor coordinate, this expression reproduces Eq.~\eqref{eq:V_imQQ}.

\section{Exterior magnetic field generated by the superconductor}
\label{Magnetic field generated by superconductor noise currents}
In the following we derive the exterior magnetic field generated by source currents inside the superconductor. The derivation follows the current-source approach established in Ref.~\cite{henkelMagnetostaticFieldNoise2005}, incorporating London screening in the superconducting medium. 

We consider a superconducting half-space occupying $z<0$, with a planar interface to vacuum at $z=0$. In the local, isotropic London description, the total current inside the superconductor consists of a prescribed source current $\mathbf{j}$ and the induced screening current $\mathbf{j}_\mathrm{scr}$. The latter satisfies
\begin{equation}
\grad\times \mathbf{j}_\mathrm{scr} =-\dfrac{\mathbf{B}}{\mu_0\lambda^2}.
\end{equation}
The source current will represent either the vortex phase contribution or thermal current-density fluctuations in the following sections. It excludes the induced screening response. We work in the low-frequency regime in which displacement currents can be neglected and London screening dominates the electromagnetic response of the normal electronic component. Combining Amp\`ere's law, $\grad\times\mathbf{B}=\mu_0(\mathbf{j}+\mathbf{j}_\mathrm{scr})$, with $\grad\cdot\mathbf{B}=0$ gives the Maxwell-London equation inside the superconductor
\begin{equation}\label{eqn:maxwell-ampere-2pde}
(\grad^2 - \lambda^{-2})\mathbf{B} = - \mu_0\grad \times \mathbf{j} ,\qquad z<0.
\end{equation}
We write $\mathbf{r}=\boldsymbol{\rho}+z\mathbf{e}_3$ and introduce the two-dimensional Fourier transform parallel to the surface, 
\begin{equation}\label{eq:app_2d_fourier}
\mathbf{b}(\mathbf{k},z,\omega) = \int\mathrm{d}^2\boldsymbol{\rho}\,\mathrm{e}^{-\mathrm{i}\mathbf{k}\cdot\boldsymbol{\rho}}\mathbf{B}(\mathbf{r}, \omega),
\end{equation}
with inverse 
\begin{equation}
\mathbf{B}(\mathbf{r}, \omega) = \int\dfrac{\mathrm{d}^2\mathbf{k}}{(2\pi)^2}\mathrm{e}^{\mathrm{i}\mathbf{k}\cdot\boldsymbol{\rho}}\mathbf{b}(\mathbf{k},z,\omega).
\end{equation}
Here, $\mathbf{k}\cdot\mathbf{e}_3=0$, $k=|\mathbf{k}|$, and the source current is transformed using the same convention. The Maxwell-London equation becomes
\begin{equation}\label{eq:app_maxwell-london_fourier}
(\partial_z^2-k_\lambda^2)\mathbf{b}=-\mu_0(\mathrm{i}\mathbf{k}+\mathbf{e}_3\partial_z)\times\mathbf{j}(\mathbf{k},z,\omega),
\end{equation}
where we defined $k_\lambda=\sqrt{k^2+\lambda^{-2}}$.  In the source-free vacuum region, the magnetic field satisfies Laplace's equation, which becomes $(\partial_z^2-k^2)\mathbf{b}=0$ after the transverse Fourier transform, such that for $k>0$, excluding the exponentially growing solution gives
\begin{equation}
\mathbf{b}(\mathbf{k},z,\omega)=\mathrm{e}^{-kz} \mathbf{b}_\mathrm{out}(\mathbf{k},\omega),
\end{equation}
where $\mathbf{b}_\mathrm{out}$ denotes the Fourier amplitude immediately above the surface, $z=0^+$. The divergence and curl conditions, $\grad\cdot\mathbf{B}=0$ and  $\grad\times\mathbf{B}=0$, determine the relative components of this amplitude
\begin{equation}\label{eq:app_B_out_3}
\mathbf{b}_\mathrm{out}= (\mathbf{e}_3\cdot\mathbf{b}_\mathrm{out})(\mathbf{e}_3-\mathrm{i}\mathbf{e}_k), \quad \mathbf{e}_k=\dfrac{\mathbf{k}}{k}.
\end{equation}
Assuming the same permeability on both sides and no singular sheet current at the interface, the normal and tangential magnetic-field components are continuous,
\begin{equation}
\mathbf{e}_3\cdot[\mathbf{b}(0^+)-\mathbf{b}(0^-)]=0,\quad \mathbf{e}_3\times[\mathbf{b}(0^+)-\mathbf{b}(0^-)]=0.
\end{equation}
Together with the condition for divergence-free fields, tangential continuity also implies continuity of $\partial_z(\mathbf{e}_3\cdot\mathbf{b})$. Consequently, the interior normal component satisfies $\mathbf{e}_3\cdot\mathbf{b}(0^-)=\mathbf{e}_3\cdot\mathbf{b}_\mathrm{out}$ and $\partial_z[\mathbf{e}_3\cdot\mathbf{b}(0^-)]=-k(\mathbf{e}_3\cdot\mathbf{b}_\mathrm{out})$. To determine its amplitude, we take the normal component of the Maxwell-London equation~\eqref{eq:app_maxwell-london_fourier}, multiply by $\exp(k_\lambda z)$, and integrate over $z<0$. Using the product rule gives
\begin{equation}
\left[\mathrm{e}^{k_\lambda z}(\partial_z-k_\lambda)(\mathbf{e}_3\cdot\mathbf{b})\right]_{-\infty}^{0^-}=-\mathrm{i}\mu_0\mathbf{e}_3\cdot[\mathbf{k}\times\mathbf{J}(\mathbf{k},\omega)],
\end{equation}
where
\begin{equation}\label{eq:app_J_weighted_integral}
\mathbf{J}(\mathbf{k},\omega)   = \int_{-\infty}^{0^-}\mathrm{d}z'\mathrm{e}^{k_\lambda z'} \mathbf{j}(\mathbf{k},z',\omega)
\end{equation}
is the depth-weighted source current. Assuming that the weighted boundary term vanishes deep inside the superconductor and substituting the interface conditions yields
\begin{equation}
\mathbf{e}_3\cdot\mathbf{b}_{\mathrm{out}}(\mathbf{k}, \omega) = \dfrac{\mathrm{i}\mu_0 k}{k +k_\lambda}(\mathbf{e}_3 \times \mathbf{e}_k) \cdot \mathbf{J}(\mathbf{k}, \omega).
\end{equation}
Using the relation between the exterior field components in  Eq.~\eqref{eq:app_B_out_3} then gives
\begin{equation}\label{eqn:b-outside-fourier-eval}
\mathbf{b}_{\mathrm{out}}(\mathbf{k}, \omega) = \dfrac{\mu_0 k}{k +k_\lambda}(\mathbf{e}_k+\mathrm{i}\mathbf{e}_3)[(\mathbf{e}_3 \times \mathbf{e}_k) \cdot \mathbf{J}(\mathbf{k}, \omega)].
\end{equation}
Thus, only the in-plane source-current component perpendicular to $\mathbf{k}$ contributes, and the real-space exterior field follows by inverse Fourier transformation,
\begin{equation}\label{eq:app_b_outside_real}
\mathbf{B}_{\mathrm{out}}(\mathbf{r}, \omega) = \int\dfrac{\mathrm{d}^2\mathbf{k}}{(2\pi)^2}\mathrm{e}^{\mathrm{i}\mathbf{k}\cdot\boldsymbol{\rho}}\mathrm{e}^{-kz}\mathbf{b}_{\mathrm{out}}(\mathbf{k}, \omega).
\end{equation}

\section{Exterior magnetic field of pinned vortices}
\label{app:vortex_field}
\subsection{Single vortex field}
We consider straight, positively oriented, singly quantized vortices perpendicular to the superconducting surface, with fixed surface positions $\rho_\nu$. The phase winding of the superconducting order parameter supplies the source current
\begin{equation}
\mathbf{j}_{\nu}(\boldsymbol{\rho,z}) =\dfrac{\Phi_0}{2\pi\mu_0\lambda^2}\boldsymbol{\nabla}_{\boldsymbol{\rho}}\varphi_{\nu},
\end{equation}
where $\varphi_{\nu}$ denotes the vortex phase, which increases by $2\pi$ around a contour encircling the vortex core once in the positive sense about $\mathbf{e}_3$. Its transverse gradient is 
\begin{equation}
\boldsymbol{\nabla}_{\boldsymbol{\rho}}\varphi_{\nu}=\dfrac{\mathbf{e}_\phi}{|\boldsymbol{\rho}-\boldsymbol{\rho}_{\nu}|}=\dfrac{\mathbf{e}_3\times\left(\boldsymbol{\rho}-\boldsymbol{\rho}_{\nu}\right)}{ \left|\boldsymbol{\rho}-\boldsymbol{\rho}_{\nu} \right|^2}.
\end{equation}
Here, $\mathbf{e}_\phi$ is the azimuthal unit vector about the vortex core, directed counterclockwise when viewed from the vacuum toward the surface. The induced screening current is already included in the Maxwell-London response derived above. We consider the static field and suppress the frequency argument throughout this section.

Using the two-dimensional Fourier convention introduced in Eq.~\eqref{eq:app_2d_fourier} gives
\begin{equation}
\mathbf{j}_{\nu}(\mathbf{k},z) =-\dfrac{\mathrm{i}\Phi_0}{\mu_0\lambda^2 k}\mathrm{e}^{-\mathrm{i}\mathbf{k}\cdot\boldsymbol{\rho}_{\nu}}(\mathbf{e}_3\times\mathbf{e}_k).
\label{eq:phase_current}
\end{equation}
Since the source current is independent of depth, its weighted integral [Eq.~\eqref{eq:app_J_weighted_integral}] is $\mathbf{J}_\nu(\mathbf{k})=\mathbf{j}_\nu(\mathbf{k})/k_\lambda$. Substituting this expression into the exterior-field response in Eq.~\eqref{eqn:b-outside-fourier-eval} yields the surface amplitude
\begin{equation}\label{eq:app_b_out_single}
\mathbf{b}_{\nu,\mathrm{out}}(\mathbf{k}) = \dfrac{\Phi_0}{\lambda^2 k_\lambda(k+ k_\lambda)} \mathrm{e}^{-\mathrm{i}\mathbf{k}\cdot\boldsymbol{\rho}_{\nu}}(\mathbf{e}_3-\mathrm{i}\mathbf{e}_k).
\end{equation}
To obtain the real-space field from Eq.~\eqref{eq:app_b_outside_real}, we introduce the lateral separation $\mathbf{s}_\nu=\boldsymbol{\rho}-\boldsymbol{\rho}_\nu$, with $s_\nu=|\mathbf{s}_\nu|$ and $\mathbf{e}_s=\mathbf{s}_\nu/s_\nu$. Choosing the in-plane polar angle $\phi$ relative to $\mathbf{e}_s$, such that $\mathbf{e}_k=\cos\phi\,\mathbf{e}_x+\sin\phi\,\mathbf{e}_y$ and $\mathbf{e}_s=\mathbf{e}_x$, the angular Fourier integral gives
\begin{equation}
\mathbf{B}_{\nu}(\mathbf{r})=\dfrac{\Phi_0}{2\pi\lambda^2} \int_0^\infty \dfrac{\mathrm{d}k\,k\,\mathrm{e}^{-kz}}{k_\lambda(k+ k_\lambda)}\left[ J_0(ks_\nu)\mathbf{e}_3+J_1(ks_\nu)\mathbf{e}_s \right],
\end{equation}
where $J_0$ and $J_1$ denote Bessel functions of the first kind of orders zero and one, respectively. 

For observation heights $z\gg\lambda$, the exponential factor $\exp(-kz)$ selects wave numbers satisfying  $k\lambda\ll1$. Using $(k_\lambda-k)(k_\lambda+k)=\lambda^{-2}$, the screening factor can be written as
\begin{equation}
\dfrac{1}{\lambda^2 k_\lambda (k+k_\lambda)}=1-\dfrac{k}{k_\lambda},
\end{equation}
which approximates to $1-k\lambda$. Replacing this factor by $\exp(-k\lambda)$ retains the leading penetration-depth correction, with an error of order $(k\lambda)^2$. The remaining integrals follow by differentiating the identity
\begin{equation}
\int_0^\infty\mathrm{d}k\,\mathrm{e}^{-a k} J_0(ks) = \dfrac{1}{\sqrt{a^2+s^2}},\quad a>0,\;s\geq 0,
\end{equation}
with respect to $a$ and $s$, yielding
\begin{equation}
\mathbf{B}_{\nu}(\mathbf{r})\simeq\dfrac{\Phi_0}{2\pi} \dfrac{\boldsymbol{\rho}-\boldsymbol{\rho}_\nu+(z+\lambda)\mathbf{e}_3 }{\left[|\boldsymbol{\rho}-\boldsymbol{\rho}_\nu|^2+(z+\lambda)^2\right]^{3/2}}.
\label{eq:app_vortex_far_field}
\end{equation}
To this order, London screening therefore shifts the effective vortex endpoint to a depth $\lambda$ below the surface. The field decreases as the inverse square of the distance from this effective endpoint. Neglecting the penetration-depth correction gives the leading far-field expression used in the vortex-noise calculations.

\subsection{Field of multiple vortices}
For several vortices, the exterior fields add linearly. Using the signed surface density $n(\boldsymbol{\rho})$ defined in Eq.~\eqref{eqn:vortex-density-theory}, with Fourier transform
\begin{equation}
n(\mathbf{k})=\int\mathrm{d}^2\boldsymbol{\rho}\,\mathrm{e}^{-\mathrm{i}\mathbf{k}\cdot\boldsymbol{\rho}}n(\boldsymbol{\rho})=\sum_\nu w_\nu\mathrm{e}^{-\mathrm{i}\mathbf{k}\cdot\boldsymbol{\rho}_\nu},
\end{equation}
the sum of the single-vortex surface amplitudes in Eq.~\eqref{eq:app_b_out_single} becomes
\begin{equation}
\mathbf{b}_{\mathrm{fr},\mathrm{out}}(\mathbf{k}) = \dfrac{\Phi_0n(\mathbf{k})}{\lambda^2 k_\lambda(k+ k_\lambda)} (\mathbf{e}_3-\mathrm{i}\mathbf{e}_k).
\end{equation}
Integrating the single-vortex far-field expression over the surface density reproduces the real-space field in Eq.~\eqref{eq:vortex-field-main}. The normal surface amplitude, $\mathbf{e}_3\cdot\mathbf{b}_{\mathrm{fr},\mathrm{out}}$, completely determines the decaying exterior field and provides the matching condition for an effective frozen-magnetization representation.

\subsection{Effective frozen-magnetization representation}
\label{app:frozen_magnetization}
The exterior vortex field can equivalently be represented by an auxiliary magnetization  confined to the superconducting half-space. More generally, we consider a magnetic field that is source-free in $z>0$ and decays at infinity. Its normal surface amplitude, denoted by $b_3(\mathbf{k})$, determines the exterior field through
\begin{equation}
\mathbf{b}(\mathbf{k},z)=b_3(\mathbf{k})\mathrm{e}^{-kz}(\mathbf{e}_3-\mathrm{i}\mathbf{e}_k), \quad k>0.
\end{equation}
We seek an effective magnetization $\mathbf{M}_\mathrm{eff}$ whose field
\begin{equation}
\mathbf{B}_{\mathbf{M}}(\mathbf{r})=\mu_0\int_{z'<0}\mathrm{d}^3\mathbf{r}'\,\mathrm{G}(\mathbf{r}-\mathbf{r}')\mathbf{M}_{\mathrm{eff}}(\mathbf{r}'),\quad z>0.
\label{eq:app_frozen_magnetization_field}
\end{equation}
matches this exterior field. The effective magnetization need not coincide with the physical magnetization inside the superconductor. 

Applying the two-dimensional Fourier transform gives
\begin{equation}
\mathbf{b}_{\mathbf{M}}(\mathbf{k},z)=\mu_0\int_{z'<0}\mathrm{d}z'\,\tilde{\mathrm{G}}(\mathbf{k},z,z')\tilde{\mathbf{M}}_{\mathrm{eff}}(\mathbf{k},z'),
\end{equation}
where, for $z\neq z'$, the Fourier-transformed Green tensor is
\begin{align}
\tilde{\mathrm{G}}(\mathbf{k},z,z')=\dfrac{k}{2}\mathrm{e}^{-k|z-z'|}&[\mathbf{e}_3-\mathrm{i}\mathbf{e}_k\,\mathrm{sign}(z-z')]\nonumber\\
&\otimes[\mathbf{e}_3-\mathrm{i}\mathbf{e}_k\,\mathrm{sign}(z-z')],
\label{eq:app_green_fourier}
\end{align}
while $\tilde{\mathbf{M}}_{\mathrm{eff}}$ transforms according to Eq.~\eqref{eq:app_2d_fourier}. Since the observation point lies above the source distribution, $z-z'>0$, and hence 
\begin{align}
\mathbf{b}_{\mathbf{M}}(\mathbf{k},z)&=\dfrac{\mu_0k}{2}\mathrm{e}^{-kz}(\mathbf{e}_3-\mathrm{i}\mathbf{e}_k)\int_{-\infty}^{0}\mathrm{d}z'\,\mathrm{e}^{kz'}\nonumber\\
&\times\left[\tilde{M}_{\mathrm{eff},3}(\mathbf{k},z')-\mathrm{i}\mathbf{e}_k\cdot\tilde{\mathbf{M}}_{\mathrm{eff},\parallel}(\mathbf{k},z')\right],
\label{eq:app_magnetization_fourier_field}
\end{align}
where the subscripts $3$ and $\parallel$ denote the normal and in-plane components, respectively. Equality with the prescribed exterior field therefore requires
\begin{equation}\label{eq:app_Meff_integral_condition}
\int_{-\infty}^{0}\mathrm{d}z'\,\mathrm{e}^{kz'}\left[\tilde{M}_{\mathrm{eff},3}(\mathbf{k},z')-\mathrm{i}\mathbf{e}_k\cdot\tilde{\mathbf{M}}_{\mathrm{eff},\parallel}(\mathbf{k},z')\right]=\dfrac{2b_3(\mathbf{k})}{\mu_0k}.    
\end{equation}

The matching condition admits a family of effective magnetizations supported within a layer $-z_0<z'<-z_0+\delta$, where $z_0\geq \delta > 0$, provided that the corresponding inverse Fourier transform exists. We write 
\begin{equation}
\tilde{\mathbf{M}}_\mathrm{eff}(\mathbf{k},z')=\dfrac{2b_3(\mathbf{k})}{\mu_0N_\mathbf{M}(\mathbf{k})}\mathbf{m}_\mathrm{eff}(\mathbf{k})\chi_\mathrm{eff}(z'),
\end{equation}
where 
\begin{equation}
\chi_\mathrm{eff}(z')=\Theta(z'+z_0)-\Theta(z'+z_0-\delta)
\end{equation}
restricts the magnetization to the layer, and 
\begin{equation}
N_\mathbf{M} = k\int_{-\infty}^0\mathrm{d}z'\,\mathrm{e}^{kz'}\chi_\mathrm{eff}(z')=\mathrm{e}^{-k(z_0-\delta)}-\mathrm{e}^{-kz_0}
\end{equation}
is a dimensionless normalization factor. The dimensionless vector
\begin{equation}
\mathbf{m}_\mathrm{eff}(\mathbf{k})=\mathbf{m}_\parallel(\mathbf{k})+[1+\mathrm{i}\mathbf{e}_k\cdot\mathbf{m}_\parallel(\mathbf{k})]\mathbf{e}_3 
\end{equation}
is parametrized by an in-plane vector $\mathbf{m}_\parallel(\mathbf{k})$, with $\mathbf{e}_3 \cdot\mathbf{m}_\parallel=0$. Its components satisfy $m_\mathrm{eff,3}-\mathrm{i}\mathbf{e}_k\cdot\mathbf{m}_{\mathrm{eff},\parallel}=1$, so that the construction fulfills the matching condition in Eq.~\eqref{eq:app_Meff_integral_condition}. Different choices of $\mathbf{m}_\parallel$ produce the same exterior field, demonstrating the nonuniqueness of the effective magnetization.

For the vortex field, consider the auxiliary magnetization $\mathbf{M}_\mathrm{vort}$ defined in Eq.~\eqref{eq:effective-vortex-magnetization}. Since this magnetization has only a normal component, its contribution to the matching condition is
\begin{equation}
\dfrac{2\Phi_0 n(\mathbf{k})}{\mu_0}\int_{-\infty}^{-\lambda}\mathrm{d}z'\,\mathrm{e}^{k z'}=\dfrac{2\Phi_0 n(\mathbf{k})}{\mu_0 k}\mathrm{e}^{-k\lambda}.
\label{eq:app_frozen_magnetization_far_field}
\end{equation}
The corresponding normal surface amplitude is therefore $b_3(\mathbf{k})=\Phi_0 n(\mathbf{k})\mathrm{exp}(-k\lambda)$, which reproduces the long-wavelength vortex field in Eq.~\eqref{eq:vortex-field-main}, including the leading penetration-depth shift.

To connect the general construction with the frozen-image model, we now consider the normal surface field generated by the particle at the superconducting transition. In the dipole approximation, a particle at $z_\mathrm{fr}\mathbf{e}_3$, with moment $\mathbf{m}$, is represented by $\mathbf{M}_\mathrm{p}(\mathbf{r}')=\mathbf{m}\delta^{(2)}(\boldsymbol{\rho}')\delta(z'-z_{\mathrm{fr}})$. Evaluating its field at $z=0$ uses the Green tensor with $z-z'=-z_\mathrm{fr}<0$, giving the normal surface amplitude
\begin{equation}
b_{\mathrm{dip},3}(\mathbf{k})=\dfrac{\mu_0 k}{2}\mathrm{e}^{-k z_\mathrm{fr}}(\mathbf{e}_3\cdot\mathbf{m}+\mathrm{i}\mathbf{e}_k\cdot\mathbf{m}_\parallel),
\end{equation}
where $\mathbf{m}_\parallel=\mathbf{m}-(\mathbf{e}_3\cdot\mathbf{m})\mathbf{e}_3$. Consider an effective point dipole at the reflected position, $-z_\mathrm{fr}\mathbf{e}_3$, with dipole moment $\mathbf{m}_\mathrm{fr}=-\mathrm{A}\mathbf{m}$, represented by
\begin{equation}
\mathbf{M}_{\mathrm{fr}}(\mathbf{r}')=\mathbf{m}_{\mathrm{fr}}\delta^{(2)}(\boldsymbol{\rho}')\delta(z'+z_{\mathrm{fr}}).
\end{equation}
Since $\mathbf{e}_3\cdot\mathbf{m}=\mathbf{e}_3\cdot\mathbf{m}_\mathrm{fr}$, and $\mathbf{m}_{\mathrm{fr},\parallel}=-\mathbf{m}_\parallel$, its contribution to the matching condition is
\begin{align}
\int_{-\infty}^{0}\mathrm{d}z'&\,\mathrm{e}^{kz'}\left[\tilde{M}_{\mathrm{fr},3}(\mathbf{k},z')-\mathrm{i}\mathbf{e}_k\cdot\tilde{\mathbf{M}}_{\mathrm{fr},\parallel}(\mathbf{k},z')\right]\nonumber\\
&=\mathrm{e}^{-kz_\mathrm{fr}}(\mathbf{e}_3\cdot\mathbf{m}+\mathrm{i}\mathbf{e}_k\cdot\mathbf{m}_\parallel)=\dfrac{2b_{\mathrm{dip},3}(\mathbf{k})}{\mu_0k}.    
\end{align}
The inverse-image dipole thus exactly reproduces the prescribed normal surface field. If the pinned-flux configuration retains this normal surface field at the superconducting transition, its source-free, decaying exterior field is therefore
\begin{equation}
\mathbf{B}_\mathrm{fr}(\mathbf{r})=\mu_0\mathrm{G}(\mathbf{r}+z_\mathrm{fr}\mathbf{e}_3)(-\mathrm{A}\mathbf{m}).
\end{equation}
By superposition, the construction extends to an arbitrary particle magnetization $\mathbf{M}_\mathrm{p}^\mathrm{fr}$ at the superconducting phase transition, namely, $\mathbf{M}_\mathrm{fr}(\mathbf{r}')=-\mathrm{A}\mathbf{M}_\mathrm{p}^\mathrm{fr}(\mathrm{A}\mathbf{r}')$. Expressing this distribution in body coordinates gives Eq.~\eqref{eqn:def-frozen-magnetisation}.

\section{Noise correlations}
\label{app:noise-correlations}
\subsection{Conducting electrons}
\label{Noise correlations due to conducting electrons}
We derive the magnetic-field and field-gradient spectra generated by thermal current-density fluctuations of the normal electronic component, using the exterior-field response derived in App.~\ref{Magnetic field generated by superconductor noise currents} and the local current correlations specified in Eq.~\eqref{eqn:j-fourier-noise-3d}. 

For observation heights $z\gg\lambda$, the exponential decay of the Fourier components selects wavevectors satisfying $k\lambda\ll1$. To leading order, we therefore approximate 
\begin{equation}
k_\lambda\simeq \dfrac{1}{\lambda},\qquad \dfrac{k}{k+k_\lambda}\simeq k\lambda. 
\end{equation}
Expressing the source current in terms of the exponentially weighted, depth-integrated in-plane fluctuation $\delta\mathbf{J}_\mathrm{c}(\boldsymbol{\rho},\omega)$ defined in the main text, the established exterior field response [Eqs.~\eqref{eqn:b-outside-fourier-eval} and \eqref{eq:app_b_outside_real}] gives
\begin{align}
\delta\mathbf{B}_\mathrm{c}(\mathbf{r},\omega)\simeq & \mu_0\lambda\int\mathrm{d}^2\boldsymbol{\rho}'\int\dfrac{\mathrm{d}^2\mathbf{k}}{(2\pi)^2}k\mathrm{e}^{\mathrm{i}\mathbf{k}\cdot (\boldsymbol{\rho}-\boldsymbol{\rho}')-kz}\nonumber\\
&(\mathbf{e}_k+\mathrm{i}\mathbf{e}_3)\left[(\mathbf{e}_3\times\mathbf{e}_k)\cdot\delta\mathbf{J}_\mathrm{c}(\boldsymbol{\rho}',\omega)\right].
\end{align}
Using the Fourier-transformed Green tensor in Eq.~\eqref{eq:app_green_fourier} and the vector identity $(\mathbf{a}\otimes\mathbf{b})(\mathbf{c}\times\mathbf{d})=\mathbf{a}[(\mathbf{b}\times\mathbf{c})\cdot\mathbf{d}]$, we obtain
\begin{equation}
\tilde{\mathrm{G}}(\mathbf{k},z,0)(\mathbf{e}_3\times \delta\mathbf{J}_\mathrm{c}) = \dfrac{k}{2}\mathrm{e}^{-k z}(\mathbf{e}_k+\mathrm{i}\mathbf{e}_3)\left[(\mathbf{e}_3\times\mathbf{e}_k)\cdot\delta\mathbf{J}_\mathrm{c}\right]
\end{equation}
Identifying the inverse Fourier transform of the Green tensor yields the field expression in Eq.~\eqref{eq:B_fluctuation_electron}. The corresponding two-point field spectrum follows from the local current correlations in Eq.~\eqref{eqn:j-fourier-noise-3d}, with the weighted depth integral contributing
\begin{equation}
    \int_{-\infty}^0\mathrm{d}z \int_{-\infty}^0\mathrm{d}z'\,\mathrm{e}^{(z+z')/\lambda}\delta(z-z')=\dfrac{\lambda}{2}.
\end{equation}

To evaluate the remaining surface integral at coincident observation points, we introduce the lateral separation $\mathbf{s}=\boldsymbol{\rho}-\boldsymbol{\rho}''$. The explicit Green tensor gives
\begin{align}
&\mathrm{G}(\mathbf{s}+z\mathbf{e}_3)(\mathbb{1}-\mathbf{e}_3\otimes\mathbf{e}_3)\mathrm{G}(\mathbf{s}+z\mathbf{e}_3)=\dfrac{1}{16\pi^2(s^2+z^2)^5}\times\nonumber\\
&\times\left[3(s^2-2z^2)\mathbf{s}\otimes\mathbf{s}+3z(2s^2-z^2)(\mathbf{s}\otimes\mathbf{e}_3+\mathbf{e}_3\otimes\mathbf{s})\right.\nonumber\\
&\,\left.+(s^2+z^2)^2(\mathbb{1}-\mathbf{e}_3\otimes\mathbf{e}_3)+9s^2 z^2 \mathbf{e}_3\otimes\mathbf{e}_3\right].
\end{align}
Using polar coordinates, $\mathbf{s}=s(\cos\phi\mathbf{e}_1+\sin\phi\mathbf{e}_2)$ and $\mathrm{d}^2\mathbf{s}=s\mathrm{d}s\mathrm{d}\phi$, the angular integration yields
\begin{align}
&\int\mathrm{d}^2\mathbf{s}\,\mathrm{G}(\mathbf{s}+z\mathbf{e}_3)(\mathbb{1}-\mathbf{e}_3\otimes\mathbf{e}_3)\mathrm{G}(\mathbf{s}+z\mathbf{e}_3)=\nonumber\\
&\dfrac{1}{16\pi}\int_0^\infty \mathrm{d}s\,s\left[\dfrac{9s^4}{(s^2+z^2)^5}-\dfrac{6s^2}{(s^2+z^2)^4}+\dfrac{2}{(s^2+z^2)^3}\right]\nonumber\\
&(\mathbb{1}-\mathbf{e}_3\otimes\mathbf{e}_3)+\dfrac{9z^2}{8\pi}\int_0^\infty \mathrm{d}s\,\dfrac{s^3}{(s^2+z^2)^5}\mathbf{e}_3\otimes\mathbf{e}_3.
\end{align}
Evaluating the remaining radial integrals yields the coincident field spectrum in Eq.~\eqref{eq:S_bb_electrons}.

The field-gradient and gradient-gradient spectra are obtained by differentiating the two-point field spectrum with respect to the appropriate observation coordinates before taking coincident points. Evaluating the resulting integrals gives Eqs.~\eqref{eq:S_dbb_eletron} and \eqref{eq:S_dbdb_eletron}. For in-plane indices $i,j\in\{1,2\}$, with $i\neq j$, the nonzero field-gradient coefficients are
\begin{equation}\label{eqn:noise-tensor-components-lab-frame-E_abc}
E_{ii3} = E_{i3i} = - E_{3ii} = -\dfrac{1}{2}, \quad E_{333} = -1.
\end{equation}
The nonzero gradient-gradient coefficients are
\begin{gather}
\label{eqn:noise-tensor-components-lab-frame-E_abcd}
E_{ii33} = E_{33ii} = - E_{\{i3\}\{i3\}} = - \dfrac{1}{2}, \nonumber \\
E_{\{iijj\}} = \dfrac{1}{8}, \quad E_{iiii} = \dfrac{3}{8}, \quad E_{3333} = 1.
\end{gather}
Here, $E_{\{iijj\}}$ denotes every distinct permutation of the four enclosed indices, whereas the two pairs in $E_{\{i3\}\{i3\}}$ are permuted independently. No summation over repeated indices is implied in these component lists, and all unlisted coefficients vanish. Combining these field and gradient spectra according to Eq.~\eqref{eq:lab-frame-noise-forcing} gives the generalized-force noise spectrum, which is transformed to the normal-mode basis using Eq.~\eqref{eqn:psd-noise-lab-to-normalmodes}.

\subsection{Vortex fluctuations}
\label{app:vortex_fluctuations}
We first derive the magnetic-field fluctuations produced by the thermal motion of pinned vortices within the rigid-displacement approximation introduced in Sec.~\ref{Thermal vortex motion}. The static exterior-field response derived in App.~\ref{app:vortex_field} is evaluated at the instantaneous vortex position. The vortex position enters the Fourier-space surface amplitude through the phase factor $\mathrm{exp}[-\mathrm{i}\mathbf{k}\cdot\boldsymbol{\rho}_\nu(t)]$. For displacements small compared with the spatial variation scale of the field, we approximate
\begin{equation}
\mathrm{e}^{-\mathrm{i}\mathbf{k}\cdot\boldsymbol{\rho}_\nu(t)}\simeq\mathrm{e}^{-\mathrm{i}\mathbf{k}\cdot\bar{\boldsymbol{\rho}}_\nu}\left[1-\mathrm{i}\mathbf{k}\cdot\delta\mathbf{u}_\nu^\mathrm{th}(t)\right].
\end{equation}
Thus, writing $\mathbf{b}_{\nu,\mathrm{out}}^{(0)}(\mathbf{k})$ for the static surface amplitude in Eq.~\eqref{eq:app_b_out_single}, evaluated at the mean vortex position, its fluctuation is
\begin{equation}
\delta\mathbf{b}_{\nu,\mathrm{out}}^\mathrm{th}(\mathbf{k},t)=-\left[\mathrm{i}\mathbf{k}\cdot\delta\mathbf{u}_\nu^\mathrm{th}(t)\right]\mathbf{b}_{\nu,\mathrm{out}}^{(0)}(\mathbf{k})
\end{equation}
to first order in the displacement. Since the displacement has zero mean, the static field at the mean position also equals the mean field to this order.

Applying the inverse transverse Fourier transform gives
\begin{equation}
\delta\mathbf{B}_\nu^\mathrm{th}(\mathbf{r},t)=-\int\dfrac{\mathrm{d}^2\mathbf{k}}{(2\pi)^2}\,\mathrm{e}^{\mathrm{i}\mathbf{k}\cdot\boldsymbol{\rho}-kz}\left[\mathrm{i}\mathbf{k}\cdot\delta\mathbf{u}_\nu^\mathrm{th}(t)\right]\mathbf{b}_{\nu,\mathrm{out}}^{(0)}(\mathbf{k}).
\end{equation}
Identifying $\mathrm{i}\mathbf{k}$ with the lateral gradient acting on the Fourier phase yields the expression for the magnetic-field fluctuation given in the main text. This relation follows from lateral translation of the field profile and applies before taking the far-field approximation.

For $z\gg\lambda$, we use the static far-field expression in Eq.~\eqref{eq:app_vortex_far_field} and neglect the penetration-depth correction, consistently with the vortex-noise calculation in the main text. Differentiation gives
\begin{equation}
\delta\mathbf{B}_\nu^\mathrm{th}(\mathbf{r},t)\simeq\dfrac{\Phi_0}{2\pi}\left[\dfrac{3(\mathbf{s}_\nu+z\mathbf{e}_3)\left[\mathbf{s}_\nu\cdot\delta\mathbf{u}_\nu^\mathrm{th}\right]}{(s_\nu^2+z^2)^{5/2}}-\dfrac{\delta\mathbf{u}_\nu^\mathrm{th}}{(s_\nu^2+z^2)^{3/2}}\right],
\label{eq:app_vortex_field_fluctuation_far_field}
\end{equation}
where retaining the leading penetration-depth shift amounts to replacing $z$ by $z+\lambda$.

The displacement statistics follow from the overdamped Langevin equation in Eq.~\eqref{eq:Langevin_u}. For independent vortices with isotropic in-plane damping, the classical equilibrium fluctuation-dissipation relation fixes the thermal-force correlation as
\begin{equation}
\mathbb{E}\left[\boldsymbol{\eta}_\nu(t)\otimes\boldsymbol{\eta}_\mu(t')\right]=2k_\mathrm{B}T c_\nu\,\delta_{\nu\mu}\delta(t-t')(\mathbb{1}-\mathbf{e}_3\otimes\mathbf{e}_3).
\end{equation}
Fourier transformation of the Langevin equation gives
\begin{equation}
\delta\mathbf{u}_\nu^\mathrm{th}(\omega)=\dfrac{\boldsymbol{\eta}_\nu(\omega)}{K_\nu-\mathrm{i}\omega c_\nu},
\end{equation}
and the resulting displacement correlation is
\begin{align}
\mathbb{E}\left[\delta\mathbf{u}_\nu^\mathrm{th}(\omega)\otimes\delta\mathbf{u}_\mu^{\mathrm{th}\ast}(\omega')\right]=&\dfrac{4\pi k_\mathrm{B}T c_\nu}{K_\nu^2+\omega^2c_\nu^2}\delta_{\nu\mu}\delta(\omega-\omega')\nonumber\\
&\times(\mathbb{1}-\mathbf{e}_3\otimes\mathbf{e}_3),
\end{align}
with relaxation rate $K_\nu/c_\nu$. For $|\omega|\ll K_\nu/c_\nu$, the displacement spectrum is approximately frequency independent, recovering the correlation stated in the main text. 

To obtain the magnetic-field spectrum, we combine these displacement correlations with the linearized field response. Summing the individual vortex-field fluctuations as in Sec.~\ref{Thermal vortex motion}, we obtain, for field components $k,\ell\in\{1,2,3\}$,
\begin{align}
&\mathbb{E}\left[\delta B_k^\mathrm{th}(\mathbf{r},\omega)\delta B_\ell^{\mathrm{th}\ast}(\mathbf{r}',\omega')\right]=\sum_{\nu,\mu}\sum_{a,b=1}^{2}[\partial_{\rho_a}B_{\nu,k}(\mathbf{r})]\nonumber\\
&\times[\partial_{\rho'_b}B_{\mu,\ell}(\mathbf{r}')]\mathbb{E}\left[\delta u_{\nu,a}^\mathrm{th}(\omega)\delta u_{\mu,b}^{\mathrm{th}\ast}(\omega')\right].
\end{align}
Independence of the vortex displacements eliminates terms with $\nu\neq\mu$, while in-plane isotropy sets $a=b$. Extracting $2\pi\delta(\omega-\omega')$ therefore yields
\begin{align}
S_{B_kB_\ell}^\mathrm{th,v}(\mathbf{r},\mathbf{r}',\omega)&=\sum_\nu\dfrac{2k_\mathrm{B}T c_\nu}{K_\nu^2+\omega^2c_\nu^2}\nonumber\\
&\times\sum_{a=1}^{2}[\partial_{\rho_a}B_{\nu,k}(\mathbf{r})][\partial_{\rho'_a}B_{\nu,\ell}(\mathbf{r}')].
\end{align}
The low-frequency limit reproduces the thermal-vortex field spectrum in Eq.~\eqref{eq:S_bb_vortex_thermal}, with the sum over $a$ forming the scalar product of the two lateral gradients. For $s_\nu\ll z$, the leading in-plane field fluctuations in Eq.~\eqref{eq:app_vortex_field_fluctuation_far_field} scale as $z^{-3}$, yielding the $z^{-6}$ spectral scaling stated in the main text.

\section{Determination of damping rates}
\label{app:damping_rates}
We derive the matrix fluctuation-dissipation relation for the linearized dynamics by requiring the thermal equilibrium distribution at temperature $T$ to be stationary under the Langevin dynamics. In terms of the canonical phase-space coordinates $\mathbf{z}=(\boldsymbol{\theta},\boldsymbol{\pi})$, the quadratic Hamiltonian~\eqref{eq:H_lin_SpinRotation} in matrix notation reads
\begin{equation}
H_\mathrm{lin}(\mathbf{z})=\dfrac{1}{2}\mathbf{z}^\mathrm{T}\mathrm{H}\mathbf{z},  
\end{equation}
where $\mathrm{H}=\mathrm{H}^\mathrm{T}$ is the Hamiltonian Hessian,
\begin{align}
\mathrm{H}=\begin{pmatrix}
\Upomega^2-\Lambda^2&-\Lambda\\
\Lambda&\mathbb{1}
\end{pmatrix}, \quad \nabla_\mathbf{z}H_\mathrm{lin}=\mathrm{H}\mathbf{z}.
\end{align}
Here, $(\Upomega^2)_{ij}=\omega_i^2\delta_{ij}$ and $\Lambda^\mathrm{T}=-\Lambda$. We include linear velocity-dependent damping and additive Gaussian white noise through
\begin{align}\Upgamma_\mathbf{z}=
\begin{pmatrix}
0&0\\
0&\Upgamma
\end{pmatrix},
\qquad
\mathrm{D}_{\mathbf z}=
\begin{pmatrix}
0&0\\
0&\mathrm{D}
\end{pmatrix},
\qquad
\boldsymbol{\eta}=
\begin{pmatrix}
0\\
\boldsymbol{\eta}^{\mathrm{th}}
\end{pmatrix}.
\end{align}
The damping matrix $\Upgamma$ and the momentum-diffusion matrix $\mathrm{D}$ are constant, real, symmetric and positive semidefinite. 
The noise has zero mean and correlations $\mathbb{E}[\boldsymbol{\eta}(t)\boldsymbol{\eta}(t')^\mathrm{T}]=2\mathrm{D}_\mathbf{z}\delta(t-t')$. The stochastic equations of motion then read
\begin{align}
\dot{\mathbf{z}}=(\mathrm{J}-\Upgamma_\mathbf{z})\mathrm{H}\mathbf{z}+\boldsymbol{\eta}(t),\quad \mathrm{J}=
\begin{pmatrix}
0&\mathbb 1\\
-\mathbb 1&0
\end{pmatrix},
\end{align}
with $\mathrm{J}$ denoting the canonical symplectic matrix.

The corresponding Fokker-Planck equation is
\begin{equation}
\partial_t f_t = \{H_\mathrm{lin},f_t\}+\nabla_\mathbf{z}\cdot[\Upgamma_\mathbf{z} (\nabla_\mathbf{z}H_\mathrm{lin})f_t+\mathrm{D}_\mathbf{z}\nabla_\mathbf{z}f_t].
\end{equation}
For the thermal equilibrium distribution
\begin{equation}
f_\mathrm{eq}(\mathbf{z})=\dfrac{1}{Z}\mathrm{e}^{-H_\mathrm{lin}(\mathbf{z})/k_\mathrm{B}T},
\end{equation}
where $Z$ denotes the partition function, the Poisson-bracket term vanishes. To impose stationarity, we substitute this distribution into the Fokker-Planck equation and set $\partial_t f_t=0$. Writing
\begin{equation}
f_\mathrm{eq}(\mathbf{z})=g[H_\mathrm{lin}(\mathbf{z})],\quad g'=-\dfrac{g}{k_\mathrm{B} T},\quad g''=-\dfrac{g'}{k_\mathrm{B} T},
\end{equation}
where primes denote derivatives with respect to the scalar argument of $g$, gives
\begin{align}
0=& g'' \nabla_\mathbf{z} H_\mathrm{lin}\cdot[(\mathrm{D}_\mathbf{z}-k_\mathrm{B}T\Upgamma_\mathbf{z})\nabla_\mathbf{z} H_\mathrm{lin}]\nonumber\\ 
&+ g'\nabla_\mathbf{z}\cdot[(\mathrm{D}_\mathbf{z}-k_\mathrm{B}T\Upgamma_\mathbf{z})\nabla_\mathbf{z} H_\mathrm{lin}].
\end{align}
For the quadratic Hamiltonian and constant damping and diffusion matrices considered here, the divergence in the second term is independent of $\mathbf{z}$. Evaluating the stationarity condition at $\mathbf{z}=0$, where $\nabla_\mathbf{z}H_\mathrm{lin}=0$ and $g'\neq0$, therefore requires this divergence to vanish. The remaining quadratic form must consequently vanish for every $\mathbf{z}$. Since the damping and diffusion matrices are symmetric and act only on the momentum coordinates, this implies the matrix fluctuation-dissipation relation
\begin{equation}
\mathrm{D}=k_\mathrm{B}T\Upgamma. 
\end{equation}
With the two-sided spectral-density convention used here, this gives $S_{\eta_i\eta_j}^{\mathrm{th}}=2D_{ij}=2k_{\mathrm B}T\Gamma_{ij}$.

\end{document}